%% file: main.tex
\documentclass[12pt]{iopart}
\usepackage[margin=1in]{geometry}
\usepackage{tikz}
\usepackage{graphicx}
\usepackage{url}
\usepackage{fancyvrb}
\usepackage{color, soul}
\usepackage{tikz}
\usepackage{hyperref}
\usepackage{float}
\usepackage{changepage}

\usetikzlibrary{shapes,arrows,positioning,shapes.geometric}
\usepackage{charter}
\usepackage{siunitx}
\usepackage{amssymb, bm}
\usepackage{graphicx}
\usepackage{color}
\usepackage{xcolor}
\usepackage{afterpage}
\usepackage{cleveref}
\usepackage{array,multirow}

\usepackage{struktex}

\usepackage{algorithm}
\usepackage{algpseudocode}
\usepackage{caption}
\usepackage{subcaption}
\usepackage{xspace}
\usepackage[colorinlistoftodos, prependcaption, textsize=tiny]{todonotes}

\usepackage{array}
\newcolumntype{P}[1]{>{\centering\arraybackslash}p{#1}}
\newcolumntype{M}[1]{>{\centering\arraybackslash}m{#1}}

\newcommand{\degree}{\ensuremath{^\circ}\xspace}

\newcommand{\BE}[0]{\begin{equation}}
\newcommand{\EE}[0]{\end{equation}}
\newcommand{\BEA}[0]{\begin{eqnarray}}
\newcommand{\EEA}[0]{\end{eqnarray}}

\DeclareCaptionFont{tenpt}{\fontsize{10pt}{12pt}\selectfont}
\mathchardef\mhyphen="2D

\newcommand{\orcid}[1]{\href{https://orcid.org/#1}{\includegraphics[height=\fontcharht\font`\B]{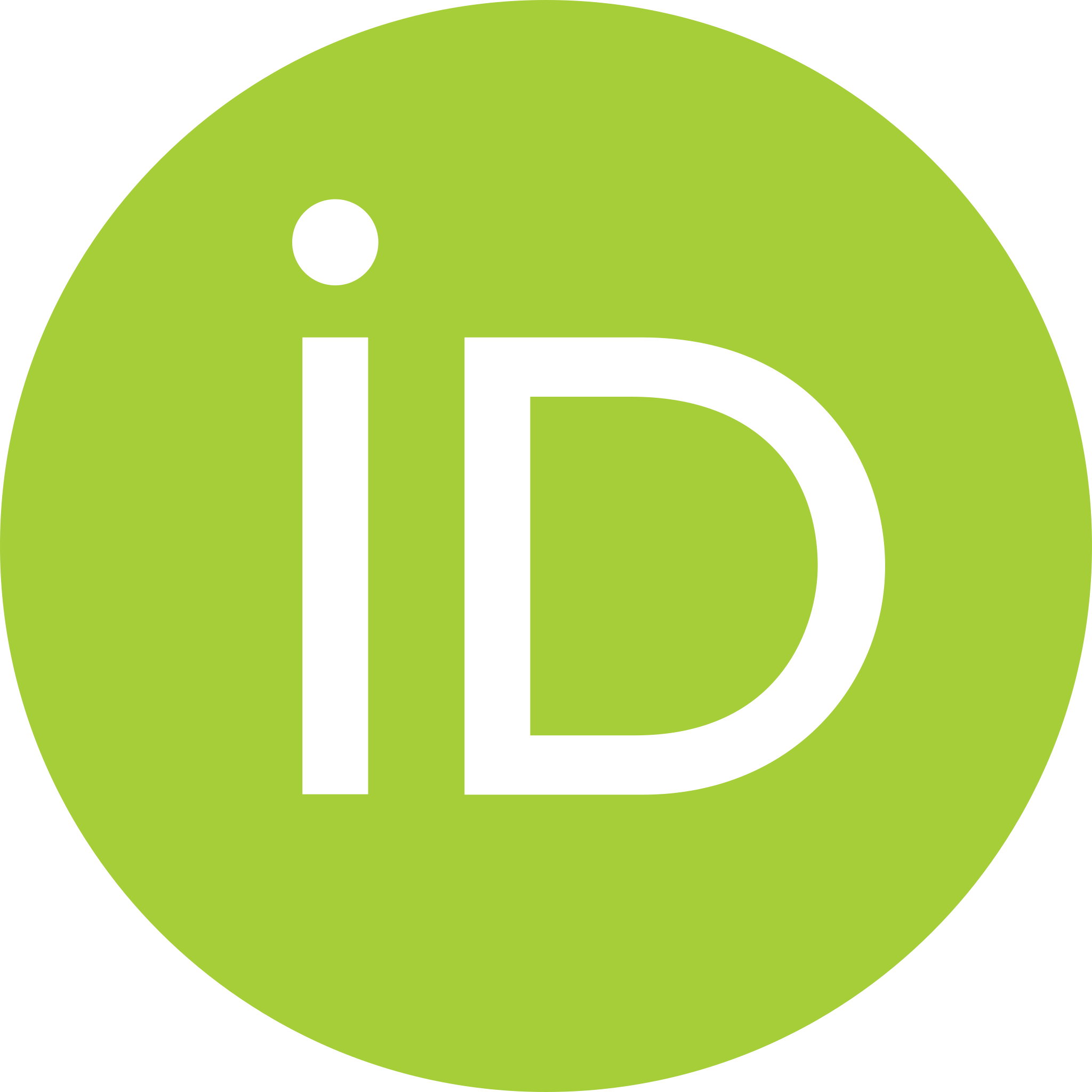}}}

\renewcommand{\hl}[1]{#1}

\begin{document}


\title[A Frequentist Simulation-Based Inference Treatment of Sterile Neutrino Global Fits]
      {A Frequentist Simulation-Based Inference Treatment of Sterile Neutrino Global Fits}

\author{Joshua Villarreal$^{*}$ \orcid{0000-0001-9690-1310}, Julia Woodward$^{*}$ \orcid{0009-0006-1636-7562}, John M. Hardin \orcid{0000-0001-8871-8065},
        Janet M. Conrad \orcid{0000-0002-6393-0438}}
\address{Department of Physics, Massachusetts Institute of Technology, 77 Massachusetts Ave, 
             Cambridge, MA 02139, USA
             
             $^*$These authors contributed equally to this work.}
\ead{villaj@mit.edu, julia785@mit.edu}

\vspace{10pt}


\begin{abstract}
A critical challenge in particle physics is combining results from diverse experimental setups that measure the same physical quantity to enhance precision and statistical power, a process known as a global fit. Global fits of sterile neutrino searches, hunts for additional neutrino oscillation frequencies and amplitudes, present an intriguing case study. In such a scenario, the key assumptions underlying Wilks' theorem, a cornerstone of most classic frequentist analyses, do not hold. The method of Feldman and Cousins, a trials-based approach which does not assume Wilks' theorem, becomes computationally prohibitive for complex or intractable likelihoods. To bypass this limitation, we borrow a technique from simulation-based inference (SBI) to estimate likelihood ratios for use in building trials-based confidence intervals, speeding up test statistic evaluations by a factor $>10^4$ per grid point, resulting in a faster, but approximate, frequentist fitting framework. Applied to a subset of sterile neutrino search data involving the disappearance of muon-flavor (anti)neutrinos, our method leverages machine learning to compute frequentist confidence intervals while significantly reducing computational expense.  In addition, the SBI-based approach holds additional value by recognizing underlying systematic uncertainties that the Wilks approach does not.  Thus, our method allows for more robust machine learning-based analyses critical to performing accurate but computationally feasible global fits. This allows, for the first time, a global fit to sterile neutrino data without assuming Wilks' theorem. While we demonstrate the utility of such a technique studying sterile neutrino searches, it is applicable to both single-experiment and global fits of all kinds.  
\end{abstract}


\maketitle


\input{Sec1_Introduction}
\input{Sec2_Methods}
\input{Sec3_Results}
\input{Sec4_Discussion}

\input{Sec5_Conclusion}
\appendix
\input{Appendix}
\section*{Acknowledgments}
JV, JW, JH and JMC thank MIT for support on this project. This material is based upon work supported in part by the National
Science Foundation Graduate Research Fellowship under Grant No. 2141064.

\newpage

\section*{References}
\bibliographystyle{iopart-num}
\bibliography{Bibliography}

\end{document}

%% file: Sec1_Introduction.tex
\section{\label{sec:introduction}Introduction}

\subsection{Overview}

Indications of new particles, fundamental forces, and symmetries can be observed via deviations of interactions and decays from the prediction of the Standard Model (SM), the highly successful present theory for particle physics, and through consistency with phenomenological models that introduce new physics. Such models depend on corresponding parameters whose measurements can naturally interpreted by classical frequentist hypothesis testing: null being consistency with the Standard Model. Robustness and significance of the fit result is usually quantified through the construction of \textit{confidence levels} (CLs).

The basic procedure entails constructing a likelihood function, evaluating it over the model parameter space, computing the log-likelihood ratio with respect to a null or signal-absent hypothesis, and interpreting this statistic using a $\chi^2$ distribution as motivated by Wilks’ theorem \cite{Wilks1938-yc}. While computationally efficient and analytically convenient, this approach can result in CLs which are potentially poorly calibrated, particularly when parameters approach boundaries,  when data is sparse, when the number of effective degrees of freedom varies, or when non-Gaussian systematic errors are present \cite{Hardin:2022qdh, Algeri2020-gv}.

In response to these and other issues, Feldman and Cousins introduced a trials-based method that more accurately reflects the fixed nature of model parameters underlying random data, offering a method to smoothly transition between setting upper limits and measuring two-sided confidence intervals by using the test statistic to build CLs  numerically \cite{feldman-cousins}. At each value of the model parameter(s), many pseudoexperiments are generated and fit, yielding a distribution of the test statistic (often the log-likelihood ratio or some variant) at each grid point. One can empirically, then, construct critical test statistic values which yield desired CLs at arbitrary confidence.  For each parameter point, outcomes are accepted into the confidence region based on their test statistic ranking when compared to the computed critical test statistic value, ensuring the total accepted probability matches the desired confidence level. This procedure is given in Alg.~\ref{alg:feldman-cousins}.

While this method has fewer biases than the simple Wilks' $\chi^2$-based method, the Feldman-Cousins approach is slow for two reasons.   First, it requires a large number of pseudoexperiments to be generated in a fine grid across the many testable parameter values of the model. Second, computations of the test statistic, which often require a likelihood minimization, can be computationally expensive. As a result, analyzers are often limited to using Feldman-Cousins-like procedures only at certain points in parameter space as a spot-checking tool.  As an example, for the IceCube experiment's recently-reported search for a new, non-interacting neutrino---a case that we will revisit in later discussion---the Feldman-Cousins probability is provided only for the best fit point \cite{icecube-sterile-prl}.  One can see that a trials-based approach, as it is implemented today, is not a 
panacea for frequentist fitting ills.

In this work, we show that the Feldman-Cousins method can be accelerated substantially by estimating its required test statistic with machine learning (ML), replacing the classic likelihood-minimization methods with a faster approximation which can alleviate Feldman-Cousins' prohibitive computational overhead. To construct a test statistic, we implement a technique from the ML field of \textit{simulation-based inference} (SBI), wherein we train a neural network classifier on Monte Carlo–simulated data and convert the network's output to a likelihood ratio between two points in parameter space. Querying the network for the best fit point, we can then compute an estimate of the log-likelihood ratio and plug-and-play into the numerical construction of CLs described in Alg.~\ref{alg:feldman-cousins}.  In the example we provide, we show this approach runs more than four orders of magnitude faster per grid point when compared to a traditional approach.

\subsection{Feldman-Cousins Applications in Sterile Neutrino Searches as an Example}

Feldman and Cousins developed their method for neutrino oscillation studies, where it remains widely used today. In a similar spirit, we introduce our simulation-based inference fitting procedure for new physics searches using muon-neutrino disappearance data. While our focus is on oscillation-driven scenarios, the method is broadly applicable to other types of beyond-Standard Model (BSM) searches as well.

The Standard Model of particle physics describes three types of neutrinos—electron, muon, and tau—each defined by the charged lepton they are associated with in charged-current weak interactions. Neutrinos propagate through space according to one of three mass eigenstates, an eigenbasis misaligned with the flavor basis, causing a quantum mechanical phenomenon known as \textit{neutrino oscillations}. Neutrino oscillations refer to the property that a neutrino (or its antimatter counterpart, an antineutrino) produced in one flavor may morph into a neutrino of a different flavor as it propagates through space and time. The survival probability of a neutrino of a particular flavor is a sinusoidal function of the ratio of the distance traveled by the neutrino (sometimes called \textit{baseline}) to its energy, $L/E$. In a simplified two-neutrino model, the survival probability of a muon neutrino, for instance, may be parameterized by a \textit{mixing angle} $\sin^2 (2 \theta)$ and a \textit{mass splitting} $\Delta m^2$, corresponding to an amplitude and a frequency, respectively:

\begin{equation}
    P_{\nu_\mu \to \nu_\mu} = 1 - \sin^2 (2 \theta) \sin^2 (1.27 \Delta m^2\,[\text{eV}^2] \, L\,[\text{km}] / E\,[\text{GeV}]).
\end{equation}
The mixing angle can also be written in terms of the mixing strength between a flavor and mass eigenstate $U_{\alpha i}$,
\begin{equation}
    \sin^2 (2 \theta) = 4|U_{\alpha i}|^2(1-|U_{\alpha i}|^2),
\end{equation}
an element of the Pontecorvo–Maki–Nakagawa–Sakata matrix in the SM picture.

Since 1993, neutrino experiments of all kinds have observed excesses or deficits in predicted neutrino events in significant tension with the existing SM \cite{wherearewe}. These anomalies, coupled with open theoretical questions in the neutrino sector, may be resolved through BSM models suggesting the existence of one or more additional \textit{sterile} neutrinos. Such a neutrino must not participate in the weak force, rendering its indirect detection exceptionally difficult. The simplest sterile neutrino oscillation model involves the addition of one sterile neutrino to the three active SM neutrino flavors, called a $3+1$ model. While it has not yet been shown to completely relieve data-SM expectation tensions, it leads to a significant 5$\sigma$ improvement \cite{wherearewe}. 

\begin{table} 

\caption{\label{tab:experiments}Experimental details and references of muon-(anti)neutrino disappearance experiments considered in this work.}

\begin{tabular*}{\textwidth}{lcccc}
\br
Name&$\nu/\overline{\nu}$&$N_{bins}$ & $L$ [km]& $E$\\
\mr
CDHS \cite{DYDAK1984281} &  $\nu$ & 15 &0.130, 0.885 & 500-7000 MeV \\
CCFR \cite{Stockdale:1984ce}&  $\nu + \overline{\nu}$ & 18  &  0.715, 1.116 & 40-230 GeV \\
MINOS/MINOS+ \cite{minos-two-detector} & $\nu$ &145 & 1.04, 735 & 3 GeV, 7 GeV \\
SciBooNE/MiniBooNE \cite{PhysRevD.85.032007, PhysRevD.86.052009} &  $\nu + \overline{\nu}$ &90 & 0.100, 0.540 & 0.8 GeV \\
\br
\end{tabular*}

\end{table}

The accelerator experiments we consider in this study, as summarized in Table ~\ref{tab:experiments}, make inferences on sterile neutrino existence using measurements of anomalous muon-flavor disappearance. In a typical accelerator experiment, a proton beam impinges on a target producing a flux of mesons which decay to primarily muon (anti)neutrinos, which undergo oscillations as they propagate toward the detector. Upon reaching the detector, neutrinos interact weakly with the target material, producing detectable charged particles. Each experiment we consider uses a combination of near and far detectors to compare neutrino fluxes at various distances, and constrain backgrounds and flux from other particle interactions. An experiment searching for sterile neutrinos will use its flux measurement as a function of baseline and energy to place constraints on the parameter space associated with oscillations between sterile and active neutrinos. For the family of muon-neutrino disappearance experiments considered here, the parameters probed are the mixing strength of the sterile neutrino with the muon neutrino, $U_{\mu 4}$, and the sterile mass splitting relative to the first three approximately-degenerate mass eigenstates, $\Delta m_{41}^2$.  MINOS and MINOS+ are treated together in this work \cite{minos-two-detector}. Although SciBooNE and MiniBooNE are separate experiments, they lie on the same beamline, allowing SciBooNE to act as a near detector ($100\/\text{m}$ away from the target) and MiniBooNE as a far detector ($540\/\text{m}$ away from the target). The SciBooNE/MiniBooNE collaboration reported exclusion regions for both $\nu_\mu$ disappearance and $\bar{\nu}_\mu$ disappearance, so we consider SciBooNE/MiniBooNE $\nu_\mu$ and $\bar{\nu}_\mu$ separately before combining all experimental results into a \textit{global fit} \cite{PhysRevD.85.032007, PhysRevD.86.052009}. None of these experiments report evidence for a sterile neutrino signal, but rather present regions of confidence that exclude certain parts of the $U_{\mu 4}, \Delta m_{41}^2$ parameter space. 

Surprisingly, IceCube, a neutrino telescope constructed at the Amundsen-Scott South Pole Station, recently reported a $3+1$ search leveraging muon-flavor disappearance that identified an allowed region at $95\%$ confidence \cite{icecube-sterile-prl, icecube-sterile-prd}. However, its region of preferred parameter space should be excluded by the combined global fit results of the experiments presented in this study. 
To interpret the meaning of the overlap of the allowed and excluded regions, one must
compare the relative probabilities of the IceCube signal best-fit to the probability of exclusion from global fit. 
Such an exercise is illustrative of the complexity and importance of accurate global fits. Given the computational limitations of high-fidelity fitting methods like Feldman-Cousins, we analyze this region of parameter space using our newly developed simulation-based inference procedure to shed some light on the problem.

\begin{algorithm}[t]
\caption{The Feldman-Cousins procedure for generating confidence intervals for a neutrino oscillation experiment \cite{feldman-cousins}.}
\label{alg:feldman-cousins}
\begin{algorithmic}[1]
\State \textbf{Input:} A grid of points in the $(\sin^2(2\theta), \Delta m^2)$ plane;  Test statistic $T(\mathbf{x}|(\sin^2 2\theta, \Delta m^2))$; Confidence level $\alpha$; Observed experimental data $\mathbf{x}_\text{obs}$
\State \textbf{Output:} Acceptance for each grid point
\For{each grid point $(\sin^2(2\theta), \Delta m^2)$}
    \State Simulate a large number of experiments $\{ \mathbf{x}_{\text{sim}, i} \}_{i=1}^N $using values at grid point 
    \For{each simulated experiment $\mathbf{x}_{\text{sim}}$}
        \State Compute the test statistic: $T(\mathbf{x}_\text{sim} | (\sin^2 (2 \theta), \Delta m^2))$
    \EndFor
    \State Determine critical $T_c$ such that a fraction $\alpha$ of simulations satisfy $T < T_c$
    \State Compute $T(\mathbf{x}_\text{obs} | (\sin^2 2 \theta, \Delta m^2))$
    \If{$T(\mathbf{x}_\text{obs} | \sin^2(2\theta), \Delta m^2) < T_c(\sin^2(2\theta), \Delta m^2)$}
        \State Accept the point into the confidence region
    \EndIf
\EndFor
\end{algorithmic}
\end{algorithm}

\subsection{\label{subsec:sbi}Simulation-Based Inference}

The difficulty of fast and accurate analysis of particle physics data is not limited to the field of sterile neutrino global fits. Besides the fact that some analyses, such as oscillation measurements, do not follow the assumptions necessary for Wilks' theorem, the calculation of likelihoods and corresponding test statistics are often computationally intractable. Alternative Bayesian analyses rely on Markov-Chain Monte Carlo (MCMC) sampling methods which may also be slow. With access to a generative model, the suite of techniques collectively known as \textit{simulation-based inference} enable inference strategies without the need for a tractable likelihood or a fast MCMC sampler.

Such likelihood-free SBI techniques include Approximate Bayesian Computation, neural density estimation of likelihoods such as normalizing flows, likelihood ratio estimation, and posterior density estimation using neural networks. In any such case, the workflow is somewhat consistent: a simulator $p(\textbf{x}|\theta)$ is used to generate sampled experimental data $\textbf{x}$ given model parameters $\theta$; a surrogate model is trained on sampled data from the simulator to approximate a likelihood, likelihood ratio, or posterior distribution; and finally, observed experimental data $\textbf{x}_\text{obs}$ is fed to the trained surrogate model to perform inference. For an overview of SBI techniques used in particle physics and otherwise, see Ref.~\cite{doi:10.1073/pnas.1912789117}.

While SBI bypasses some of the computational difficulties associated with likelihood or posterior computations, it is not without its challenges; in high-dimensional settings, large samples of training data are required to exhaust the possible model parameter space. Validation can also be difficult in settings such as sterile neutrino global fits, since exact solutions are often unknown. In this case, we present qualitative performance of the method by presenting fits to known-parameter Monte Carlo data to show agreement with expectations, and show the CL construction's calibration through plots of empirical coverage versus expected confidence.

%% file: Sec2_Methods.tex
\section{\label{sec:methods}Methods}

\subsection{\label{subsec:datagen}Data Generation}
Training data was generated using Monte Carlo simulations with the \texttt{sblmc} software suite, described in detail in Refs.~\cite{wherearewe, Hardin2023}. \texttt{sblmc} uses either covariance matrices or pull parameters from each experiment’s published $3+1$ fits and data releases. For each $\nu_\mu / \bar{\nu}_\mu$ disappearance experiment, parametrized by $U_{\mu 4}$ and $\Delta m_{41}^2$, model predictions are combined with these experiment-specific cost functions to produce synthetic datasets. When covariance matrices are available, correlated residuals are sampled in a diagonalized basis. For experiments using pull parameters, realizations are generated by sampling these parameters (with any provided correlations) and reconstructing the model predictions.

Training data is generated on a $50 \times 50$ grid of points linear in $\log U_{\mu 4}, \log \Delta m_{41}^2$. At each grid point, we generate $1000$ realizations. Then, $70\%$ of those are used for network training, with $10\%$ reserved for validation for the hyperparameter optimization, and the remaining $20\%$ for the out-of-sample test set. For CDHS, however, a greater training/testing split ($80/10$) was used to improve stability of neural network-computed test statistic comparison to critical values.

\subsection{\label{subsec:dnre}Direct Amortized Neural Likelihood Ratio Estimation}

\begin{algorithm}[t]
\caption{The neural network training strategy for a Direct Neural Ratio Estimator (DNRE) \cite{10.1609/aaai.v38i18.30018}.}
\label{alg:dnre}
\begin{algorithmic}[1]
\State \textbf{Input:} Loss criterion $l$ (Binary Cross-Entropy); Implicit generative model $\mathbb{P} (\mathbf{x}|\theta)$; Prior $\mathbb{P} (\theta)$; Number of training steps $N$; Batch size $M$
\State \textbf{Output:} Parameterized classifier $\mathbf{d}_\phi (\mathbf{x}, \theta, \theta')$
\For{$i=1$ to $N$}
    \State $\theta \leftarrow \{ \theta_j \sim \mathbb{P} (\theta) \}_{j=1}^M $
    \State $\theta' \leftarrow \{ \theta_j' \sim  \mathbb{P} (\theta) \}_{j=1}^M $
    \State $\mathbf{x} \leftarrow \{ \mathbf{x}_j \sim \mathbb{P} (\mathbf{x} | \theta) \}_{j=1}^M $
    \State $\mathcal{L} \leftarrow l(\mathbf{d}_\phi (\mathbf{x}, \theta, \theta'), 1) + l(\mathbf{d}_\phi (\mathbf{x}, \theta', \theta)$
    \State $\phi \leftarrow \textrm{OPTIMIZER} (\phi, \nabla_\phi \mathcal{L})$
\EndFor \\
\Return $\mathbf{d}_\phi (\mathbf{x}, \theta, \theta')$
\end{algorithmic}
\end{algorithm}

In the following sections, we succinctly refer to the parameter space relevant for neutrino oscillation experiments as \hl{$\theta \equiv (\sin^2 2 \theta, \Delta m^2)$, in line with common notation in SBI literature (despite the clash with the mixing angle).} A test statistic often chosen in a traditional Feldman-Cousins framework is the log-likelihood ratio
\begin{equation}
\label{eq:llhr}
T(\mathbf{x}|\theta ) = \log r (\mathbf{x} | \theta, \theta^*) = \log \mathbb{P} (\mathbf{x} |\theta) - \log \mathbb{P} (\mathbf{x} | \theta^*),
\end{equation}
with $\theta^*$ a best-fit point such as the \textit{maximum likelihood estimator} $\theta^* = \textrm{argmax}_\theta \mathbb{P} (\mathbf{x}| \theta)$. Given the  computational intractability of a likelihood $\mathbb{P} (\mathbf{x}|\theta)$ for a global fit (or in some cases even for a single experiment), a procedure which can estimate the likelihood ratio in a faster amount of time would find utility. 

Cobb \textit{et al.} propose the scheme in Alg.~\ref{alg:dnre} to train a doubly-parameterized neural network classifier with inputs $(\mathbf{x},\theta, \theta')$, which they refer to as a direct neural ratio estimator (DNRE) \cite{10.1609/aaai.v38i18.30018}. In short, realizations from one or more experiments $\mathbf{x}$ are simulated from grid points in the model parameter space $\theta$, extended with randomly selected grid points $\theta'$, and assigned a label $1$. Swapping $\theta$ and $\theta'$ produces samples with label $0$, and the neural network classification strategy  $\mathbf{d}_\phi (\mathbf{x}, \theta, \theta')$ can be directly converted into an estimate of the likelihood ratio via Eq. 6 of Ref.~\cite{10.1609/aaai.v38i18.30018}:
\begin{equation}
    \label{eq:llhr-dnre}
    \hat{r} (\mathbf{x} | \theta, \theta') = \frac{\mathbf{d}_\phi (\mathbf{x}, \theta, \theta')}{1 - \mathbf{d}_\phi (\mathbf{x}, \theta, \theta')}.
\end{equation}

To build the test statistic necessary for Alg.~\ref{alg:feldman-cousins}, a best fit point $\theta^*$ must be computed. Eq. 7 of Ref.~\cite{10.1609/aaai.v38i18.30018} provides a prescription for posterior density estimation with knowledge of the prior and the neural network classifier:
\begin{equation}
    \label{eq:posterior-estimate}
    \log \mathbb{P} (\theta | \mathbf{x} ) \approx - \mathrm{LogSumExp} \big{[} - \log \hat{r} (\mathbf{x} | \theta, \theta_i') \big{]} + \log M + \log \mathbb{P} (\theta),
\end{equation}
for $M$ a Monte Carlo subsample size. A maximum \textit{a posteriori} (MAP) estimator $\theta^*_\textrm{MAP}$ or a minimum mean squared error (MMSE) estimator can be computed via $\theta^*_\textrm{MAP} = \mathrm{argmax}_\theta \log \mathbb{P} (\theta | \mathbf{x}) $ or $\theta^*_\textrm{MMSE} = \mathbb{E} [\theta | \mathbf{x}] $. We use the former to define an SBI analog to Eq.~\ref{eq:llhr}:
\begin{equation}
    \label{eq:dnre-test-statistic}
    \hat{T} (\mathbf{x} | \theta) = \log \hat{r} (\mathbf{x} | \theta, \theta^*_\text{MAP}).
\end{equation}
Such a test statistic satisfies the ranking requirement given in Sec. 5B of Ref.~\cite{feldman-cousins}.

\subsection{\label{subsec:freq-cls}Estimating Frequentist Confidence Levels}

Our construction of frequentist confidence levels exactly parallels  the Feldman-Cousins method \cite{feldman-cousins}, wherein the test statistic is replaced with the DNRE equivalent (Eq.~\ref{eq:dnre-test-statistic}). We construct a DNRE with training and testing data described in Sec. \ref{subsec:datagen}, and with $\theta’$ spanning a slightly greater range than $\theta$ to account for edge effects. Then, for every realization $\mathbf{x}$, we find the best fit point $\theta_{\text{MAP}}^*$ by maximizing the posterior distribution estimate (Eq.~\ref{eq:posterior-estimate}) over the grid taking $M=600$ (chosen as a compromise between runtime considerations and ensuring sufficient representation of the prior). We can then compute $\hat{r}(\mathbf{x}|\theta, \theta_{\text{MAP}}^*))$ from the output of the DNRE (Eq. \ref{eq:llhr-dnre}). 

We then compute a critical test-statistic $\hat{r}_c(\theta)$ for each point $\theta$ in the parameter space. For each fixed parameter point $\theta$, we find $\hat{r}(\mathbf{x}|\theta, \theta^*_\text{MAP})$ for each realization $\mathbf{x}$ generated from model parameters $\theta$ belonging to the training set, yielding a distribution with around 800 samples for most experiments. The threshold value $\hat{r}_c(\theta)$ corresponding to a confidence level $\alpha$ is defined such that a fraction $\alpha$ of the samples satisfy $\hat{r}(\mathbf{x}|\theta, \theta^*)<\hat{r}_c(\theta)$. An acceptance region at confidence level $\alpha$ for one realization $\mathbf{x}$ consists of all points $\theta$ that satisfy $\hat{r}(\mathbf{x}|\theta, \theta^*_\text{MAP}) < \hat{r}_c(\theta)$.  We refer to this procedure as Simulation-Based-Inference-Feldman-Cousins, SBI-FC, for short.

\subsection{\label{subsec:sensitivity}Estimating Experimental Sensitivities}
The \textit{sensitivity} of a particle physics experiment corresponds qualitatively to the region of parameter \hl{space} about which an experiment's measurements can make conclusive determinations. Ref.~\cite{feldman-cousins} describes a method to compute the sensitivity of an oscillation experiment, wherein the median $\alpha$ sensitivity is defined by the following procedure:
\begin{enumerate}
    \item Choose a value of the model parameters $\theta$ consistent with null (SM)
    \item Generate many realizations under $\theta$
    \item For each realization generated from $\theta$:
    \begin{enumerate}
        \item Generate the $\alpha$-confidence acceptance region using the procedure in Sec \ref{subsec:freq-cls}.
        \item Find the maximum accepted $\log U_{\mu 4}$ (``upper limit”) for each value of $\log \Delta m_{41}^2$.
    \end{enumerate}
    \item For each value of $\log \Delta m_{41}^2$, calculate the median upper limit across the realizations under $\theta$. The union of these points comprises the median sensitivity at $\alpha$ confidence.
\end{enumerate}

\subsection{\label{subsec:globalfit}Extending from Single-Experimental to Global Fits}
Unlike a classical frequentist global analysis, requiring the costly minimization of a joint likelihood, the DNRE neural network trained by Alg.~\ref{alg:dnre} is amenable to the inclusion of multiple experiments simultaneously. The only task an analyzer has in upgrading a single experiment to a global fit  is to stack measurements or simulations from the experiments corresponding to the same grid points. That is, a global fit of experiments $A, B$ with realizations $\mathbf{x}_A , \mathbf{x}_B $ assuming the same model parameters $\theta$ requires partial input $\mathbf{x} =  \mathbf{x}_A || \mathbf{x}_B$. While this increases the number of network parameters and therefore the network training time, it is negligible compared to the increase in runtime for a joint fit of multiple experiments.

\subsection{\label{subsec:hyerparams}Hyperparameter Optimization}
For optimal classifier performance, we fine-tune our neural network hyperparameters by conducting a random search across a grid of testable hyperparameter values. The initial three network layers had widths of 100, 100, and 50, respectively, while all subsequent layers had a width of 25. We eliminated network widths in the hyperparameter search after noticing little to no improvement in network performance and results. We use the ROC-AUC score to measure the capability of our model in distinguishing inputs of Monte Carlo data and true parameters from inputs of Monte Carlo data and random parameters. After 40 trials, the parameters corresponding to the maximum ROC-AUC score, for each experiment and combined, are shown in Table \ref{tab:hyperparam}. The minimal variation in the ROC-AUC values over the tested hyperparameters confirms the robustness of the model architecture.

\begin{table} 
\caption{\label{tab:hyperparam} Optimal neural network hyperparameters determined from ROC-AUC score maximization.}

\begin{tabular*}{\textwidth}{lccccl}
\br
Experiment&Dropout Rate&Learning Rate& Batch Size& Depth\\
\mr
CDHS & 0.1833 & 1e-4 & 1024 & 7\\
CCFR & 0.0966 & 0.77e-4 & 512 & 6\\
SciBooNE/MiniBooNE $\nu_\mu$ & 0.0533 & 2.78e-4 & 256 & 3\\
SciBooNE/MiniBooNE $\bar{\nu_\mu}$ & 0.0966 & 2.15e-4 & 1024 & 7\\
MINOS/MINOS+ & 0.01 & 1.3e-4 & 1024 & 4\\
All & 0.0966 & 0.77e-4 & 128 & 4\\
\br
\end{tabular*}

\end{table}

\subsection{\label{subsec:evaluation}Evaluation Strategies}

\subsubsection{Neural Network Performance} To assess the network's predictive accuracy, we report the ROC-AUC score across the parameter space grid. We expect scores to be greater than 0.5, indicating random predictions, and close to 1 in regions where the experiment has sensitivity, where the network should accurately identify the parameter point at which the realization was generated. 

\subsubsection{\label{subsubsec:coverage}Coverage Checks}

In the development of any SBI procedure, checking model calibration is an important step towards building credence when looking to a surrogate model as an inference tool. SBI-FC requires calibration of the frequentist confidence intervals introduced by the numerical CL construction, and the coverage of the posterior density estimation described by Eq.~\ref{eq:posterior-estimate} for use in computing the point of best fit. While both frequentist CLs and Bayesian credibility levels fundamentally express uncertainty about an unknown parameter $\theta$, they have different philosophical interpretations and offer nuanced guarantees.

A frequentist confidence interval at level $\alpha$ is constructed such that, under repeated sampling, it contains the parameter $\theta$ with probability $\alpha$. Importantly, this guarantee is made for each fixed value of $\theta$: if one were to generate many sets of synthetic data assuming constant oscillation model parameters $\theta$, a fraction $\alpha$ of the built confidence regions will cover $\theta$. Randomness lies in the computed confidence region and the simulated data: not in $\theta$ itself, which is treated as fixed.

In contrast, a Bayesian credibility region at level $\alpha$ is defined as a region in parameter space which contains posterior probability mass $\alpha$, given the observed data. $\theta$ is modeled as an unknown random variable with prior $\mathbb{P} (\theta)$, and inference is performed by updating the prior with observed data $\mathbf{x}$. The $\alpha$-level credibility region contains the true value of $\theta$ with probability $\alpha$ under the defined prior distribution; that is, drawing many $\theta \sim \mathbb{P} (\theta)$ and computing corresponding posterior credibility regions for realizations generated by each would contain the injected model parameters a fraction $\alpha$ of the time \cite{4d625c73-e285-3ead-ae1f-415e8ad60311, Talts:2018zdk}. The guarantee is over the joint randomness of both $\theta \sim \mathbb{P}(\theta)$ and the data, not over repeated sampling with a fixed $\theta$.

Evaluation of the empirical coverages of the constructed frequentist confidence intervals (on $25$ uniformly sampled parameter points for each experiment and the global fit) and Bayesian credibility regions (on $1000$ uniformly sampled parameter points for each experiment and the global fit) are presented in Sec.~\ref{sec:coverage} and \ref{sec:appB_coverages} respectively. Consistency with the $45^\circ$ line indicates a well-calibrated model. Furthermore, in the frequentist case, one expects that for sampled parameter values $\theta_i$, $X_{i, \alpha}$, the number of times in $M$ trials that $\theta = \theta_i$ is covered by the $\alpha$ CL is distributed as $X_i \sim \text{Binom}(M,\alpha)$. For large $M$, one expects the empirical coverage $\hat{p}_i \equiv X_{i,\alpha} / M \sim \mathcal{N}(\alpha, \alpha (1-\alpha) / M) $, and the coverage residual $\hat{p}_i - \alpha \sim \mathcal{N} (0, \alpha (1-\alpha) / M)$. We verify these residual distributions graphically in Fig.~\ref{fig:exp_coverages}, taking $M=100$ for CDHS (due to reduced test size) and $M=200$ for other experiments and the global fit.

\subsection{Code Availability}

We make available an implementation of the presented SBI-FC fitting technique in Ref.~\cite{toygithub} for the simplified case of two-neutrino oscillations.

%% file: Sec3_Results.tex
\section{\label{sec:results}Results}

\subsection{Neural Network Optimization and Performance}
We observe strong neural network performance (ROC-AUC score $> 0.97$) for samples generated from $\theta$ corresponding to obvious sterile neutrino signals. As expected, the ROC-AUC scores are closer to $0.5$ in regions of parameter space where the experiment (or combination of experiments) lose sensitivity. We show the ROC-AUC score distribution across the parameter space for each experiment and the global fit in \ref{sec:appA_ROC-AUC}. 

\subsection{Sample Fits}
\begin{figure}[htbp]
    \centering
    \begin{subfigure}[b]{0.45\textwidth}
        \includegraphics[width=\linewidth]{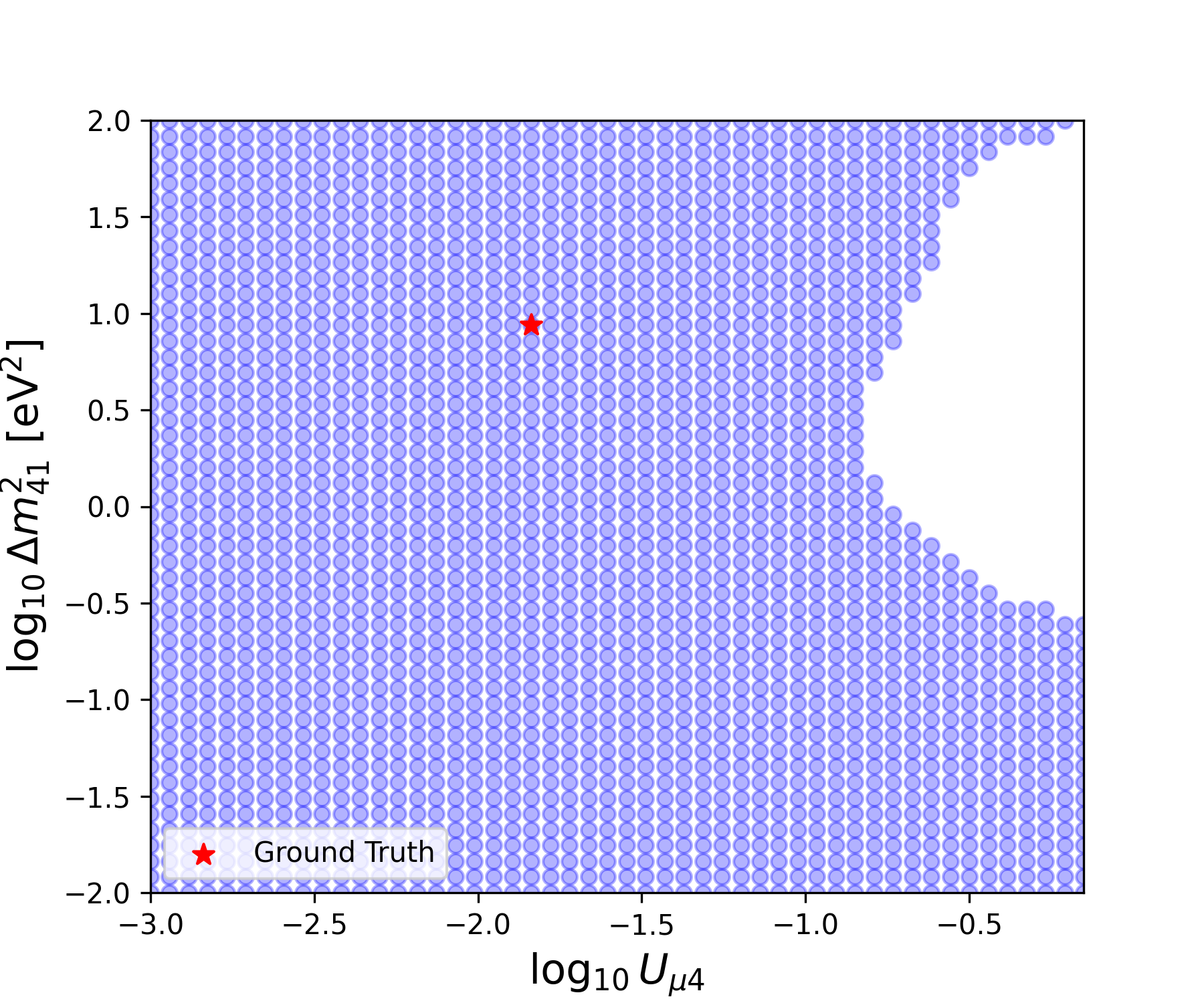}
        \label{fig:cdhs_null}
        \caption{CDHS}
    \end{subfigure}
    \begin{subfigure}[b]{0.45\textwidth}
        \includegraphics[width=\linewidth]{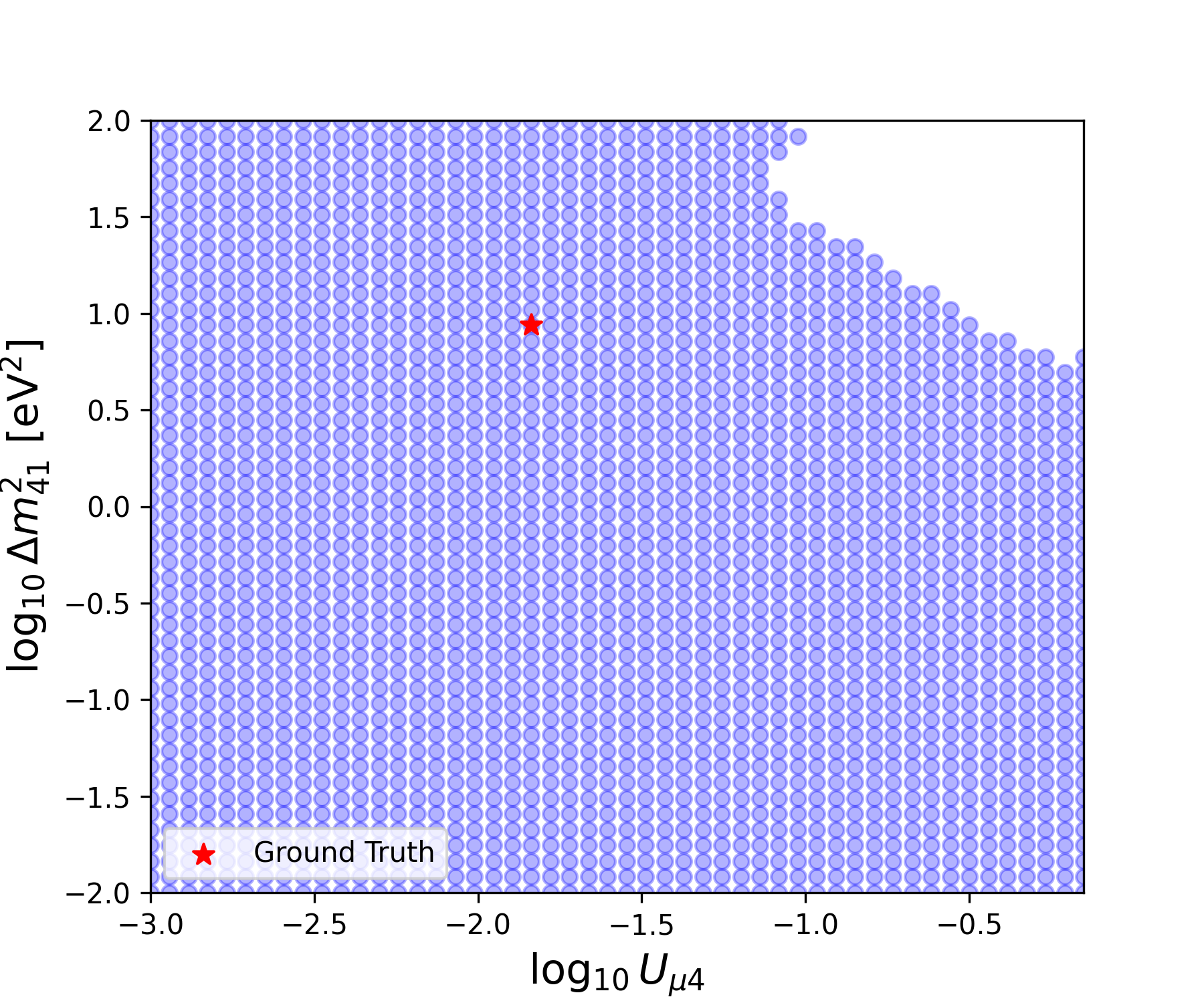}
        \label{fig:ccfr_null}
        \caption{CCFR}
    \end{subfigure}
    \begin{subfigure}[b]{0.45\textwidth}
        \includegraphics[width=\linewidth]{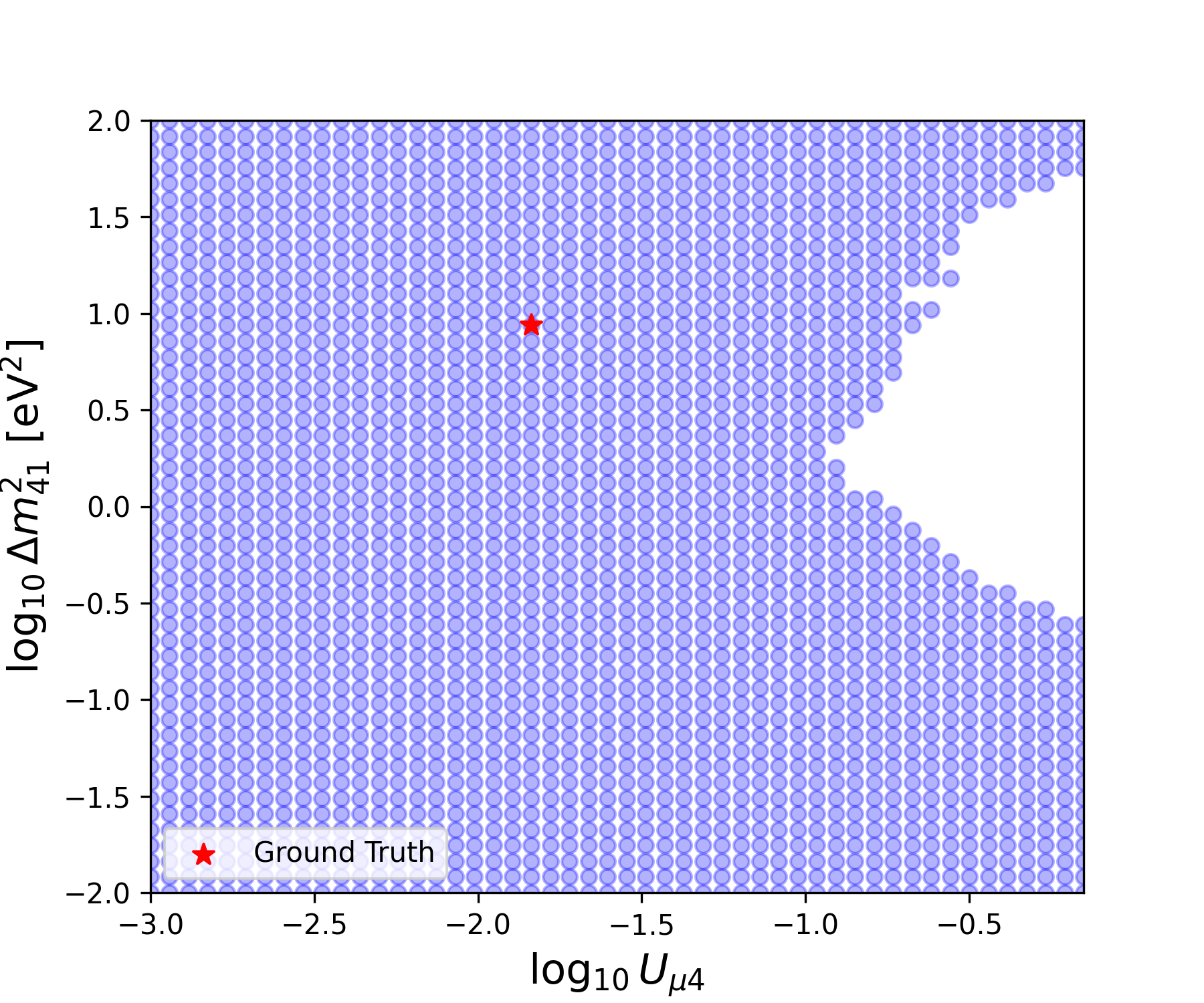}
        \label{fig:sbmb_null}
        \caption{SciBooNE-MiniBooNe $\nu_\mu$}
    \end{subfigure}
    \begin{subfigure}[b]{0.45\textwidth}
        \includegraphics[width=\linewidth]{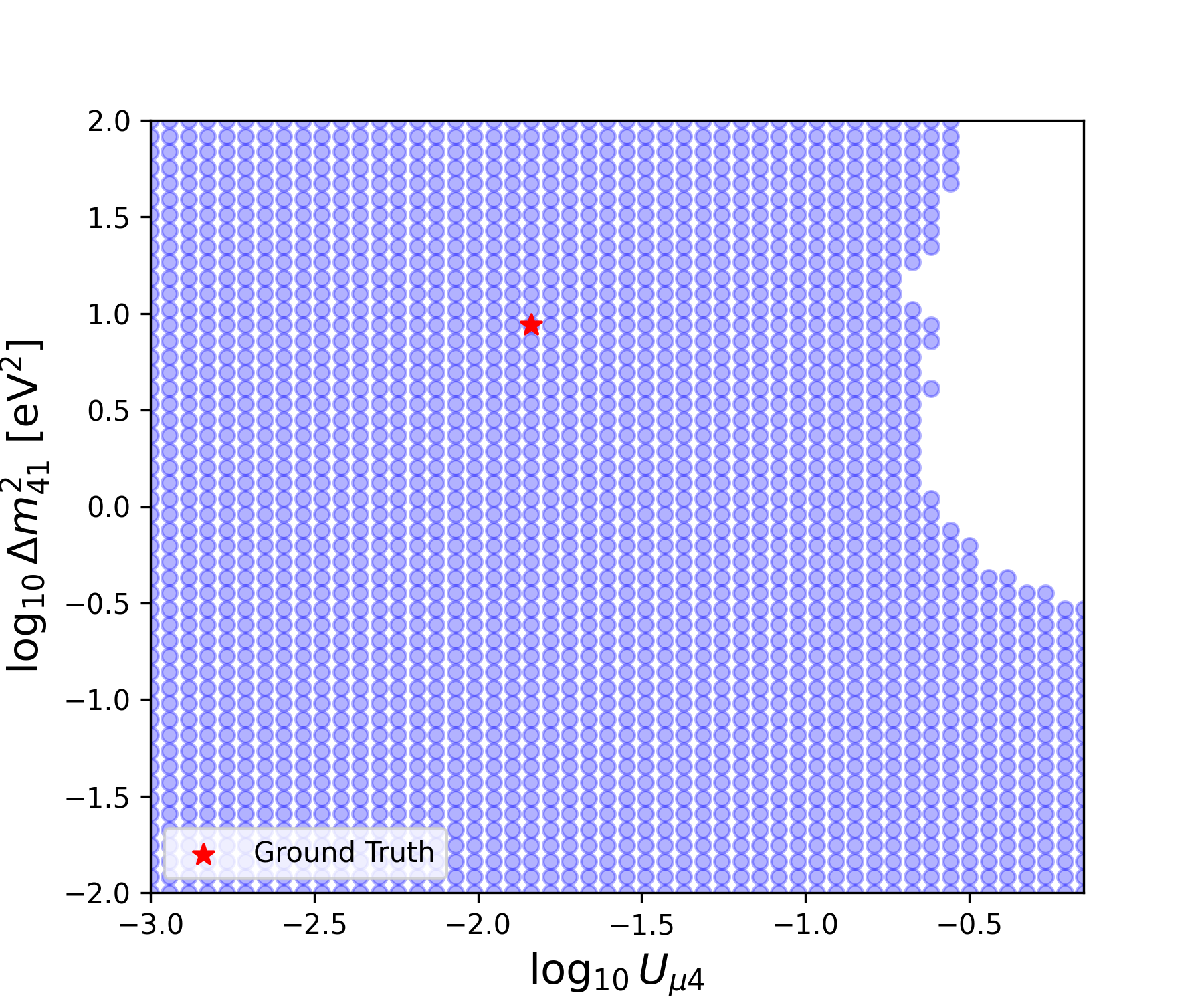}
        \label{fig:sbmbbar_null}
        \caption{SciBooNE-MiniBooNe $\bar{\nu}_\mu$}
    \end{subfigure}
    \begin{subfigure}[b]{0.45\textwidth}
        \includegraphics[width=\linewidth]{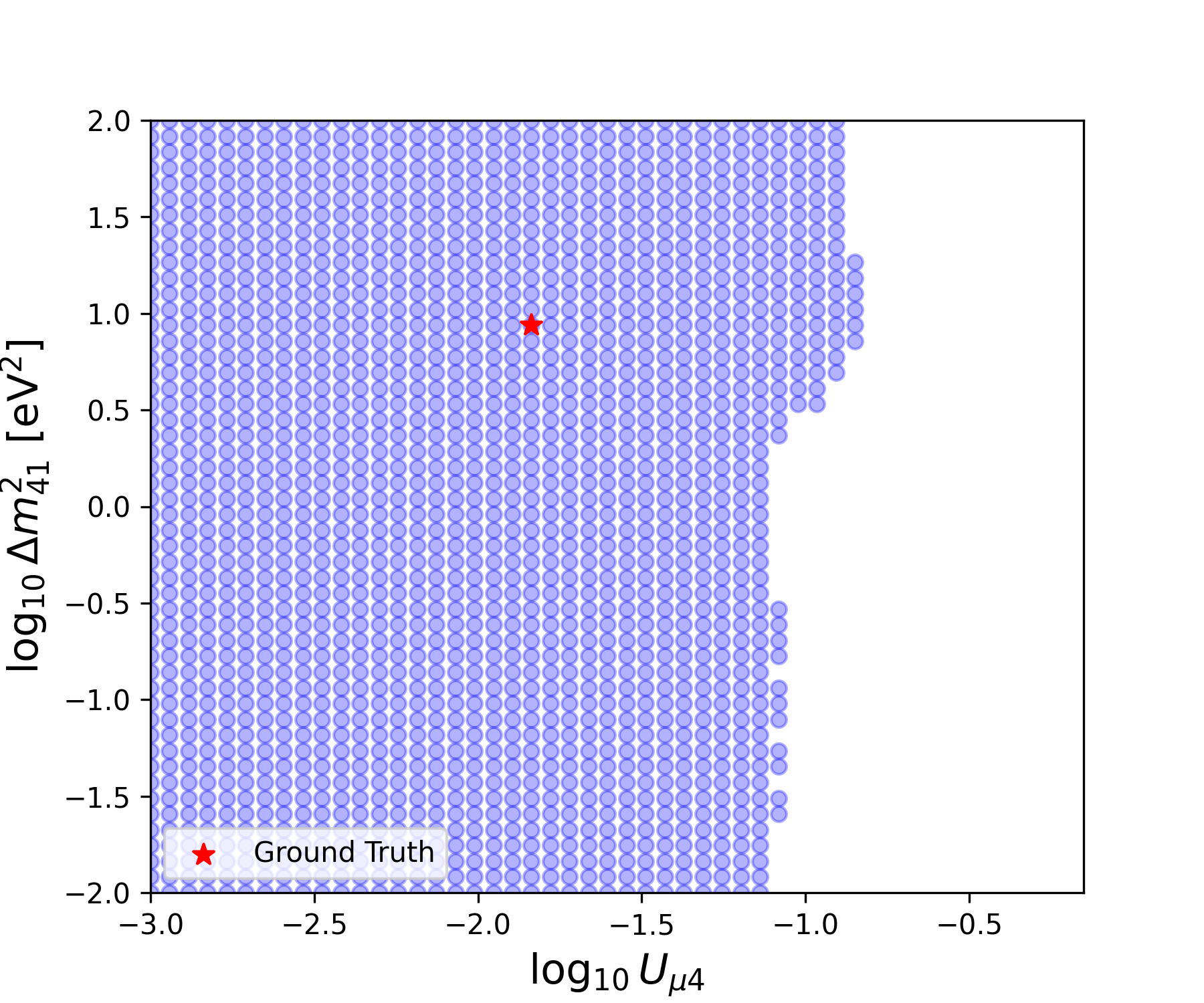}
        \label{fig:minos_null}
        \caption{MINOS}
    \end{subfigure}
    \begin{subfigure}[b]{0.45\textwidth}
        \includegraphics[width=\linewidth]{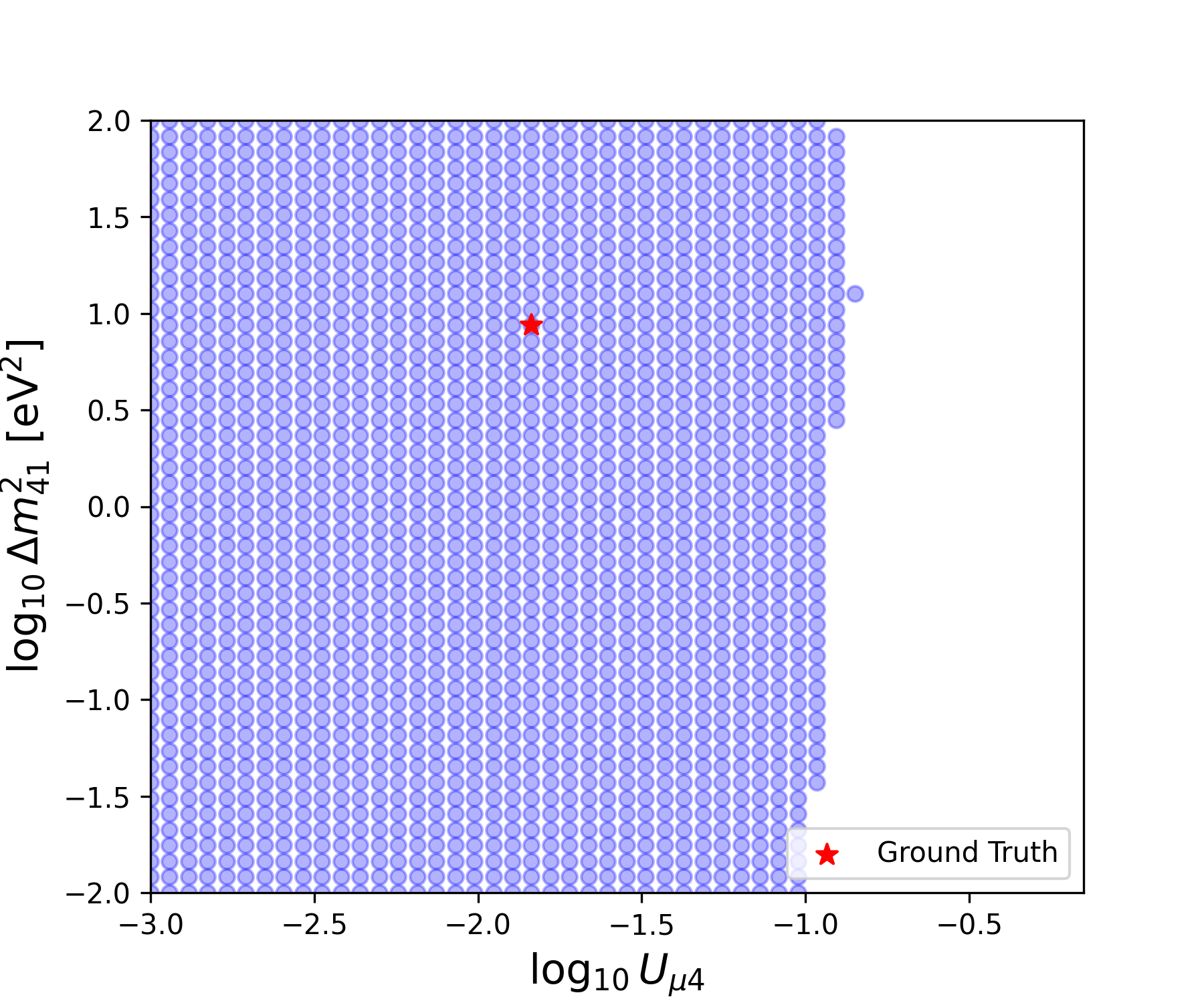}
        \label{fig:all_exp_null}
        \caption{All experiments}
    \end{subfigure}
    \caption{\hl{SBI-generated acceptance regions at $90\%$ confidence for a null-like realization. The qualitative form of these acceptance regions resemble familiar exclusion curves, where small values of mixing $U_{\mu 4}$ are contained in the region of confidence, demonstrating consistency with the SM. Points that remain white are excluded at $90\%$ confidence.}}
    \label{fig:sample_null_fits}
\end{figure}

\begin{figure}[htbp]
    \centering
    \begin{subfigure}[b]{0.45\textwidth}
        \includegraphics[width=\linewidth]{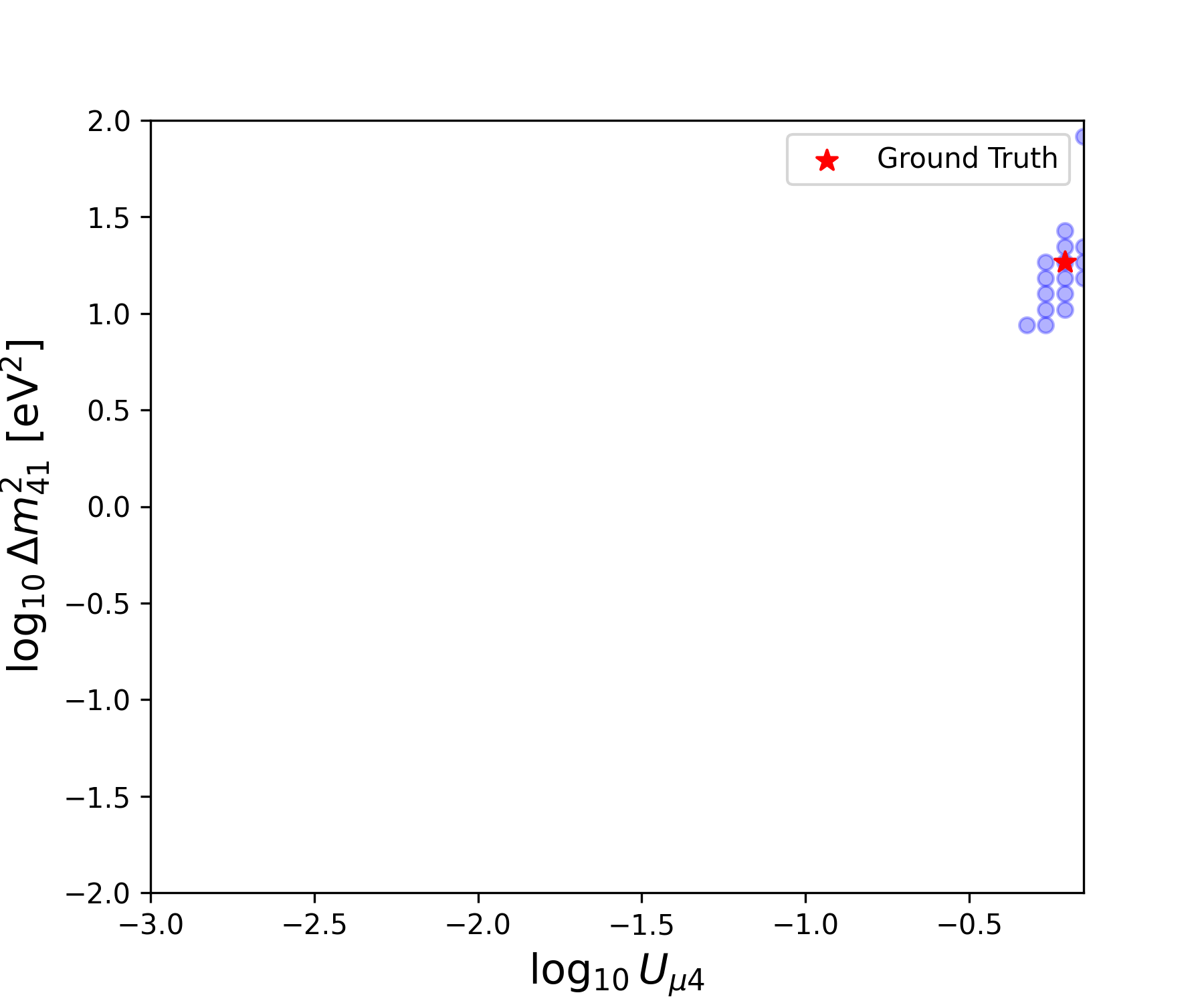}
        \label{fig:cdhs_sig}
        \caption{CDHS}
    \end{subfigure}
    \begin{subfigure}[b]{0.45\textwidth}
        \includegraphics[width=\linewidth]{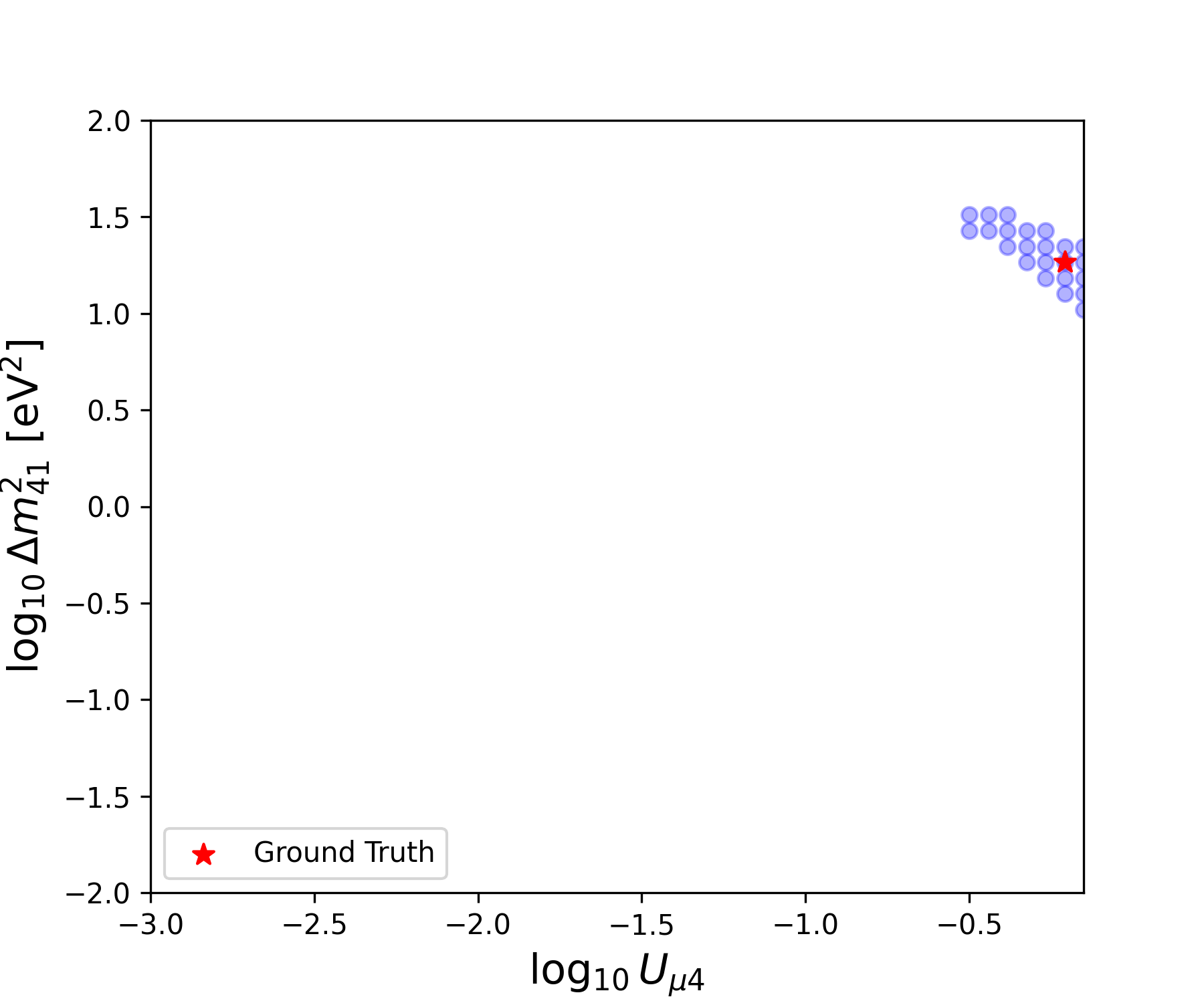}
        \label{fig:ccfr_sig}
        \caption{CCFR}
    \end{subfigure}
    \begin{subfigure}[b]{0.45\textwidth}
        \includegraphics[width=\linewidth]{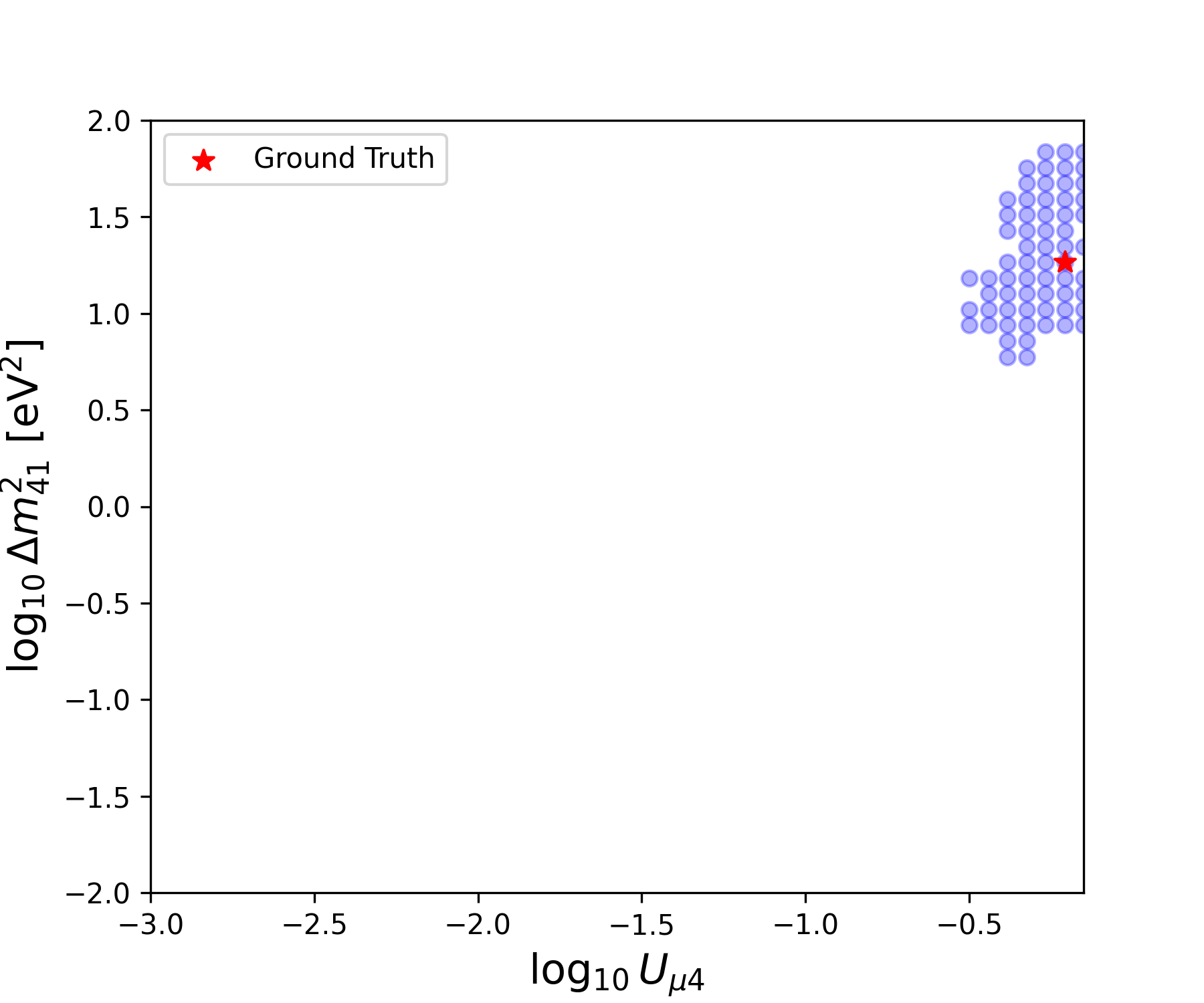}
        \label{fig:sbmb_sig}
        \caption{SciBooNE-MiniBooNe $\nu_\mu$}
    \end{subfigure}
    \begin{subfigure}[b]{0.45\textwidth}
        \includegraphics[width=\linewidth]{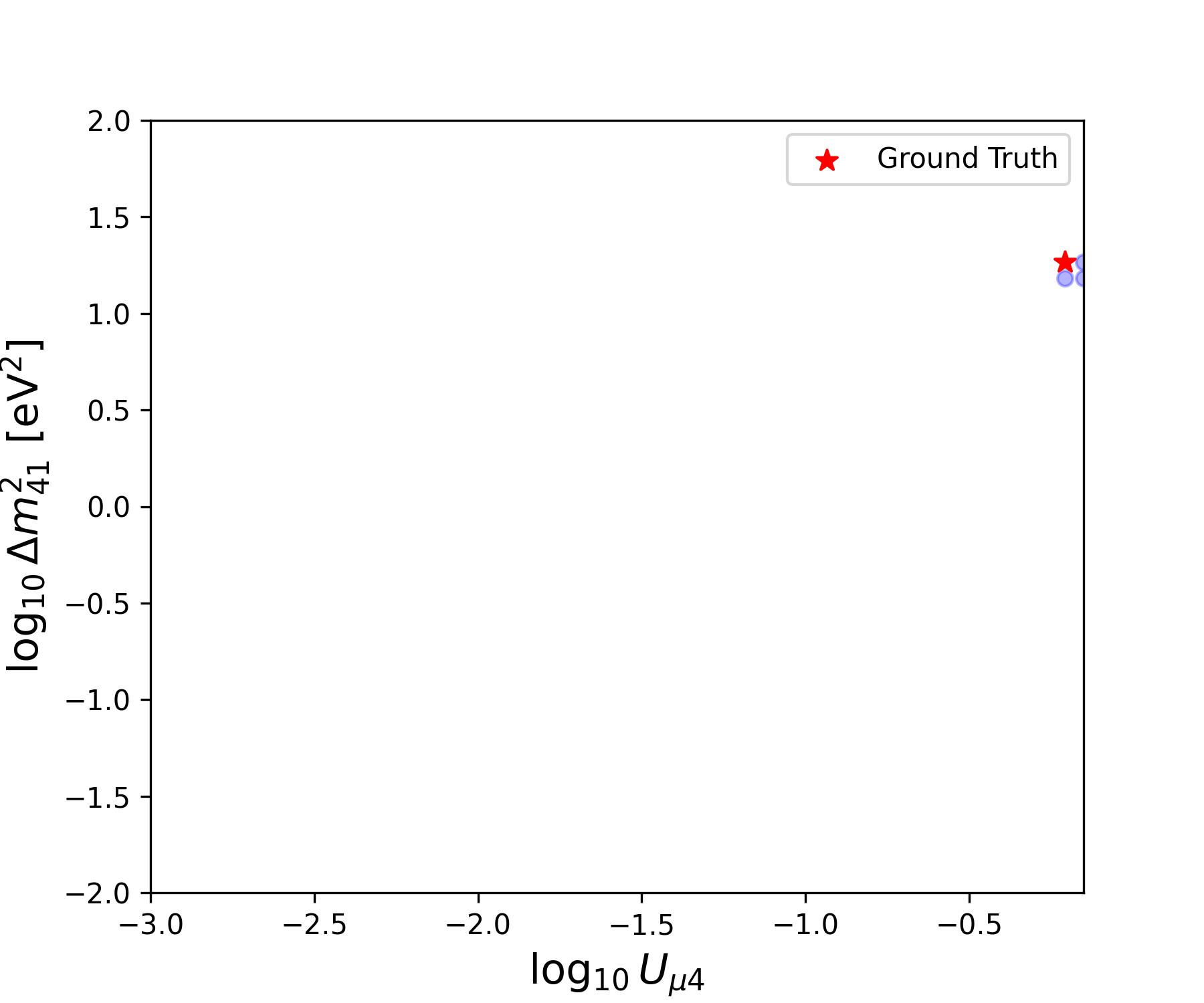}
        \label{fig:sbmbbar_sig}
        \caption{SciBooNE-MiniBooNe $\bar{\nu}_\mu$}
    \end{subfigure}
    \begin{subfigure}[b]{0.45\textwidth}
        \includegraphics[width=\linewidth]{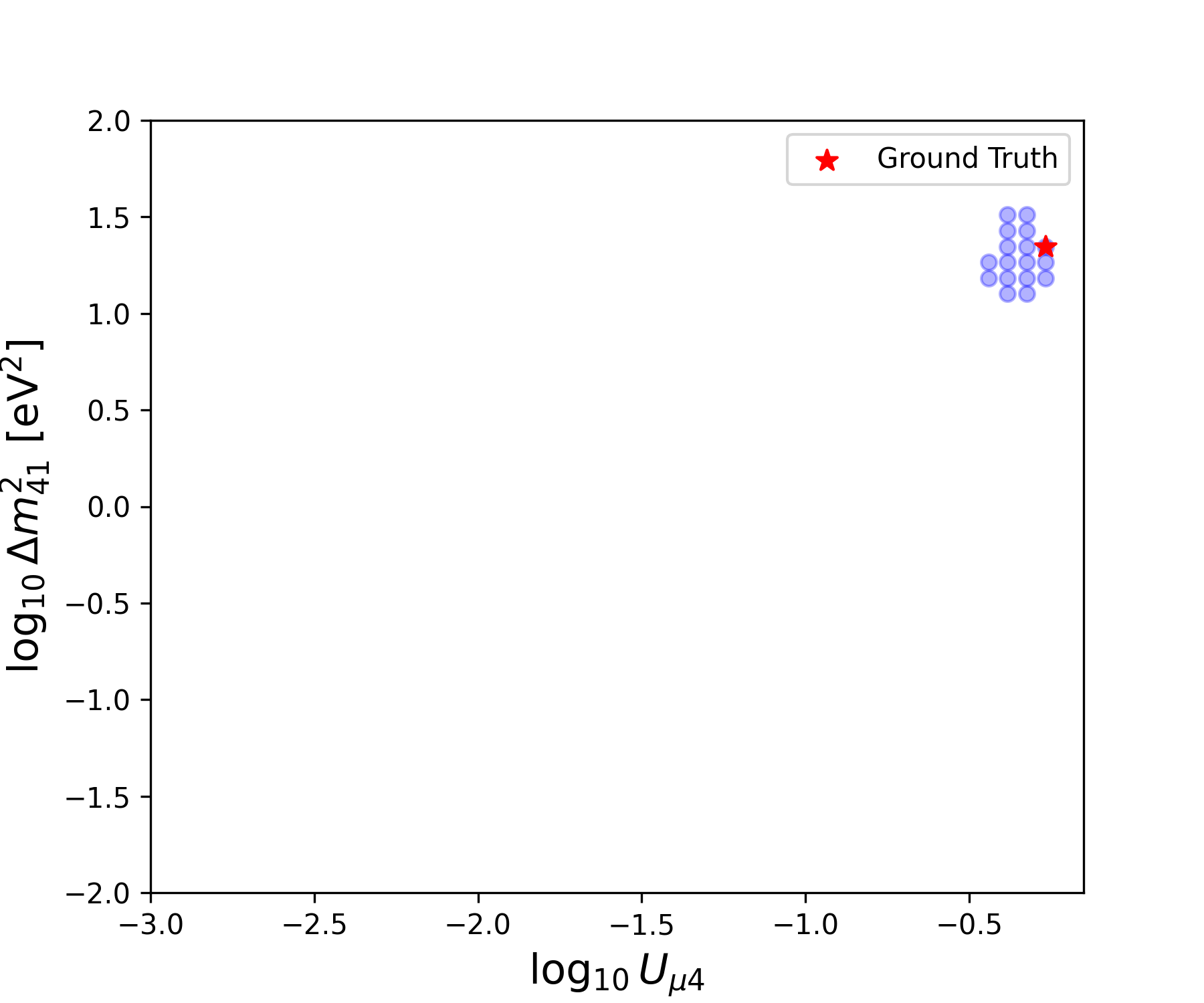}
        \label{fig:minos_sig}
        \caption{MINOS}
    \end{subfigure}
    \begin{subfigure}[b]{0.45\textwidth}
        \includegraphics[width=\linewidth]{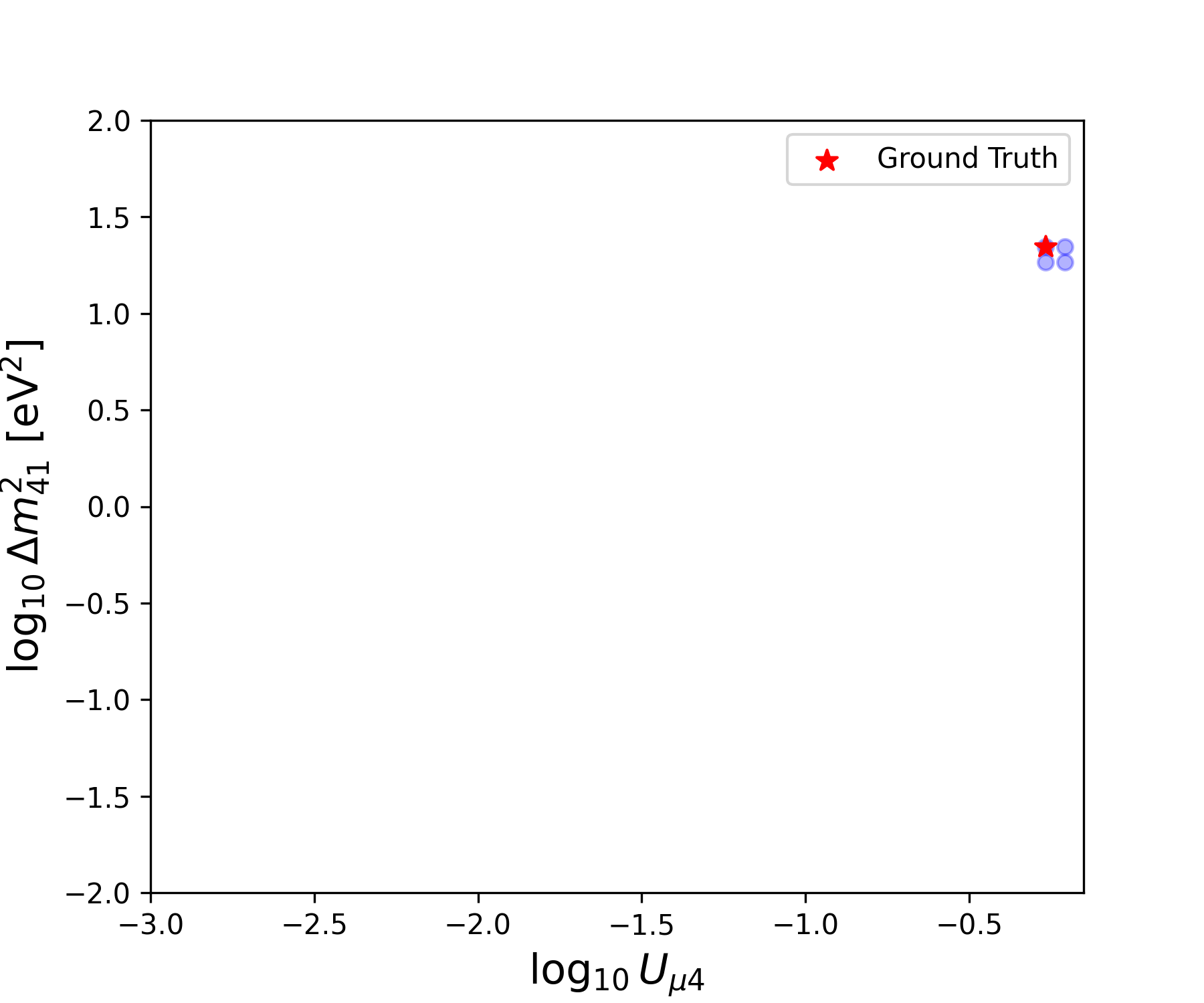}
        \label{fig:all_exp_sig}
        \caption{All experiments}
    \end{subfigure}
    \caption{\hl{Same as Fig.} \ref{fig:sample_null_fits}\hl{, but for ground truth model parameter injections in signal-like model parameter space. Closed contours around injected parameter values at $90\%$ confidence rule out regions of null parameter space, indicating a preference for sterile neutrino oscillations.}}
    \label{fig:sample_sig_fits}
\end{figure}

We use the combined SBI+Feldman-Cousins (SBI-FC) fitting procedure to generate acceptance regions, ensuring qualitative reasonableness in shape. These acceptance regions for null-like and signal-like model parameter injections are shown in Figs. \ref{fig:sample_null_fits} and \ref{fig:sample_sig_fits}, at a $90\%$ confidence level. For null-like samples, we expect the accepted region to approximately exclude the parameter space where the experiment is sensitive, and generate acceptance regions that are consistent with null (small $U_{\mu 4}$). For signal-like samples, the network should accurately identify the true parameter values used to generate the sample, yielding tight, closed contours around the injected ground truth parameters. 

\subsection{Estimated Experimental Sensitivities}
\begin{figure}
    \centering
    \begin{subfigure}[b]{0.45\textwidth}
        \includegraphics[width=\linewidth]{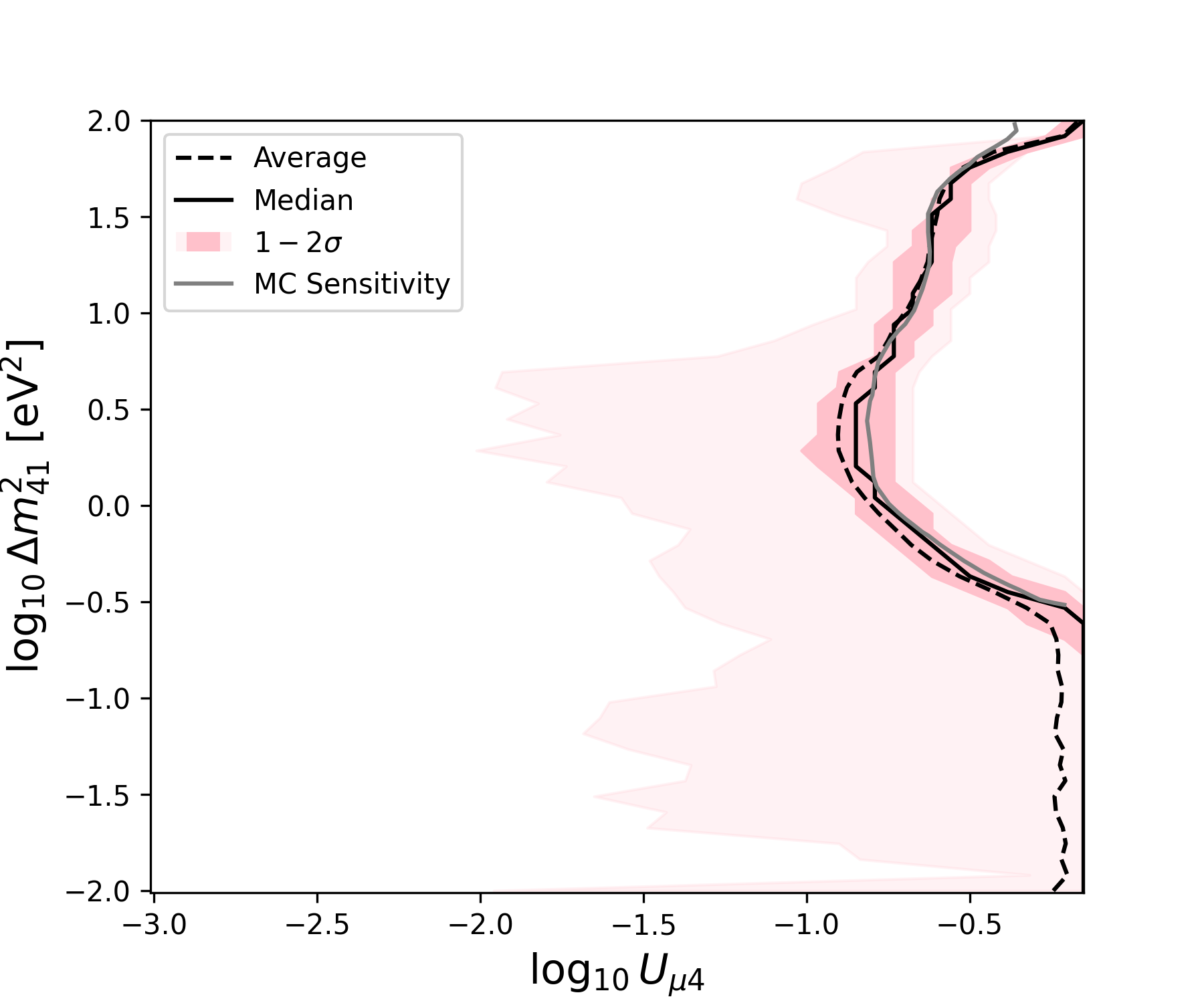}
        \label{fig:cdhs_sensitivity}
        \caption{CDHS}
    \end{subfigure}
    \begin{subfigure}[b]{0.45\textwidth}
        \includegraphics[width=\linewidth]{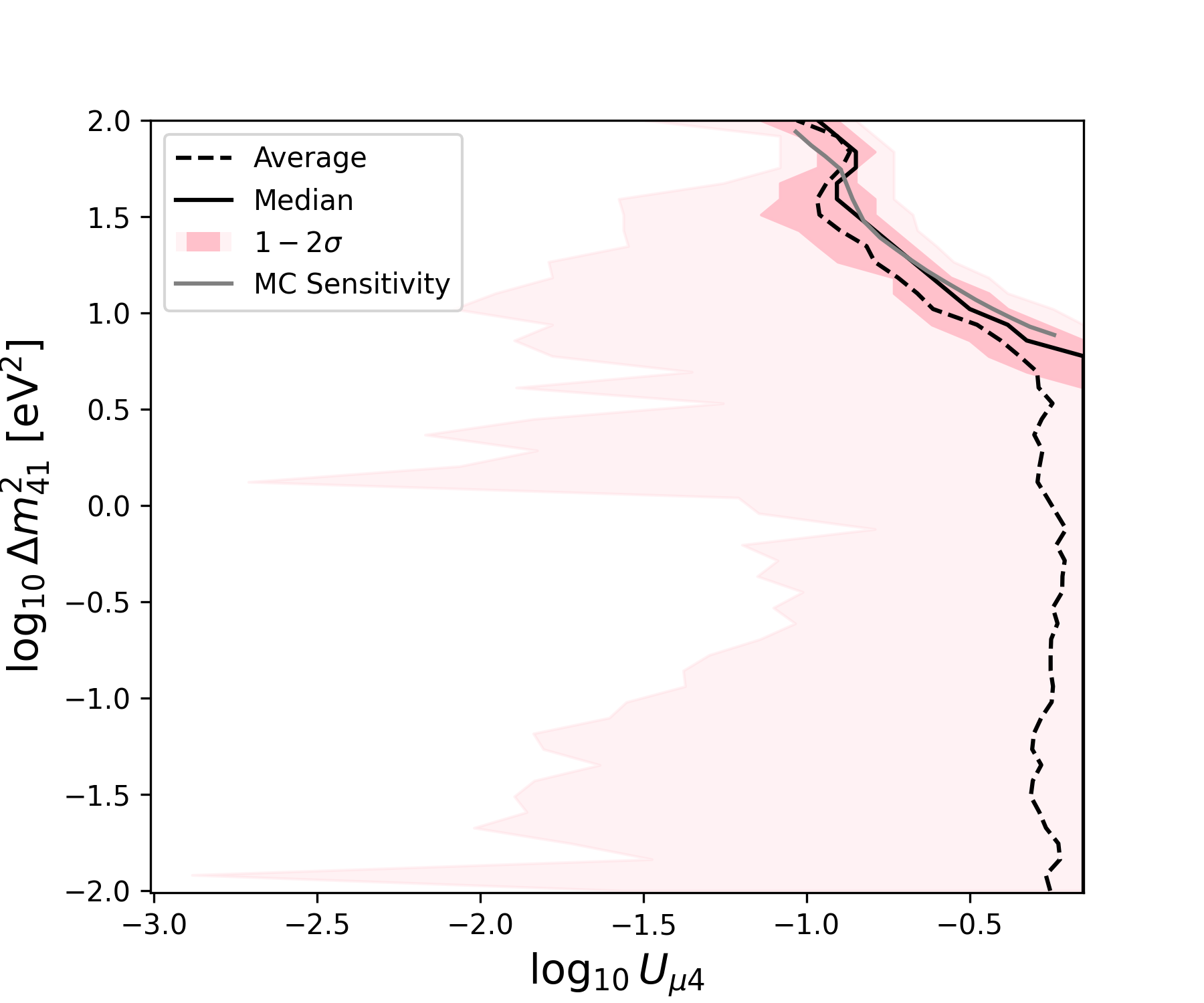}
        \label{fig:ccfr_sensitivity}
        \caption{CCFR}
    \end{subfigure}
       \begin{subfigure}[b]{0.45\textwidth}
        \includegraphics[width=\linewidth]{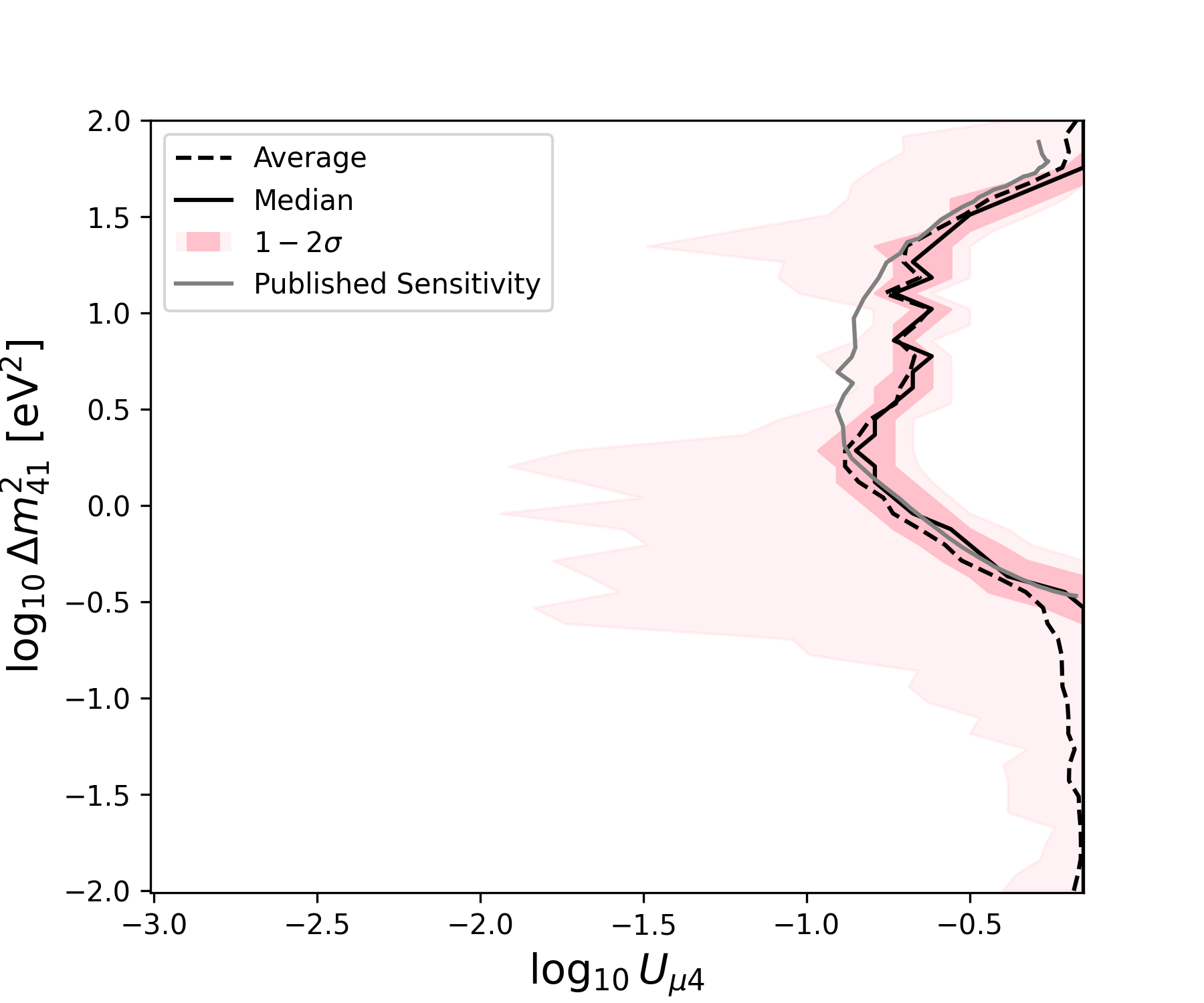}
        \label{fig:sbmb_sensitivity}
        \caption{SciBooNE-MiniBooNE $\nu_\mu$}
    \end{subfigure}
    \begin{subfigure}[b]{0.45\textwidth}
        \includegraphics[width=\linewidth]{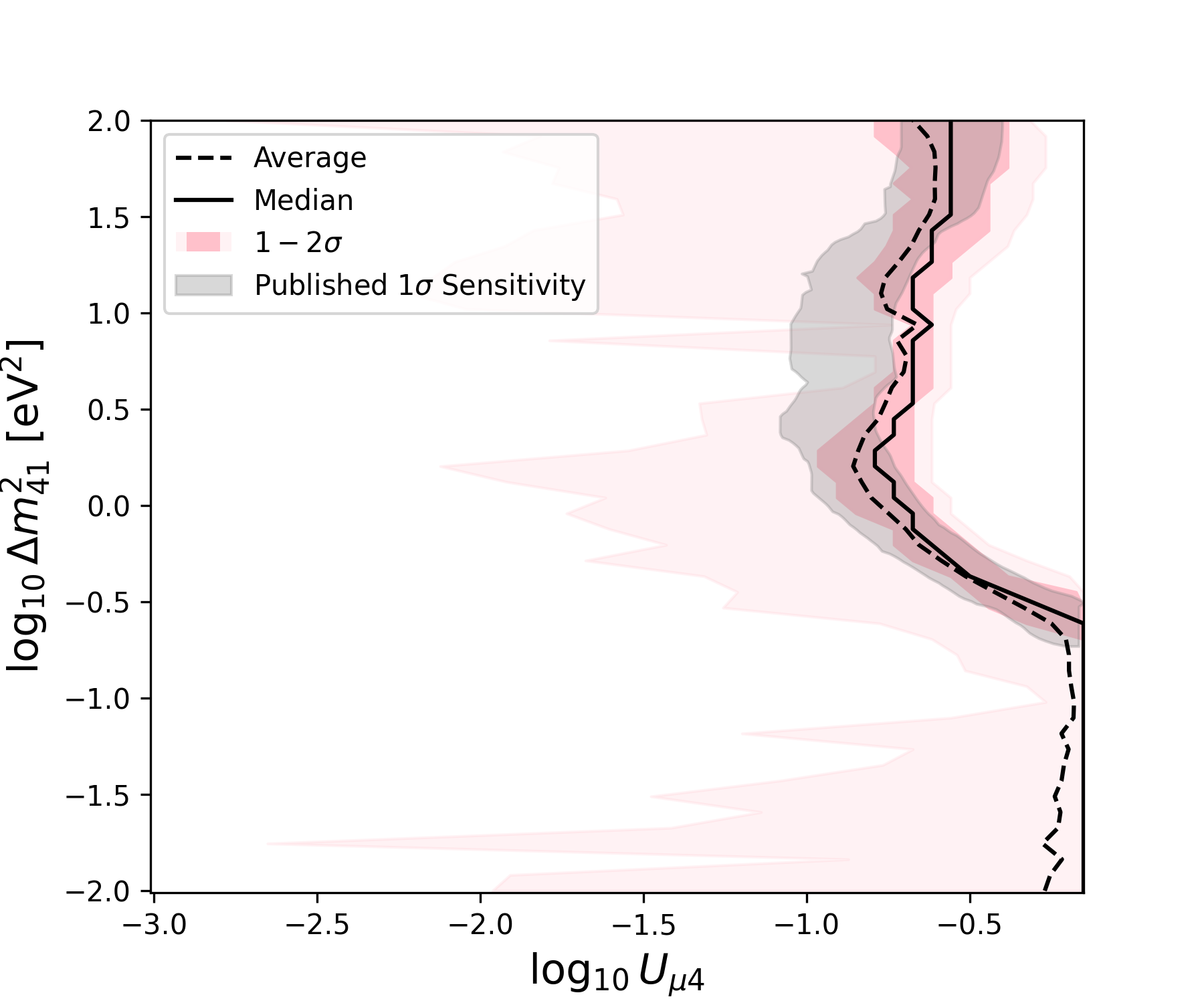}
        
        \caption{SciBooNE-MiniBooNE $\bar{\nu}_\mu$}
        \label{fig:sbmbbar_sensitivity}
    \end{subfigure}
       \begin{subfigure}[b]{0.45\textwidth}
        \includegraphics[width=\linewidth]{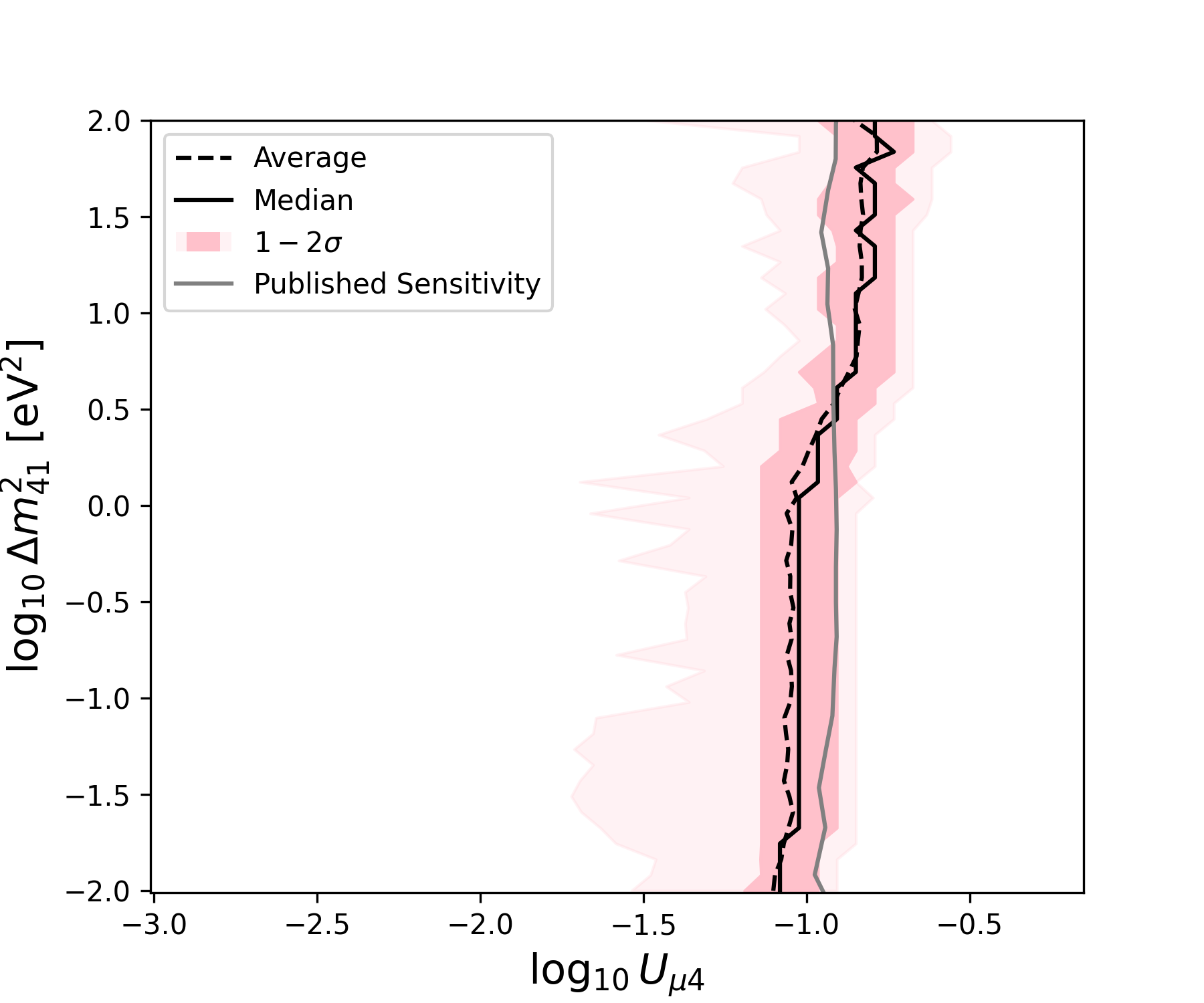}
        \label{fig:minos_sensitivity}
        \caption{MINOS}
    \end{subfigure}
    \begin{subfigure}[b]{0.45\textwidth}
        \includegraphics[width=\linewidth]{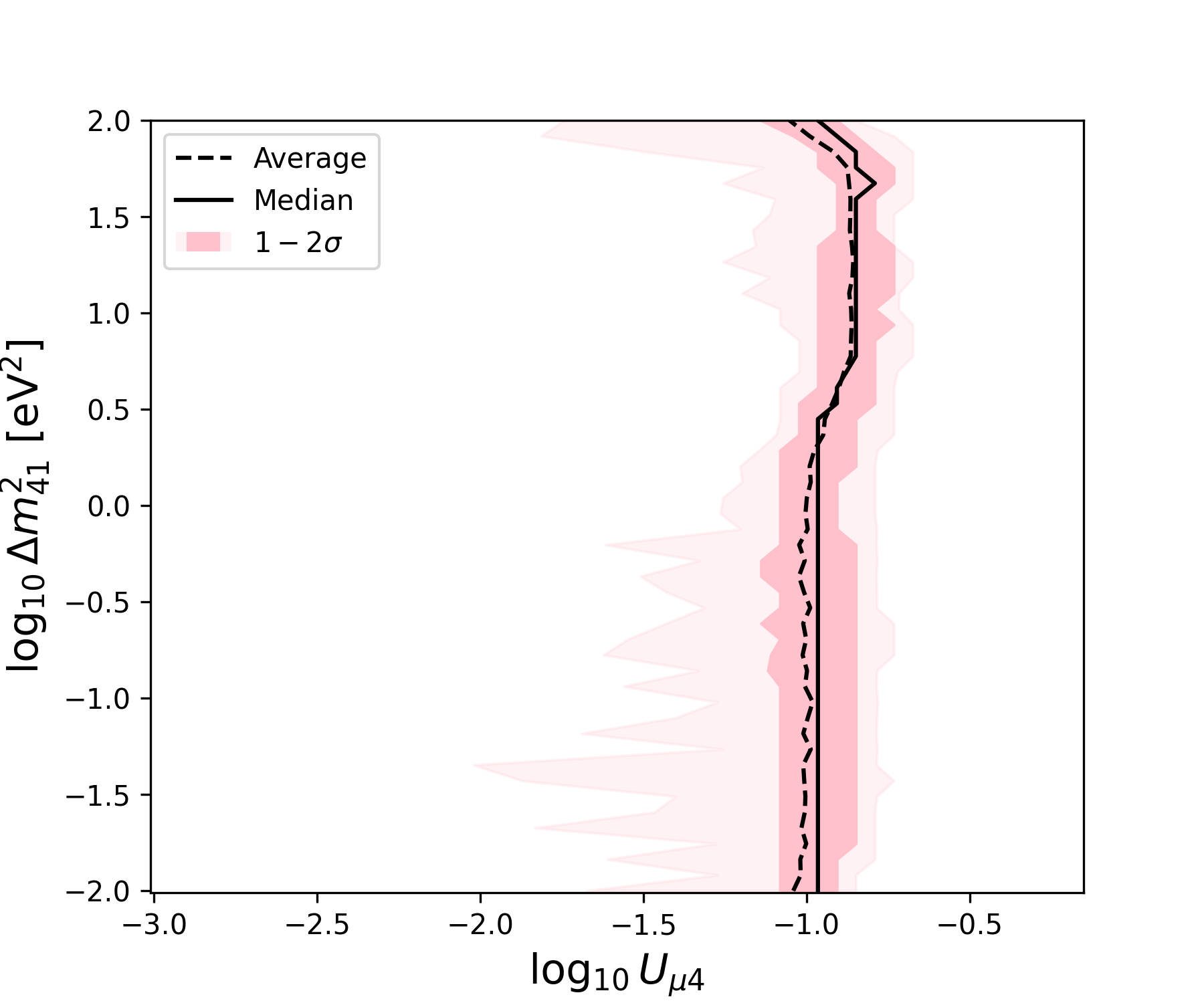}
        \label{fig:all_sensitivity}
        \caption{All Experiments}
    \end{subfigure}
    
    \caption{Generated SBI-FC sensitivities generated for each experiment and the global fit. For experiments that publish a sensitivity from fake-data studies, the reported sensitivity is overlaid. Otherwise, we overlay a Monte Carlo (MC) sensitivity by generating many null realizations and looking at the median exclusion assuming Wilks' theorem.}
    \label{fig:sensitivities}
\end{figure}

Next, we calculate sensitivities for each experiment using the process outlined in Sec. \ref{subsec:sensitivity}. We show the average, median, and $1-2\sigma$ distributions of the upper limits for each experiment and the global fit in Fig. \ref{fig:sensitivities}. For the single experiments, we report excellent agreement between the published or Monte Carlo computed sensitivity and the median sensitivity generated from SBI. The Monte Carlo (MC) sensitivity sometimes used as a point of comparison is produced by generating many null realizations and finding the median point above which we expect $90\%$ exclusion assuming Wilks' theorem. For CDHS and CCFR, Wilks' theorem is a good approximation; for other experiments, the published sensitivities can account for relevant statistical effects.

\subsection{\label{sec:coverage}Coverage Checks}
\begin{figure}[htbp]
    \centering
    \begin{subfigure}[b]{0.42\textwidth}
        \includegraphics[width=\linewidth]{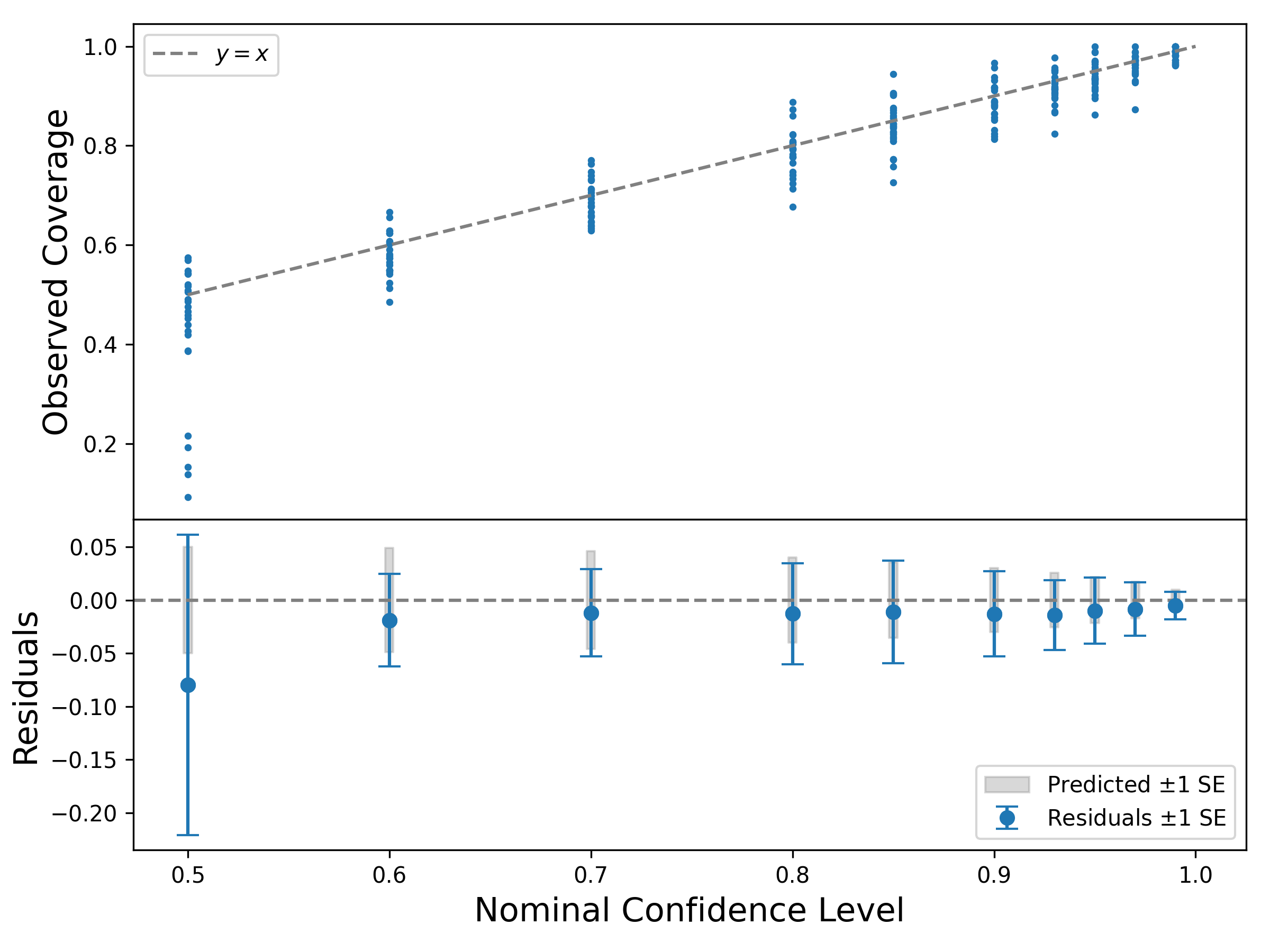}
        \label{fig:cdhs_cov}
        \caption{CDHS}
    \end{subfigure}
    \hspace{0.04\textwidth}
    \begin{subfigure}[b]{0.42\textwidth}
        \includegraphics[width=\linewidth]{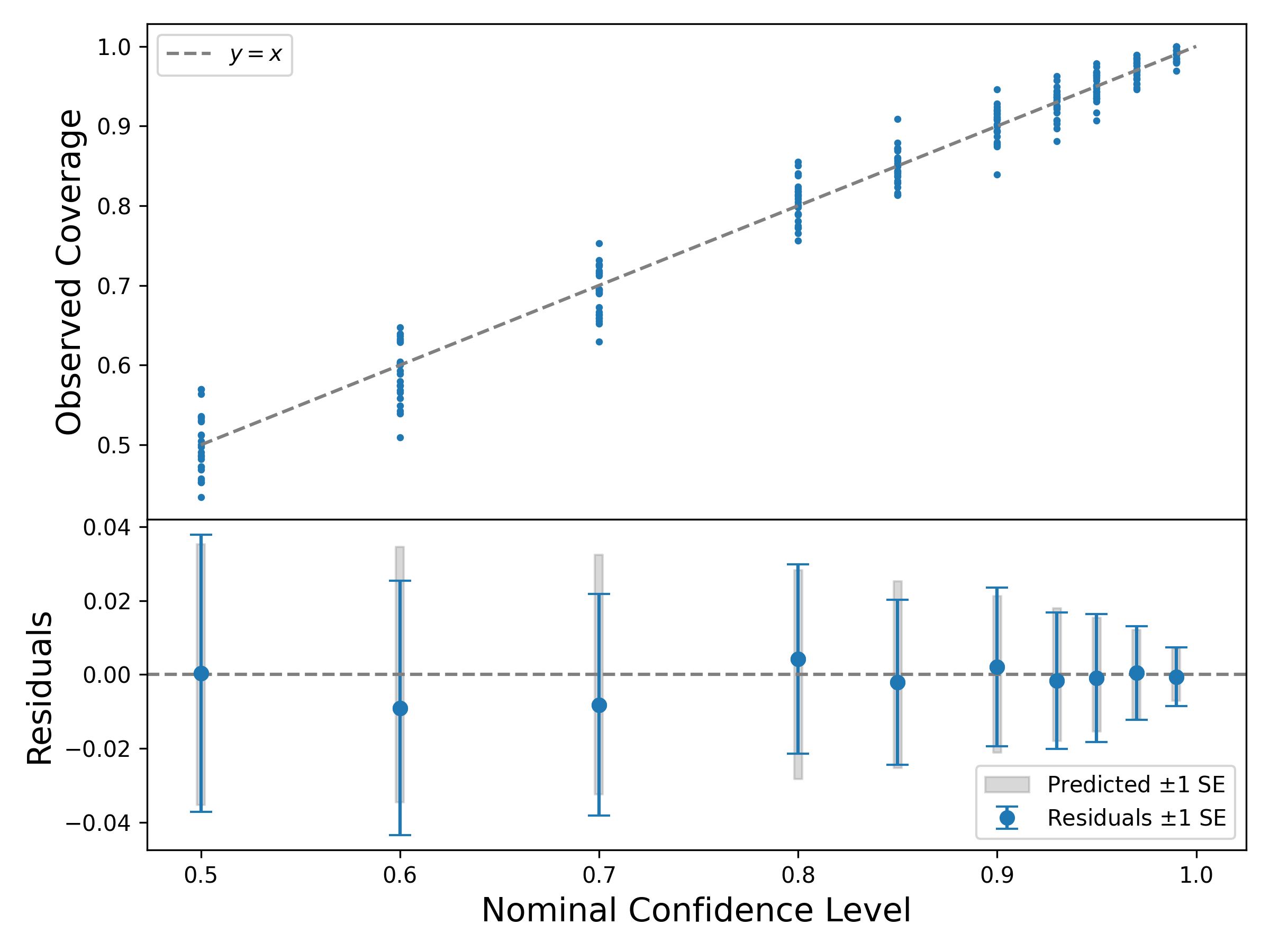}
        \label{fig:ccfr_cov}
        \caption{CCFR}
    \end{subfigure}

    \vspace{0.05\textwidth}
    
    \begin{subfigure}[b]{0.42\textwidth}
        \includegraphics[width=\linewidth]{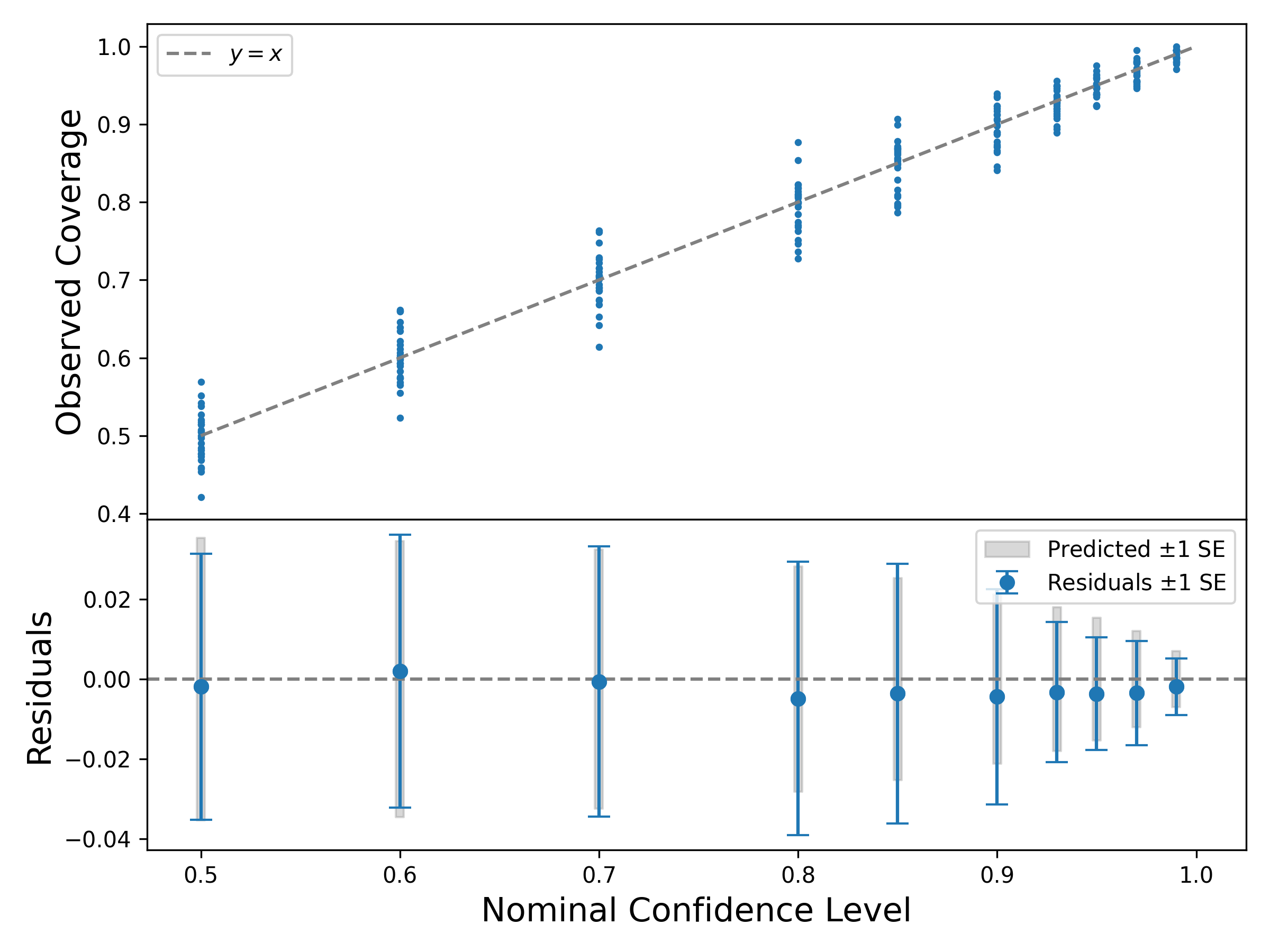}
        \label{fig:sbmb_cov}
        \caption{SciBooNE-MiniBooNe $\nu_\mu$}
    \end{subfigure}
    \hspace{0.04\textwidth}
    \begin{subfigure}[b]{0.42\textwidth}
        \includegraphics[width=\linewidth]{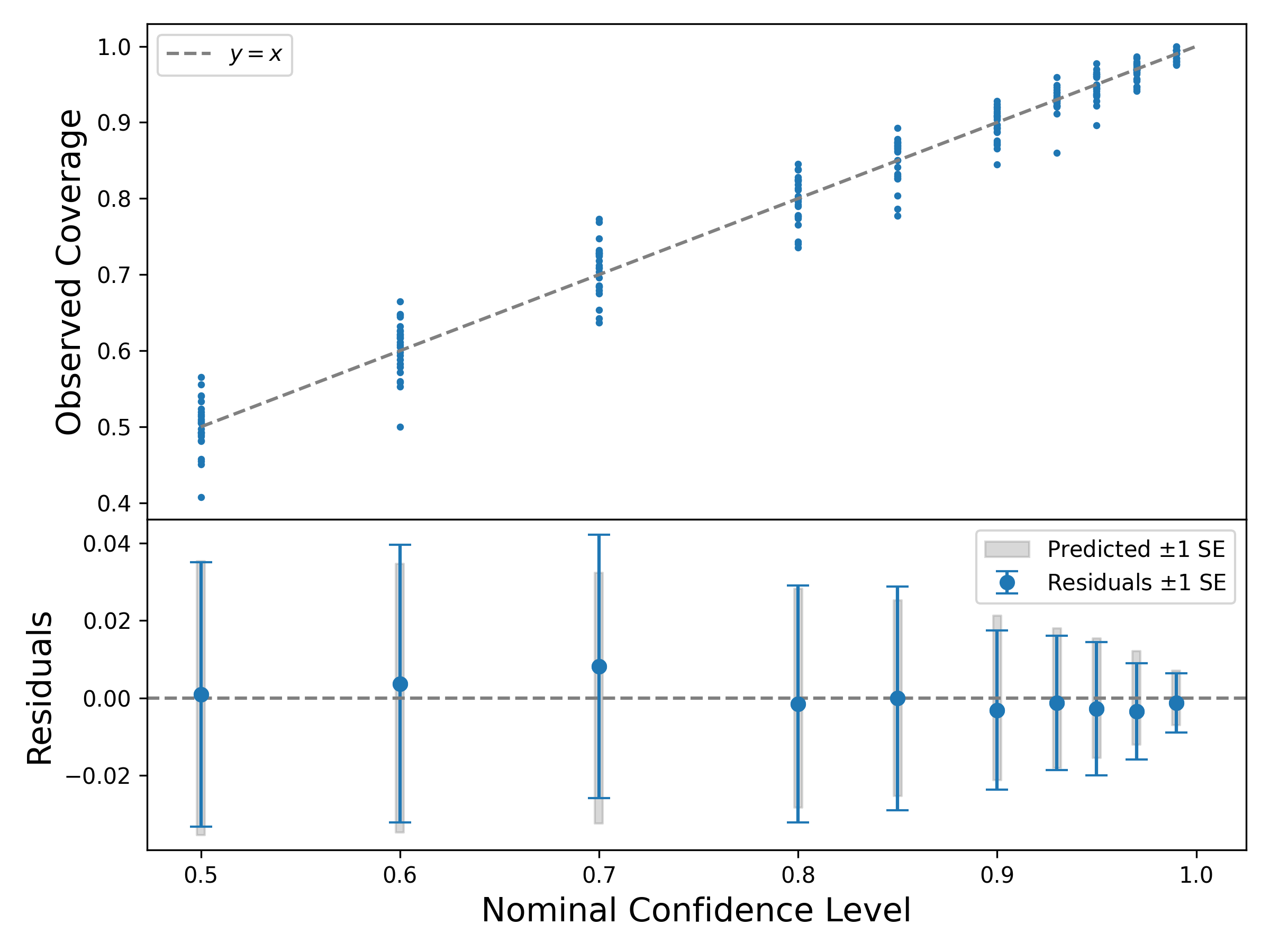}
        \label{fig:sbmbbar_cov}
        \caption{SciBooNE-MiniBooNe $\bar{\nu}_\mu$}
    \end{subfigure}

    \vspace{0.05\textwidth}
    
    \begin{subfigure}[b]{0.42\textwidth}
        \includegraphics[width=\linewidth]{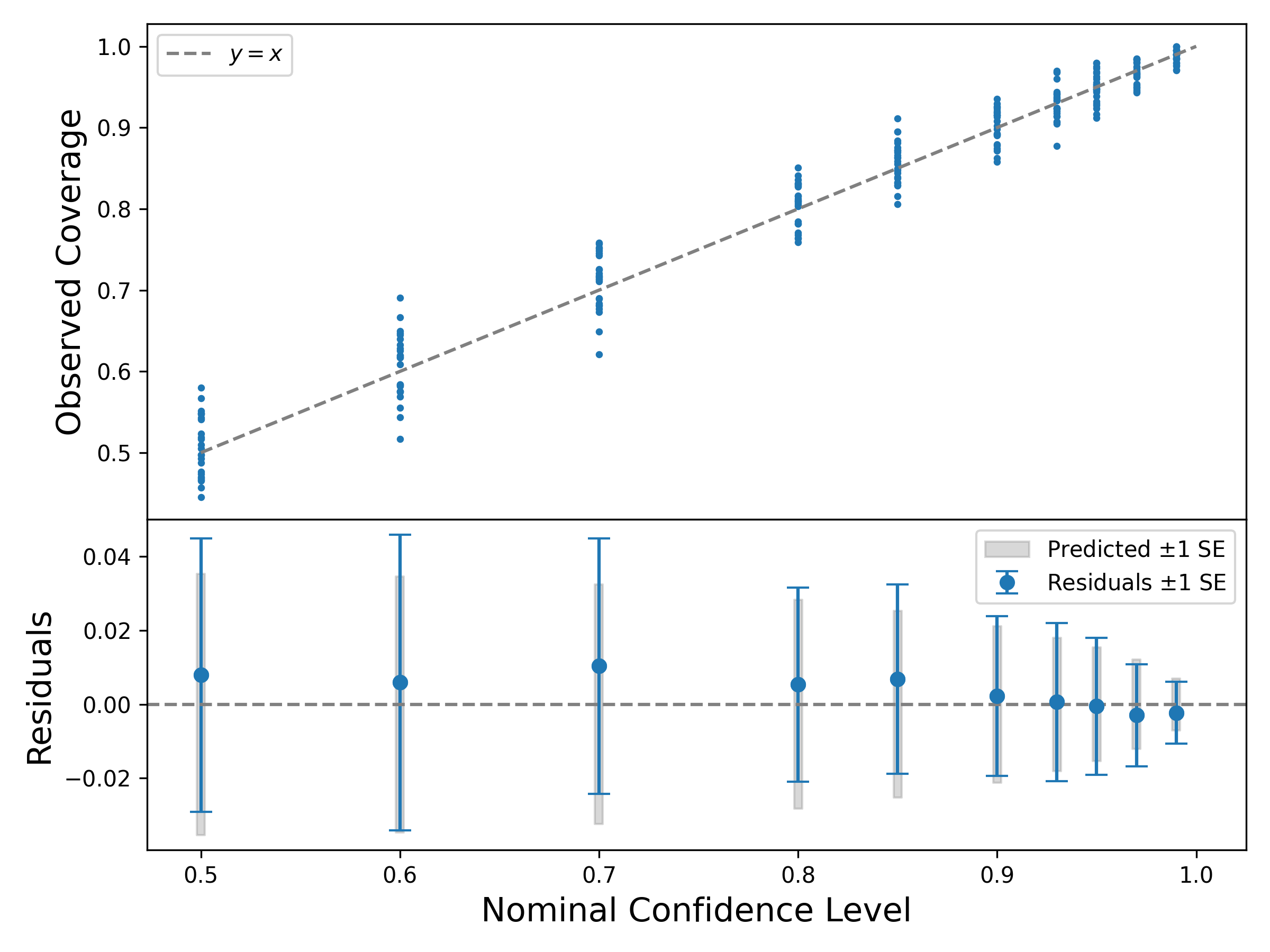}
        \label{fig:minos_cov}
        \caption{MINOS}
    \end{subfigure}
    \hspace{0.04\textwidth}
    \begin{subfigure}[b]{0.42\textwidth}
        \includegraphics[width=\linewidth]{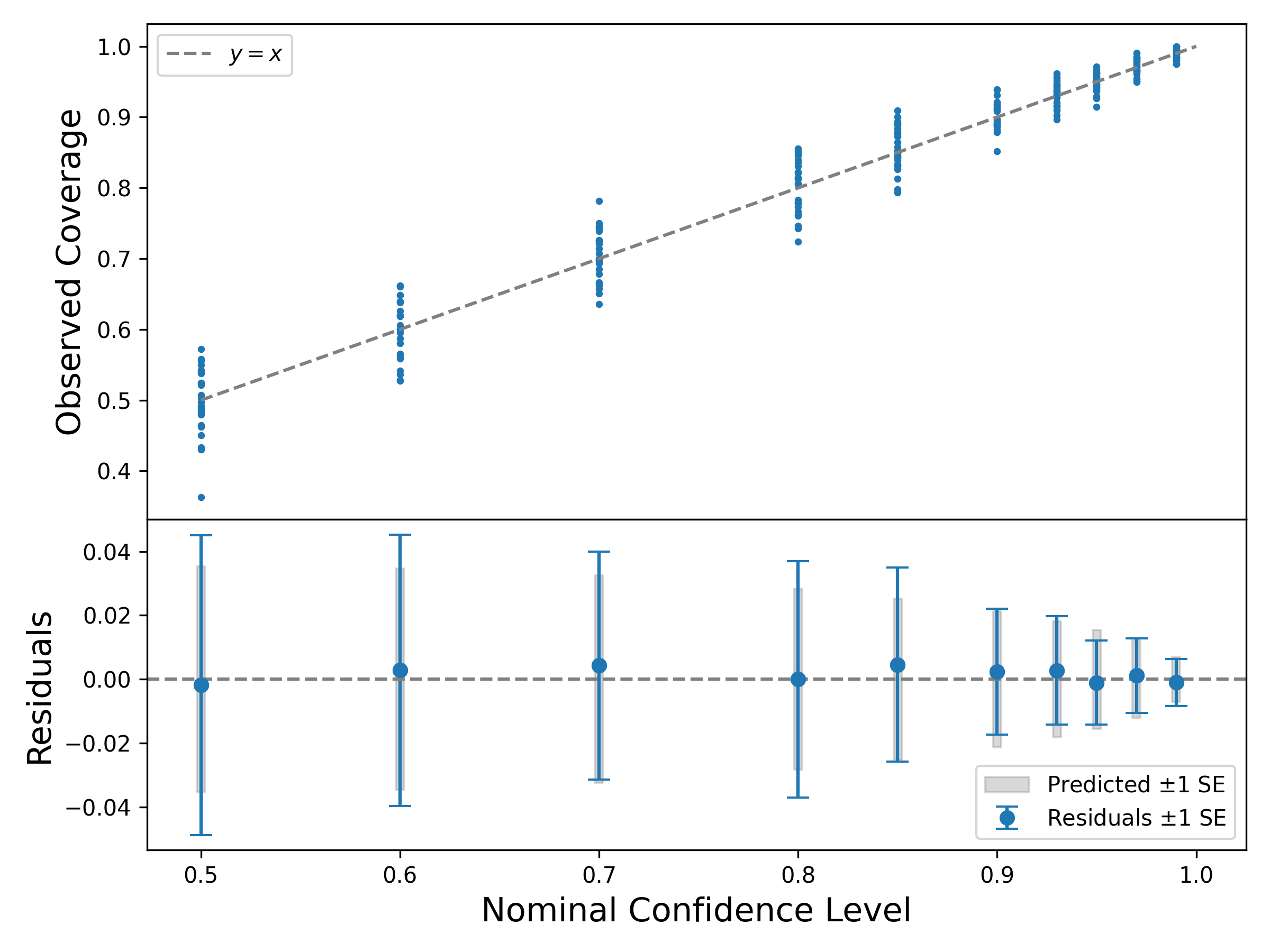}
        \label{fig:all_exp_cov}
        \caption{All experiments}
    \end{subfigure}
    \caption{Verified coverage of SBI-computed frequentist CLs for $25$ uniformly sampled points in the testable $3+1$ parameter space. For each, the top plot shows observed vs. expected frequentist coverage (with a well-calibrated fit showing clustering around the $45\degree$ line); the bottom plot shows the spread of columnwise residuals compared to approximate expectation of a well-calibrated fitter, see Sec.~\ref{subsubsec:coverage}.}
    \label{fig:exp_coverages}
\end{figure}

In Fig.~\ref{fig:exp_coverages}, we show the frequentist coverage of the SBI-FC CLs following the methods outlined in Sec. \ref{subsec:freq-cls}. Given the discussion in Sec.~\ref{subsubsec:coverage}, we observe excellent observed versus expected coverage agreement, with clustering around the $45\degree$ line and residual spreads consistent with expectations. Although the CDHS $50\%$ confidence level shows significant deviation from the predicted coverage, given domain-specific preferences for reporting CLs of at least $90\%$ confidence, we believe the calibration of the SBI-based fitting technique presented here is sufficient in the worst case and outstanding in the best.

\subsection{Fits to Observed Experimental Data}
\begin{figure}[htbp]
    \centering
    \begin{subfigure}[b]{0.45\textwidth}
        \includegraphics[width=\linewidth]{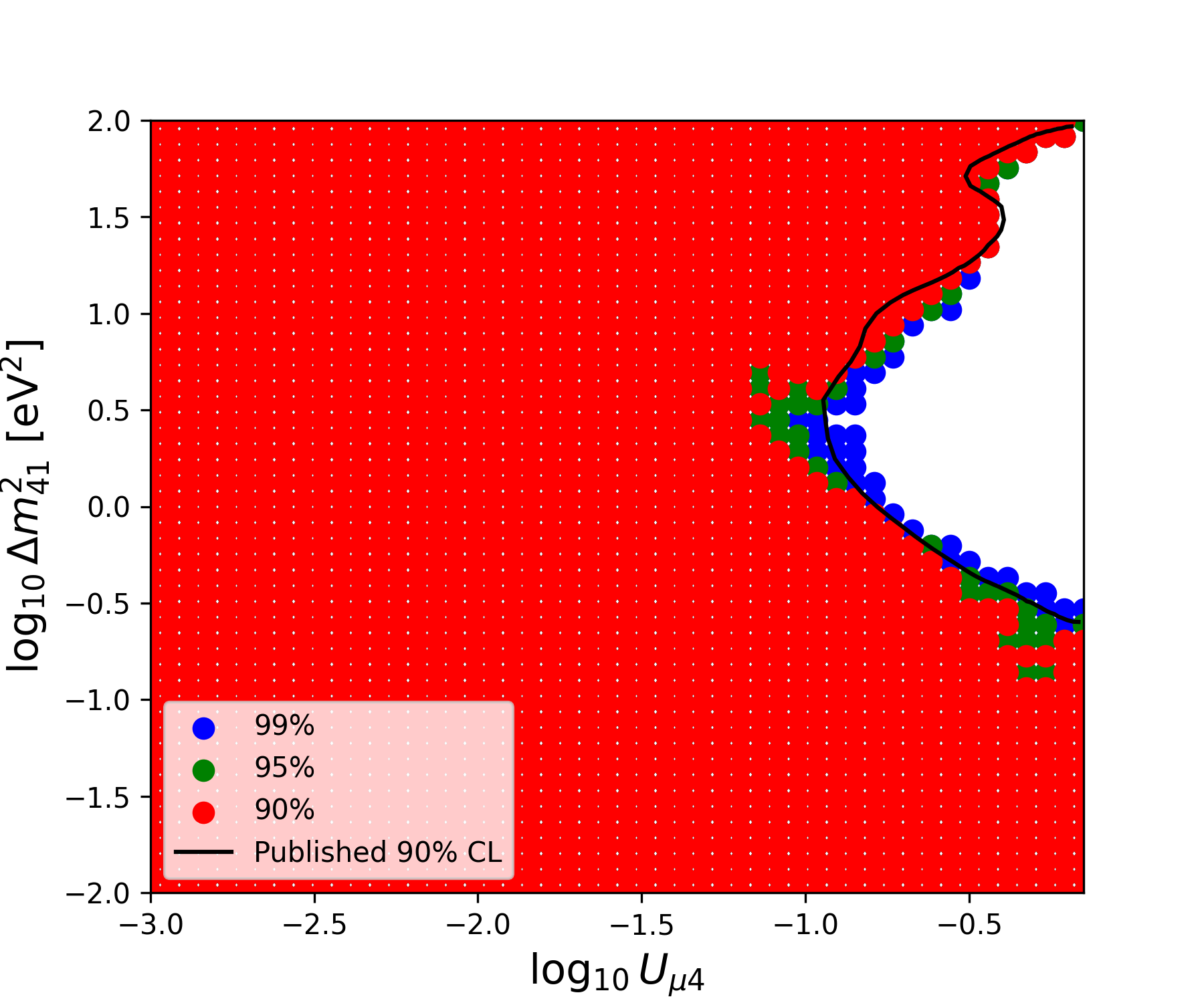}
        \caption{CDHS}
        \label{fig:cdhs_real_data}
    \end{subfigure}
    \begin{subfigure}[b]{0.45\textwidth}
        \includegraphics[width=\linewidth]{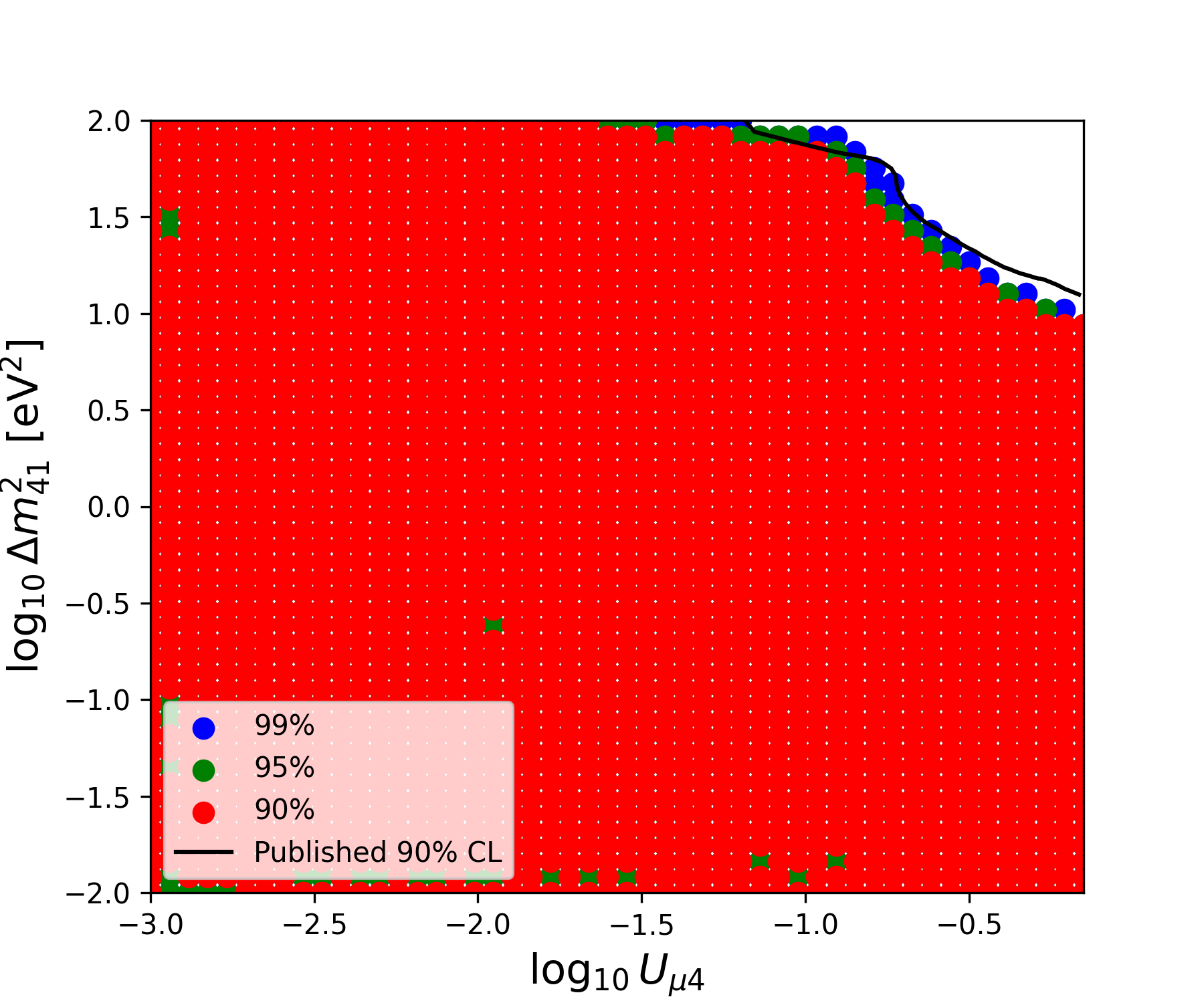}
        \caption{CCFR}
        \label{fig:ccfr_real_data}
    \end{subfigure}
    \begin{subfigure}[b]{0.45\textwidth}
        \includegraphics[width=\linewidth]{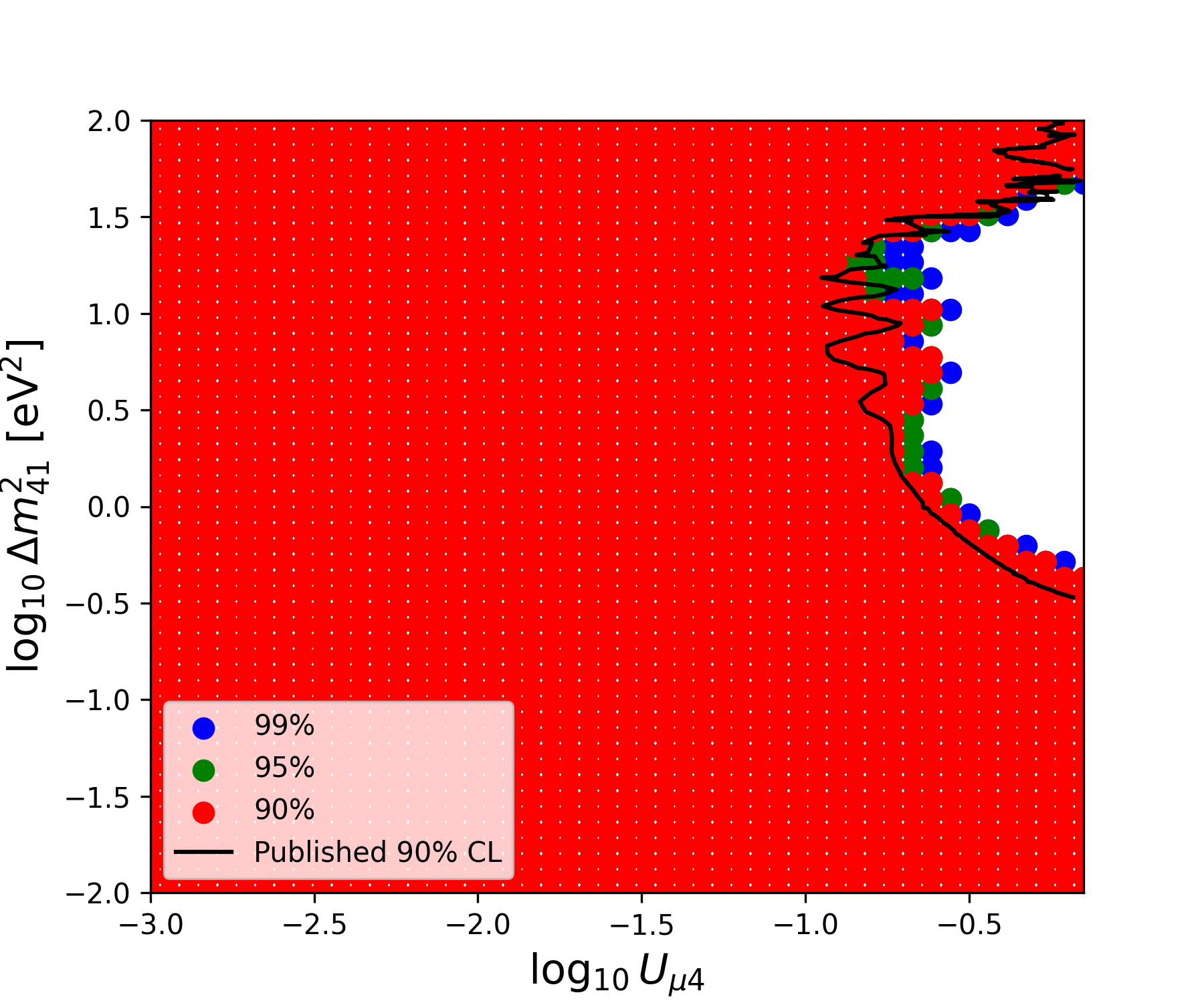}
        \caption{SciBooNE-MiniBooNe $\nu_\mu$}
        \label{fig:sbmb_real_data}
    \end{subfigure}
    \begin{subfigure}[b]{0.45\textwidth}
        \includegraphics[width=\linewidth]{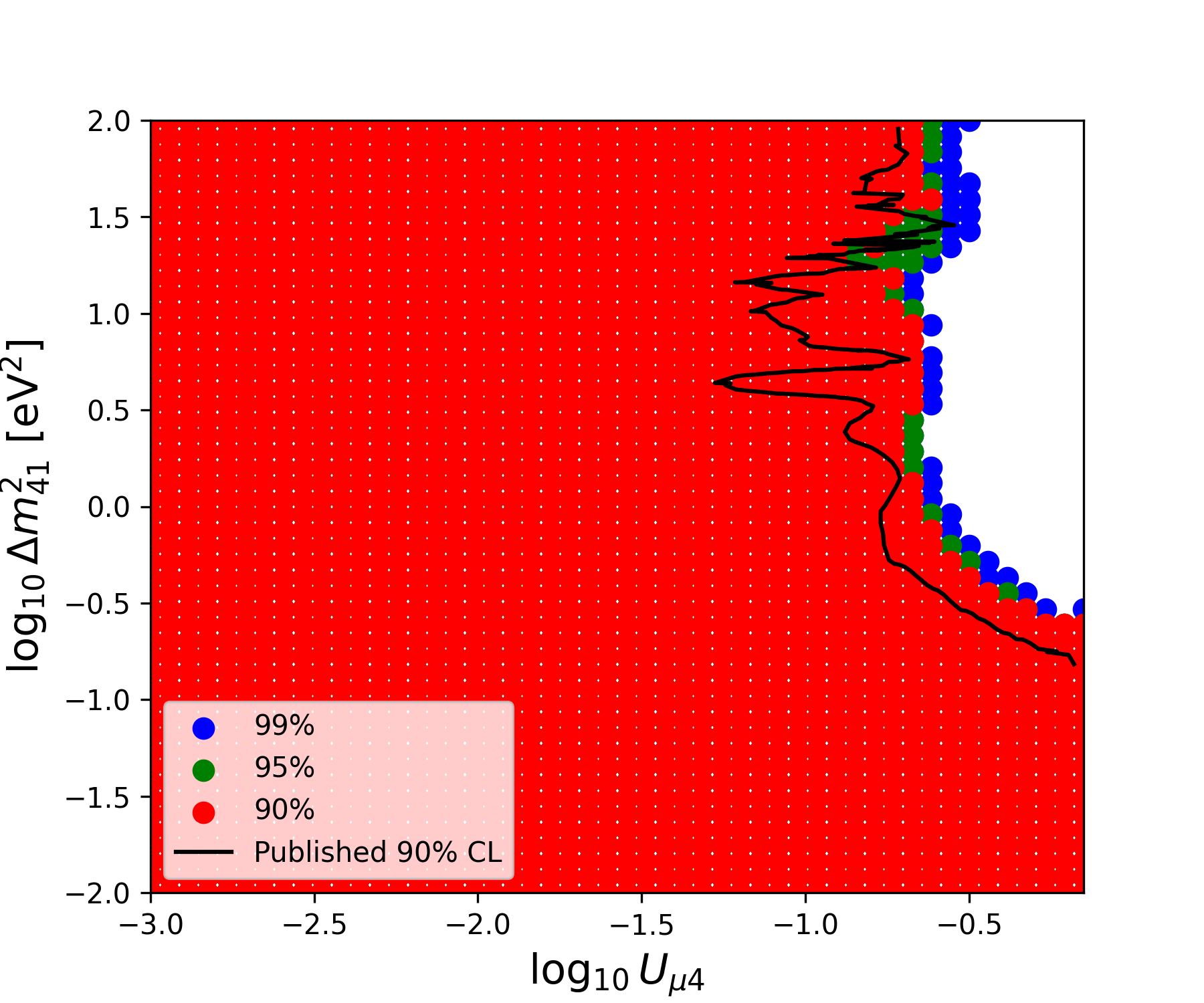}
        \caption{SciBooNE-MiniBooNe $\bar{\nu}_\mu$}
        \label{fig:sbmbbar_real_data}
    \end{subfigure}
    \begin{subfigure}[b]{0.45\textwidth}
        \includegraphics[width=\linewidth]{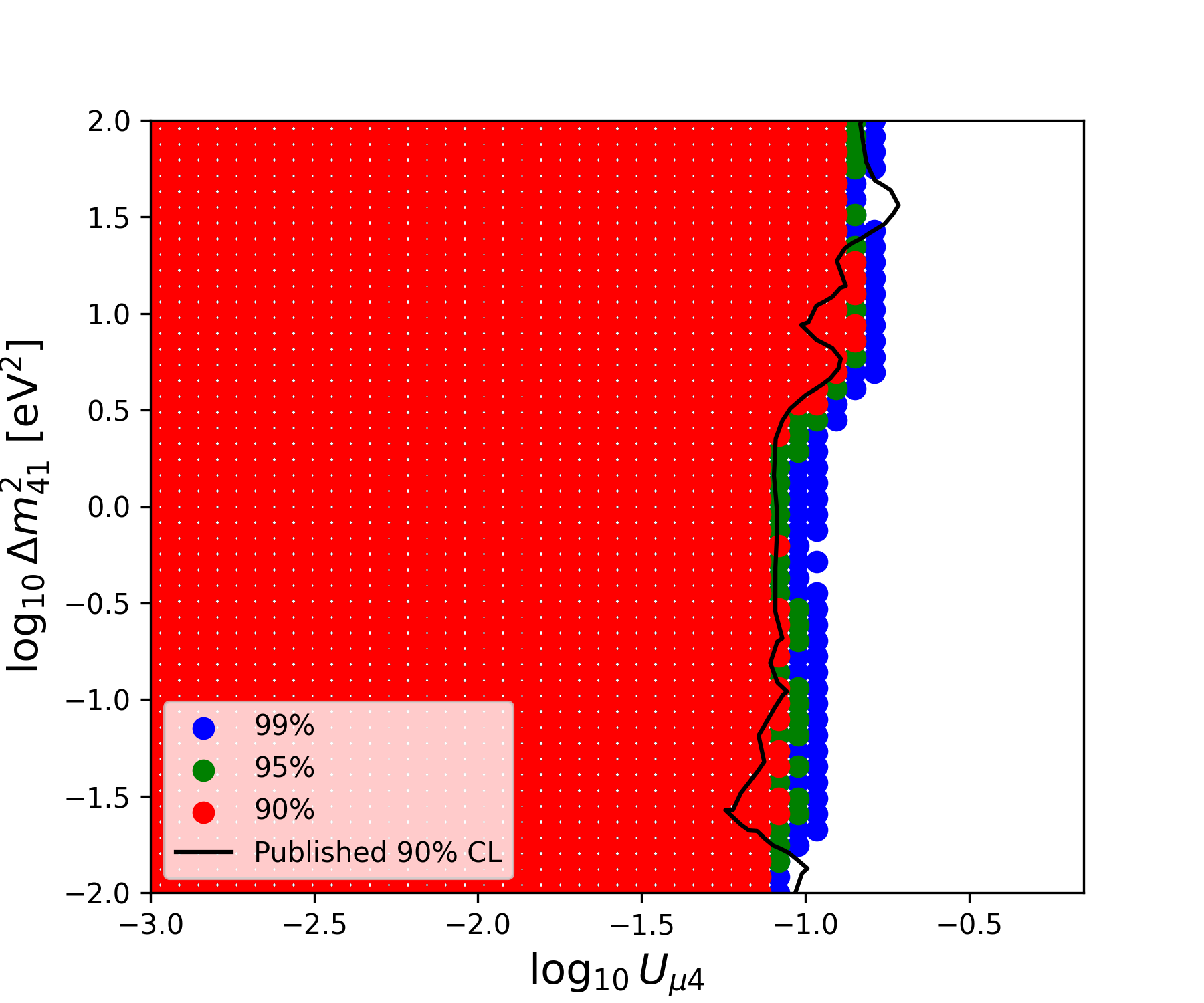}
        
        \caption{MINOS}
        \label{fig:minos_real_data}
    \end{subfigure}
    \begin{subfigure}[b]{0.45\textwidth}
        \includegraphics[width=\linewidth]{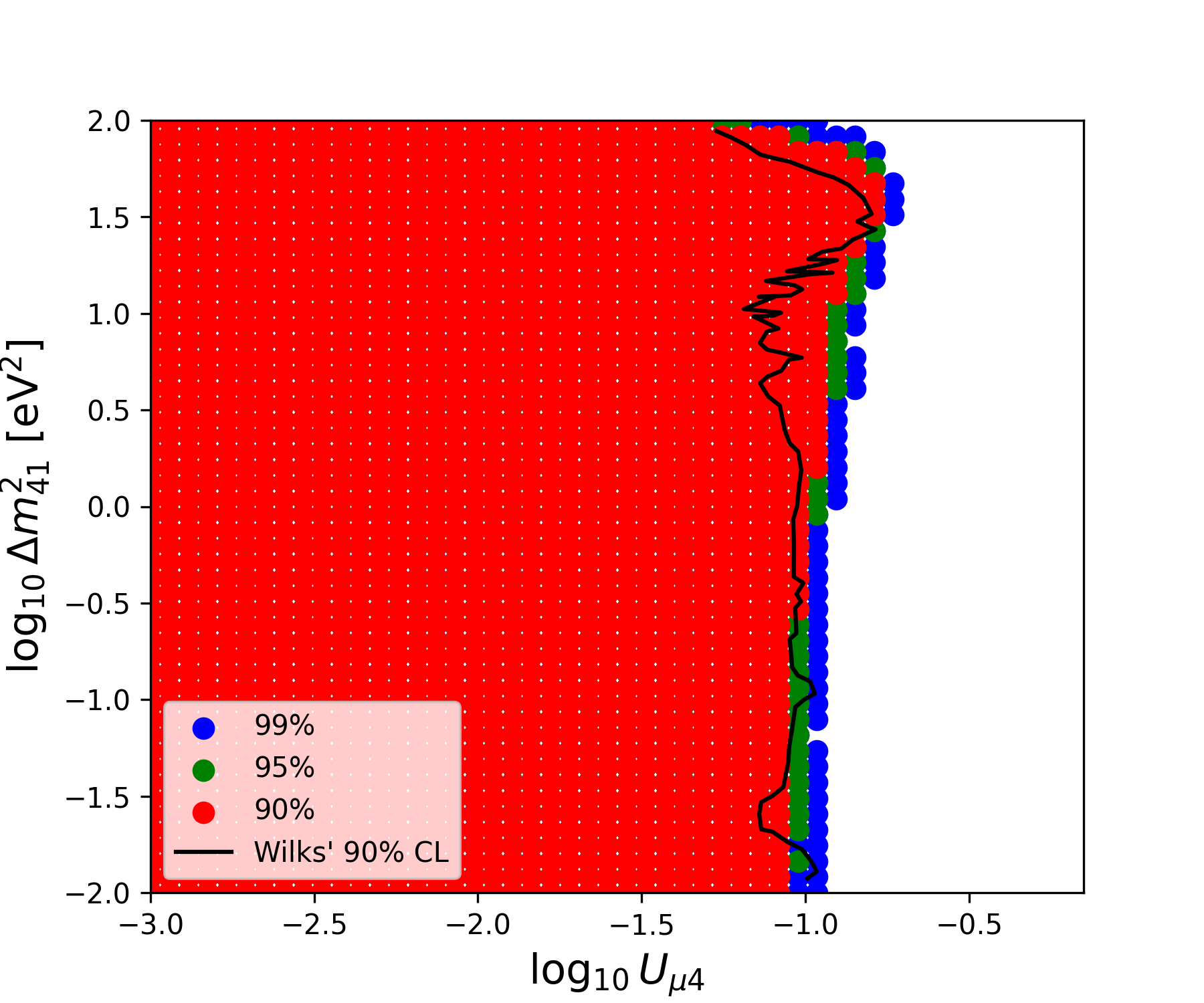}
        \caption{All experiments}
        \label{fig:all_exp_real_data}
    \end{subfigure}
    \caption{Allowed regions for each experiment and combined using real experimental data. The red/green/blue regions indicate the 90\%/95\%/99\% confidence regions.}
    \label{fig:allowed_regions}
\end{figure}

After confirming that the sensitivities computed by the SBI-FC on simulated data reasonably agree with expectations, we proceed to fit real experimental data. Fig. \ref{fig:allowed_regions} shows these results for each of the experiments considered here, as well as the global fit, with the red/green/blue regions corresponding to $90\%/95\%/99\%$ confidence regions respectively. The single-experiment $90\%$ confidence regions are compared to published results, and we observe relatively strong agreement between the two. Note that the MINOS $90\%$ CL drawn from real data sets stronger limits than the sensitivity, matching reported findings \cite{minos-two-detector}.
We discuss the CDHS result in Sec.~\ref{subsec:CDHS} and SciBooNE-MiniBooNE results in Sec. \ref{subsec:SBMB}. A global fit assuming Wilks' theorem is used as a point of comparison in Fig.~\ref{fig:sensitivities}f.

This example represents the first global fit to neutrino oscillation data that is trials-based.


%% file: Sec4_Discussion.tex
\section{\label{sec:discussion}Discussion}

\subsection{Limitations of Our Approach}
The Neyman-Pearson lemma states that the statistical test of size $\alpha$ having the most power in comparing a simple null hypothesis against a simple alternative hypothesis is one using the likelihood ratio as a test statistic \cite{Neyman1933-uw}. Plainly stated, any test using a neural network-based approximation of the likelihood ratio results in that test having less power. We acknowledge slight discrepancies in the SBI-constructed confidence regions and those published by experiments using a high-fidelity approach in Fig.~\ref{fig:allowed_regions}. That said, qualitative agreement between constructed confidence regions and validation of coverages provides evidence for the fact that an SBI-based method for interpreting sterile neutrino global fits is still useful, particularly given that leveraging such machine learning techniques made the joint evaluation of this muon-neutrino disappearance data possible for the first time, reducing computational overhead and avoiding reliance on Wilks' theorem. 

What's more, the Monte Carlo generation scripts used in this work do not provide a complete end-to-end picture of each experiment studied. Full-picture experimental simulations are almost never made open-source at the time of data release, and global fitters must resort to simulators which draw summary statistics of experimental outcomes, such as counts per baseline and energy bin. Unavoidably, there is less information contained in a Monte Carlo constructed by stitching together reported expectations and covariance matrices than contained by an in-house simulation suite. Altogether, the quality of an SBI-based fit is fundamentally limited by the quality of the simulator one uses to train.

\subsection{A New SBI-Informed Interpretation of CDHS Data\label{subsec:CDHS}}
\begin{figure}[htbp]
    \centering
    \includegraphics[width=0.6\linewidth]{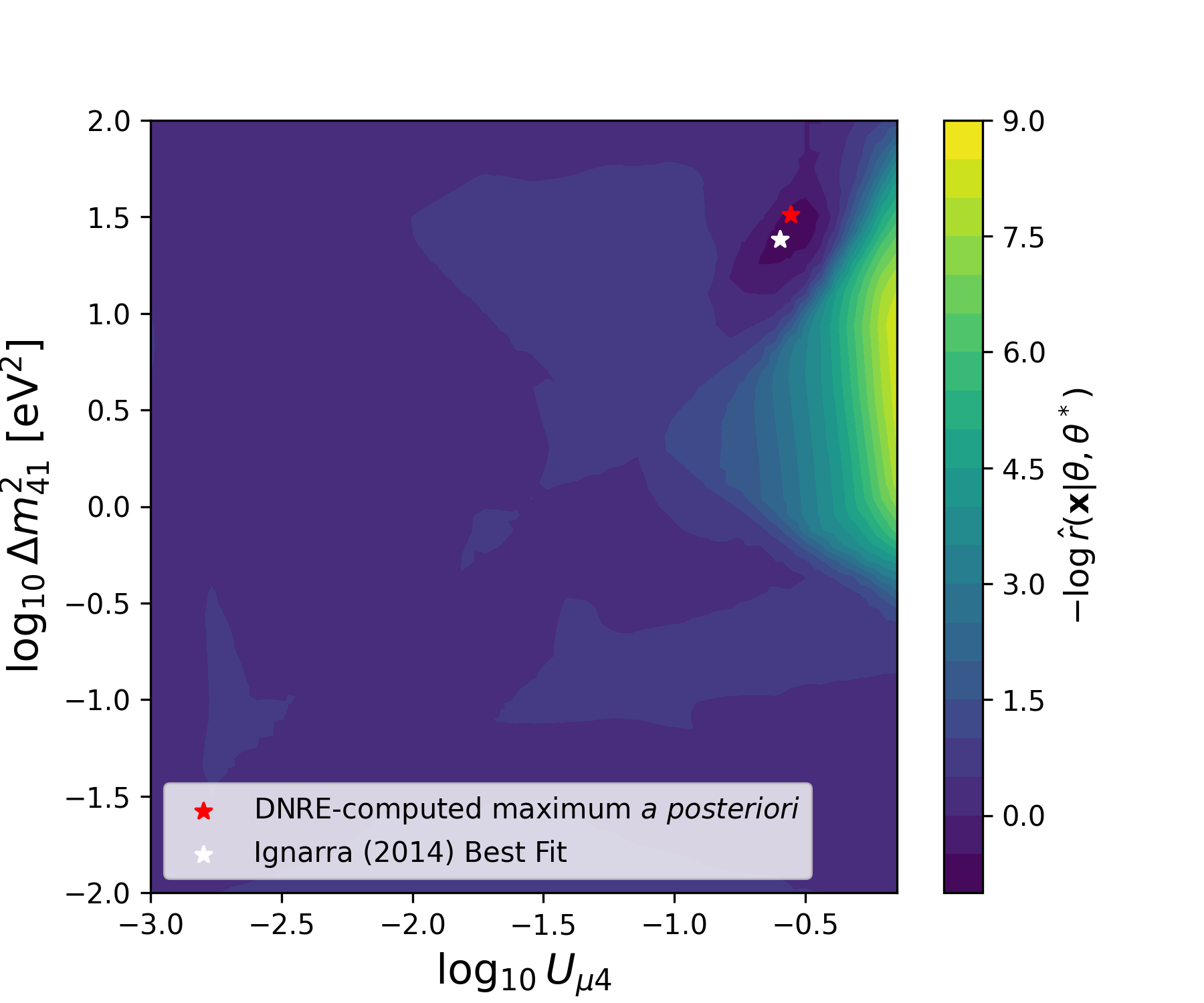}
    \caption{Distribution of the SBI-computed test statistic across $3+1$ parameter space using real data from CDHS. ``Ignarra (2014) best fit" refers to the best fit reported in Fig.~5-2 of Ref.~\cite{Ignarra:2014yqa}. The DNRE-computed maximum \textit{a posteriori} corresponds to the point $\sin^2 2\theta_{\mu \mu} = 0.284$, $\Delta m_{41}^2 = 32.4$ eV$^2$, while the best fit from Ref.~\cite{Ignarra:2014yqa} is at $\sin^2 2\theta_{\mu \mu} = 0.238$, $\Delta m_{41}^2 = 24.1$ eV$^2$.}
    \label{fig:CDHS_bestfit}
\end{figure}

Past reanalysis of the CDHS experiment \cite{Ignarra:2014yqa} reports a closed contour at $90\%$ confidence, indicating a slight preference for $3+1$, despite the fact that the CDHS collaboration itself reports an exclusion at $90\%$ confidence \cite{DYDAK1984281}. This discrepancy is attributed to the ``flip-flopping" discussed in Ref.~\cite{feldman-cousins}. The CDHS collaboration reports an exclusion curve by way of performing a \textit{raster scan}: for each value of $\Delta m_{41}^2$, $\sin^2 2 \theta_{\mu \mu}$ are admitted into the one-dimensional CL such that each has $\Delta \chi^2 < 2.71$ (the two-sided 90\% CL for a $\chi^2$ distribution having one degree of freedom), with the best fit $\sin^2 2 \theta_{\mu \mu}$ possibly changing as $\Delta m_{41}^2$ varies. The union of these one-dimensional CLs therefore forms a two-dimensional CL over testable model parameter space. Ref.~\cite{Ignarra:2014yqa} opts for the two-dimensional approach where a global minimum likelihood is determined over all $(\sin^2 2 \theta_{\mu \mu}, \Delta m_{41}^2)$, and points are admitted to the $90\%$ CL so long as they have a $\Delta \chi^2 < 4.61$, the two-sided $90\%$ CL for a $\chi^2$ with two degrees of freedom. Both techniques assume Wilks' theorem, and as Feldman and Cousins point out, both are biased, since it is the decision of the analyzer whether to opt for the raster or two-dimensional scan based on the analyzer's preference, resulting in a CL which may appear more like a closed contour (indicating a signal) or an exclusion (consistency with the SM) \cite{feldman-cousins}. Ambiguity of whether to apply a raster or global scan is exactly one of the motivating factors that Feldman and Cousins cite in the development of a trials-based approach to CL construction. Admitting parameter points based instead on comparisons to a properly-covering critical test statistic value is agnostic to such a choice.

The SBI-FC method presented in this work follows the recommendation of Feldman and Cousins to construct confidence intervals based on trials, which bypasses the need for an analyzer to make a biased choice of whether to report evidence for a signal or an exclusion based on their personal preference for how the data look. We point out that the maximum \textit{a posteriori} best fit point determined by DNRE seems to agree with the best fit point reported in Ref.~\cite{Ignarra:2014yqa}, as shown in Fig. \ref{fig:CDHS_bestfit}. Discrepancies in its exact location can be attributed to approximation errors of DNRE and the fact that the two best fit points are slightly different estimators fundamentally.

\subsection{A New SBI-Informed Interpretation of SciBooNE-MiniBooNE Disappearance Data\label{subsec:SBMB}}

\begin{figure}[htbp]
    \centering
    \includegraphics[width=0.8\linewidth]{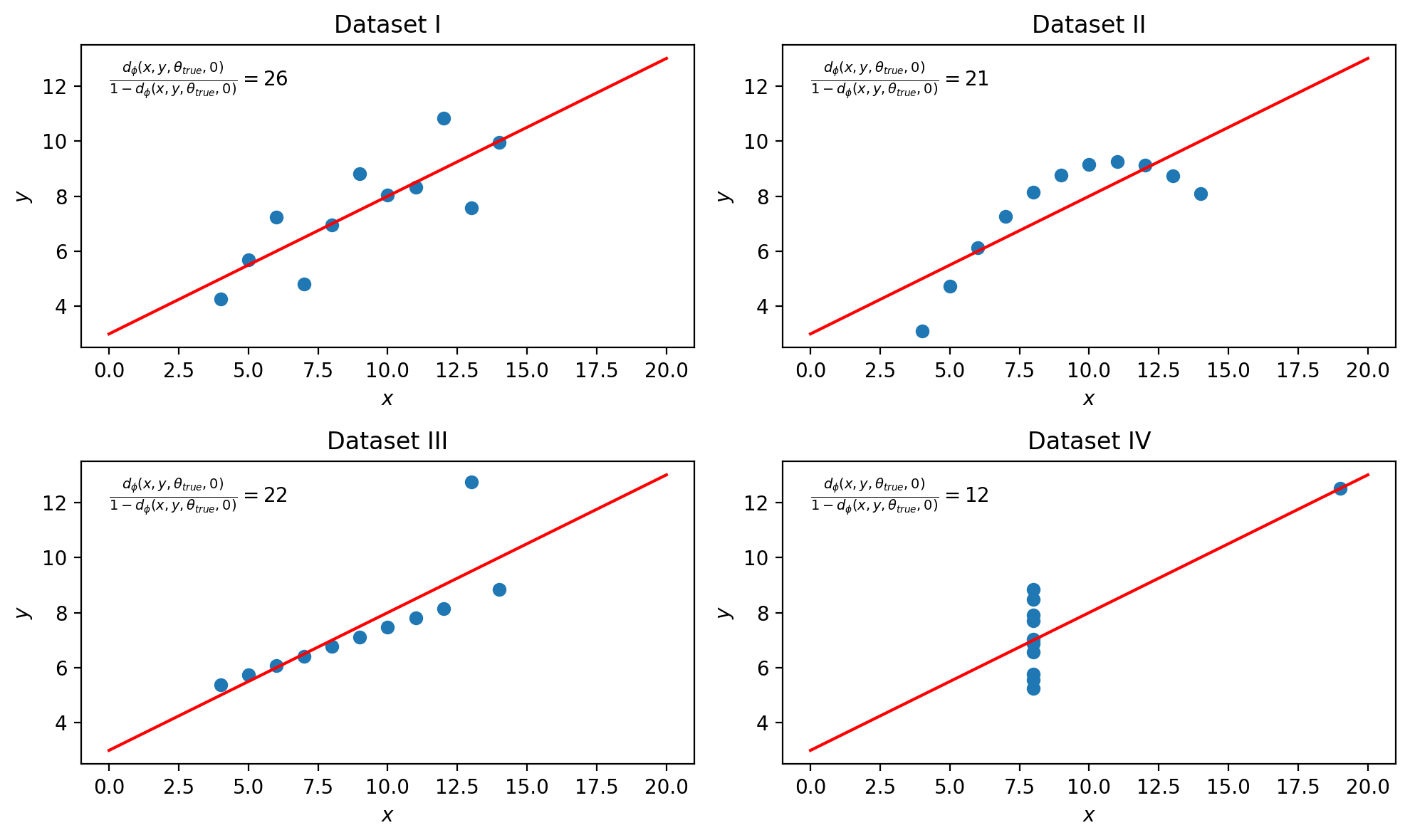}
    \caption{\hl{Anscombe's quartet of datasets, and network-estimated test statistics for each.}}
    \label{fig:sbi-anscombe}
\end{figure}

In general, the SBI-computed exclusions shown in Fig.~\ref{fig:allowed_regions} are in good agreement with the results published by the experiments. In the SciBooNE-MiniBooNE disappearance searches, however, one observes significant disagreement for both the $\nu_\mu$ and $\overline{\nu}_\mu$ cases. Given the same apparatus for the $\bar \nu$ and $\nu$ run, we expect common origins for these similar discrepancies.  We will refer to the poorer-matching $\bar \nu_\mu$ disappearance search for our discussion. First, the published exclusion for $\bar \nu_\mu$ disappearance shows wildly varying values of $\sin^2 2 \theta_{\mu\mu}$ in the region $0 \lessapprox \log (\Delta m_{41}^2 / \text{eV}^2) \lessapprox 1.5$ \cite{PhysRevD.86.052009}
that are not seen in the SBI-FC result.   Second, the SBI-FC exclusion covers less parameter space than its published equivalent.   We discuss each of these differences in detail to determine whether the SBI-FC result should be accepted.  

First, consider the published exclusion's variations in $\sin^2 2 \theta_{\mu\mu}$, which are much larger than those seen in the other experiments and not present in the published sensitivity or the sensitivity that we obtain (Fig.~\ref{fig:sbmbbar_sensitivity}).   Typically, such variations appear in a $\chi^2$-based fit when random fluctuations in the data are misattributed to an oscillation signal.  In the case of the SciBooNE-MiniBooNE $\bar \nu_\mu$ analysis, computational burdens of the experimental Monte Carlo forced analyzers to limit the number of samples used in modeling uncertainties via a covariance matrix. In fact, a domain rule-of-thumb is to have an order of magnitude more Monte Carlo samples than expected events, but the SciBooNE-MiniBooNE $\bar \nu_\mu$ disappearance analysis was only able to generate a factor of three more MC than real events \cite{GaryChengThesis}. A covariance built from too few Monte Carlo samples can destabilize the fit, introducing additional (but physically meaningless) jitter in an exclusion curve. These wildly varying exclusions are sufficiently common that Feldman and Cousins themselves advise analyzers to report their experimental sensitivity alongside the exclusion, and emphasize the caution that must be taken in interpreting such variations as physically meaningful \cite{feldman-cousins}.   In contrast, the SBI-FC method does not suffer from this problem; the DNRE network is trained to recognize and ignore such random jitter, avoiding strong variations which can impact fit interpretations.

The second difference, the more conservative limit set by the SBI-FC method, is due to another  feature specific to the joint SciBooNE-MiniBooNE disappearance search:  significant correlated structure in the data can yield a weaker estimated test statistic since the effect is unrecognized by the SBI method.  To see this, let us first introduce how a DNRE procedure responds to data that is unexpected and different from those seen in training.   We then argue that the SciBooNE-MiniBooNE data has such a feature.

To examine how a SBI method responds to unexpected deviations of data, we will make use of four datasets called Anscombe's quartet \cite{Anscombe1973-lj}, seen in  Fig.~\ref{fig:sbi-anscombe}.  These datasets have the extraordinary property that, despite clearly appearing different by eye, they share summary statistics, including the $\chi^2$ and the  parameters of their lines of best fit, both of which agree to two significant digits \cite{Anscombe1973-lj, Conrad:2016sve}. Thus, a $\chi^2$ test for a straight line does not differentiate between the datasets, even though dataset I clearly reflects random jitter about a line while datasets II to IV have significant systematic (non-random) effects that deviate. If one trains a DNRE network on datasets showing random jitter about lines (for details, see \ref{appendixc:anscombe}), like dataset I, the network \hl{may identify} dataset I as more likely to be generated from a line than datasets II to IV. 
One sees that in the presence of systematic effects not included in network training data (like those seen in datasets II–IV), the network-estimated likelihood ratio is smaller than in dataset I, indicating that dataset I is most consistent with a true linear structure.    Thus, such an SBI method can recognize and flag underlying systematic effects that were not included in the training, a feature that sets such an inference method apart from those based on classic statistics like a $\chi^2$.

\begin{figure}[htbp]
    \centering
    \includegraphics[width=0.8\linewidth]{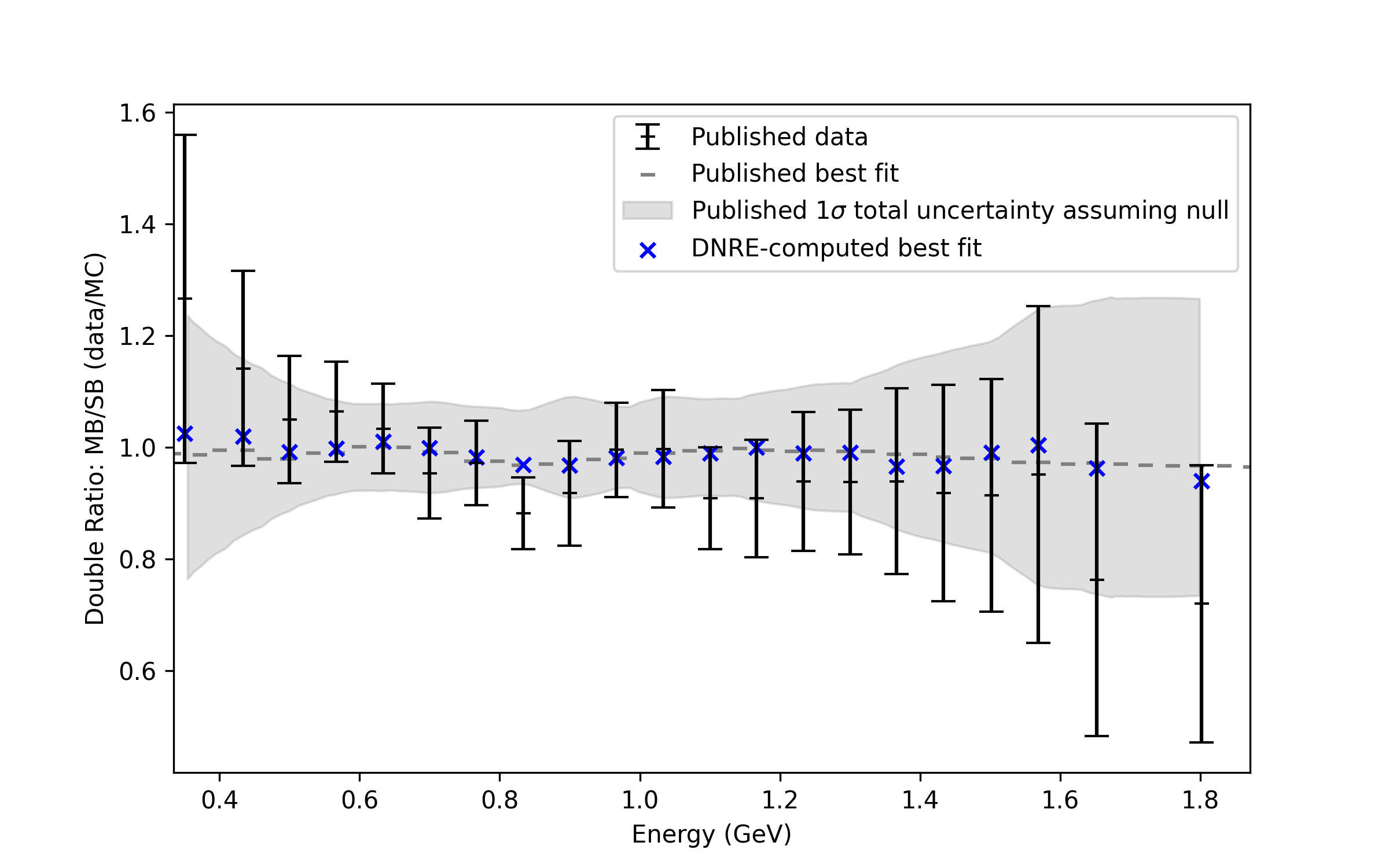}
    \caption{Double ratio ($\frac{\text{MiniBooNE data/MC}}{\text{SciBooNE data/MC}}$) as a function of neutrino energy for $\bar{\nu}_\mu$ disappearance, where MC is generated from null. Published best fit, $1\sigma$ systematic and statistical uncertainties assuming SM oscillations, and experimental data are extracted from Fig. 17 in Ref.~\cite{PhysRevD.86.052009}. Points marked with a blue $\times$ correspond to energy-binned means of simulated realizations generated from the best fit $3+1$ oscillation parameters found by the DNRE maximum \textit{a posteriori}: $\sin^2 2\theta_{\mu \mu} = 0.061$, $\Delta m_{41}^2 = 1.93$ eV$^2$. The best fit from Ref.~\cite{PhysRevD.86.052009} is reported to be $\sin^2 2\theta_{\mu \mu} = 0.086$, $\Delta m_{41}^2 = 5.9$ eV$^2$.}
    \label{fig:SBMBnubar_doubleratio}
\end{figure}

The treatment of systematic uncertainties for the SciBooNE-MiniBooNE dataset is provided by the form of the published covariance matrix used to throw the training data for this study; however, there is evidence that there are residual systematic effects not well described by this uncertainty.   To see this, Fig.~\ref{fig:SBMBnubar_doubleratio} shows collected experimental data for $\bar{\nu}_\mu$ disappearance, reported as the double ratio of MiniBooNE data/MC to SciBooNE data/MC, as a function of neutrino energy \cite{PhysRevD.86.052009, GaryChengThesis}. In the double ratio, shared systematic effects cancel, with the remaining $1\sigma$ systematic uncertainty indicated by the shaded region in the figure.  The data show correlated behavior exceeding this remaining systematic uncertainty in places, indicating a potential underlying systematic effect not included in the covariance matrix, and therefore not included in the training samples provided to SBI-FC.  As with the examples from Anscombe’s quartet, the underlying systematic confuses the DNRE, weakening the SciBooNE-MiniBooNE exclusion as a result.

These arguments make the case that the smoother and weaker exclusion from the SBI-FC method is likely to be a better representation of the true confidence level associated with these data. In the worst case, such a treatment of oscillation data is informative,  playing an important role in flagging underlying systematic effects in experimental data which could impact the interpretation of clues for or against BSM physics.

\subsection{\label{subsec:icecube}A Case Study of the Value of Fast Global Fits: Interpreting the IceCube $3+1$ Result}
\begin{figure}[htbp]
    \centering
    \includegraphics[width=0.6\linewidth]{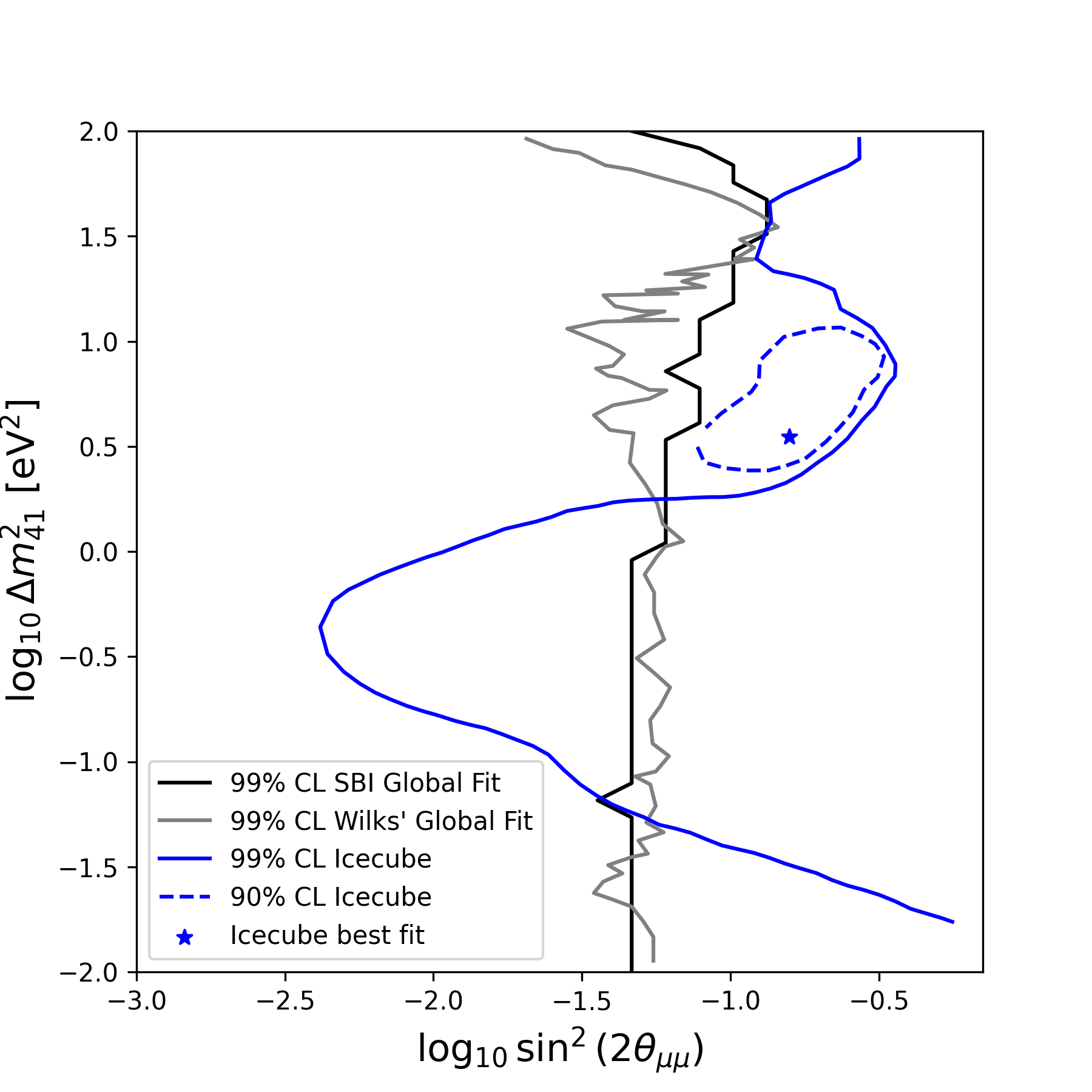}
    \caption{The SBI-FC $99\%$ exclusion from the global fit overlaid with the $99\%$ exclusion and $90\%$ allowed region drawn from IceCube data, taken from \cite{icecube-sterile-prl}. For comparison, the $99\%$ exclusion from a global fit using Wilks' theorem is also shown. The IceCube best fit is reported at $\sin^22\theta_{\mu \mu} = 0.16$, $\Delta m_{41}^2 = 3.5$ eV$^2$.}
    \label{fig:icecube_comparison}
\end{figure}

In this section we illustrate the value of producing fast and accurate trials-based global fits to interpret new data sets.  
Recently, the IceCube neutrino telescope reported a $3+1$ search \cite{icecube-sterile-prl, icecube-sterile-prd} with a markedly different experimental setup to those considered in Tab.~\ref{tab:experiments}. At IceCube, cosmic rays collide with matter in Earth's atmosphere, producing mesons that decay to neutrinos and antineutrinos. These traverse the Earth to the IceCube Neutrino Telescope located at the South Pole.   Similar to the Mikheyev–Smirnov–Wolfenstein effect---where neutrinos propagating through matter of varying density undergo resonant oscillations \cite{Mikheyev:1985zog, Mikheev:1986wj, MSWPhysRevD.17.2369}---the presence of one or more sterile neutrinos can cause $\nu_\mu$ to exhibit a resonant disappearance as they travel through the Earth. This behavior is governed by the same model parameters as vacuum oscillations observable by accelerator-based experiments.  As such, IceCube's search for a sterile neutrino offers an important complementary addition to the discussion of a $3+1$ sterile neutrino in a global picture.  

It is surprising, then, to observe in 
Fig.~\ref{fig:icecube_comparison} that IceCube reports an allowed region at the $90\%$ CL (blue) using the Wilks method with coverage spot-checks provided by Feldman-Cousins.    The best fit is reported at $\sin^22\theta_{\mu \mu} = 0.16$, $\Delta m_{41}^2 = 3.5$ eV$^2$. 
At the time that these data were published, there was no available $\nu_\mu$ disappearance global fit for the experiment to cite, making the result difficult to interpret.
To understand the implications of IceCube's allowed region, we are now able to ask if the IceCube result agrees with the data from existing $\nu_\mu$ disappearance searches using the global fit (the method for which is described in  Sec.~\ref{subsec:globalfit}).

In order to test agreement, we will compare the significance of the IceCube best fit point to the level of exclusion from our SBI-FC global fit (Fig.~\ref{fig:icecube_comparison}, black line).  Using Feldman-Cousins to evaluate the test statistic with respect to the null (no-sterile) hypothesis, the IceCube experiment rejects the null at $2.2 \sigma$ \cite{icecube-sterile-prl}.  Our SBI-FC analysis excludes sterile neutrino oscillations for the IceCube best fit parameters at $\sim3\sigma$ (see \ref{appendixc:icecube-sig}).  Thus the exclusion from our global fit is stronger than the significance of the IceCube result, implying that this model is unlikely to be the correct explanation.  

For comparison, Fig.~\ref{fig:icecube_comparison} also presents a traditional Wilks-based global fit (gray line) to the same data sets calculated using the code described in Ref.~\cite{wherearewe}.   The Wilks-based global fit gives a stronger exclusion than the SBI-FC trials-based approach.  This highlights an important point of this paper: a Wilks-based global fit can be misleading compared to a trials-based result, which is the better representation.

\subsection{\label{subsec:timing}Timing Considerations}
\begin{table} 
\centering
\caption{\label{tab:timing}End-to-end timing comparisons for the high-fidelity and SBI-charged Feldman-Cousins methods for a global fit to muon-(anti)neutrino disappearance data. The Monte Carlo generation is very fast and depends only on the number of bins.  We represent the speed as $\mathcal{O}(1)$ because it is independent of the difficulty of minimizing on the parameters of a given model.  It depends only on the dimension of the model via number of grid points and the number of bins in the experiment.}

\begin{tabular*}{0.95\textwidth}{@{}lcc}
\br
Step & Feldman-Cousins & SBI-FC \\
\mr
Monte Carlo generation & $\mathcal{O}(1)$ & $\mathcal{O}(1)$ \\
Network training & N/A & $\approx 1 \text{ hr}$ \\
Best fit point determination (per realization) & $\approx 10 \text{ hrs}$ & $\approx 1 \text{ s}$ \\
Test statistic computation (per realization) & $\approx 1 \text{ s}$ & $\approx 1 \text{ s}$ \\
Critical test statistic computation (per grid point) & $\approx 1 \text{ s}$ & $\approx 1 \text{ s}$ \\
\br
\end{tabular*}
\end{table}

Despite amortized up-front costs of Monte Carlo pseudodata generation and the network training required by DNRE, an SBI-FC fit to data (once a network is trained) is at least $\mathcal{O}(10^4)$ faster per grid point in parameter space when compared to the high-fidelity global fit. The bottleneck for a complete Feldman-Cousins treatment of global fit data is the fit itself for reasons discussed in Sec.~\ref{subsec:globalfit}. A network-powered fitting framework does not suffer nearly as much from the computational burdens of joint likelihood minimization; the inclusion of more experiments is straightforward and only slightly impacts network training time. A proper global fit of $3+1$ data would involve the inclusion of perhaps twenty or more experiments which not only span muon-flavor disappearance, but also electron-flavor disappearance and electron-flavor appearance, making measurements in a three-dimensional model parameter space rather than the two-dimensional one presented in this proof-of-principle. For this reason, a Feldman-Cousins treatment of these data in a true, proper global fit is impossible with existing computational resources.

%% file: Sec5_Conclusion.tex
\section{\label{sec:Conclusion}Conclusion}

In this paper, we have presented, for the first time, a frequentist framework for building confidence intervals leveraging simulation-based inference in the context of sterile neutrino global fits. Computational limitations of higher-fidelity fitting frameworks like the Feldman-Cousins method make the proper analysis of these data, until this point, impossible. We have presented a ML-based method with per-grid point factor of $>10^4$ improvement in fit runtime compared to that of Feldman-Cousins.   Using the example of a $\nu_\mu$ disappearance global fit, we have shown that Wilks-based approaches differ from trials-based approaches like the Feldman-Cousins method and its SBI extension, emphasizing the value of our technique.    We have also shown that the SBI-based approach has additional value in recognizing underlying systematic effects compared to a $\chi^2$ fit.   We provide a toy model SBI-FC fit example in Ref.~\cite{toygithub} that will allow users to make use of our method.  We anticipate that the continued use of such ML techniques will answer yet-untackled questions about the existence of one or more sterile neutrinos and other beyond-Standard Model scenarios.

%% file: Appendix.tex
\section{\label{sec:appA_ROC-AUC}ROC-AUC Scores}

\begin{figure}[H]
    \centering
    
    \begin{subfigure}[b]{0.45\textwidth}
        \includegraphics[width=\linewidth]{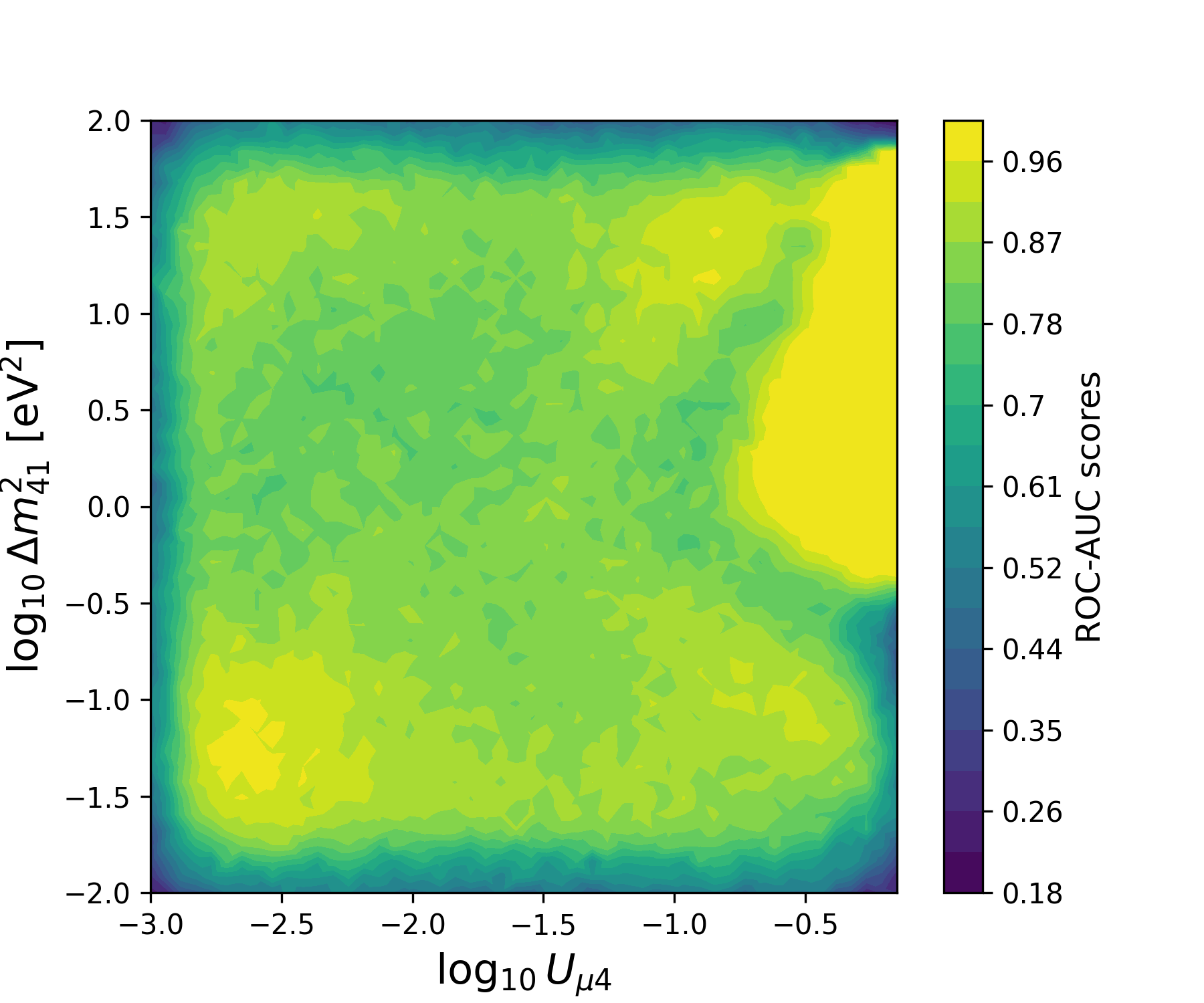}
        \label{fig:cdhs_roc-auc}
        \caption{CDHS}
    \end{subfigure}
    \begin{subfigure}[b]{0.45\textwidth}
        \includegraphics[width=\linewidth]{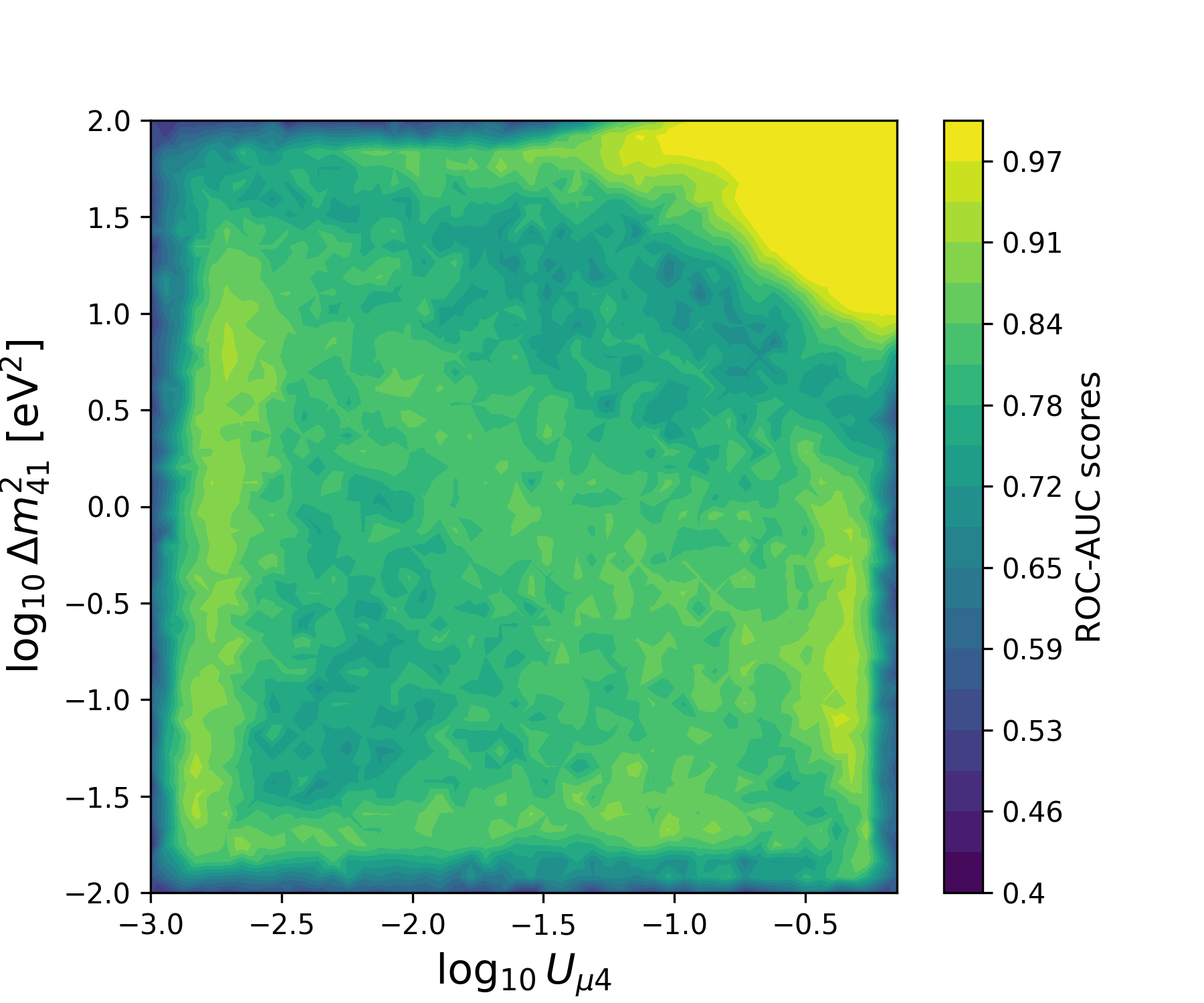}
        \label{fig:ccfr_roc-auc}
        \caption{CCFR}
    \end{subfigure}
    \begin{subfigure}[b]{0.45\textwidth}
        \includegraphics[width=\linewidth]{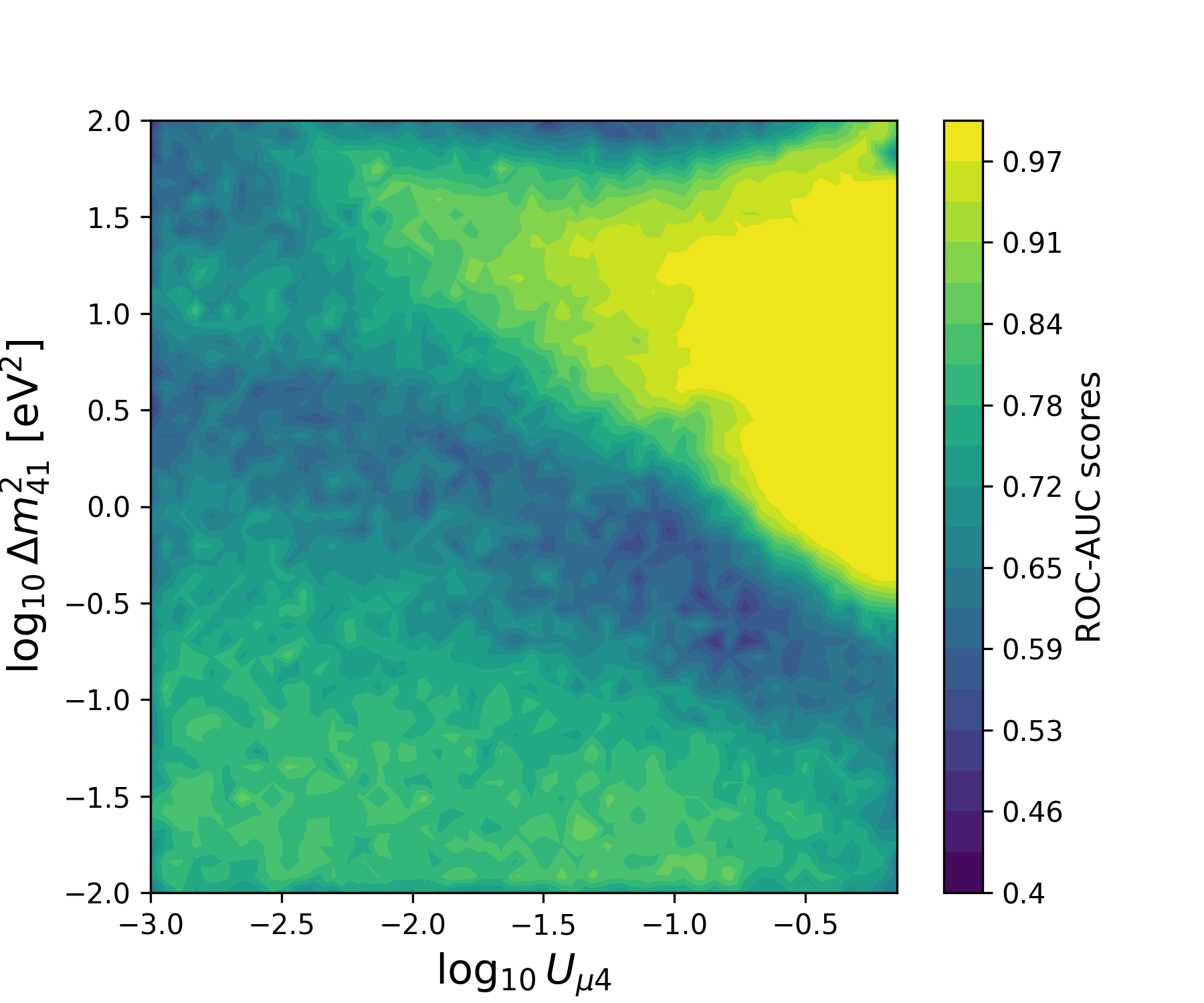}
        \label{fig:sbmb_roc-auc}
        \caption{SciBooNE-MiniBooNe $\nu_\mu$}
    \end{subfigure}
    \begin{subfigure}[b]{0.45\textwidth}
        \includegraphics[width=\linewidth]{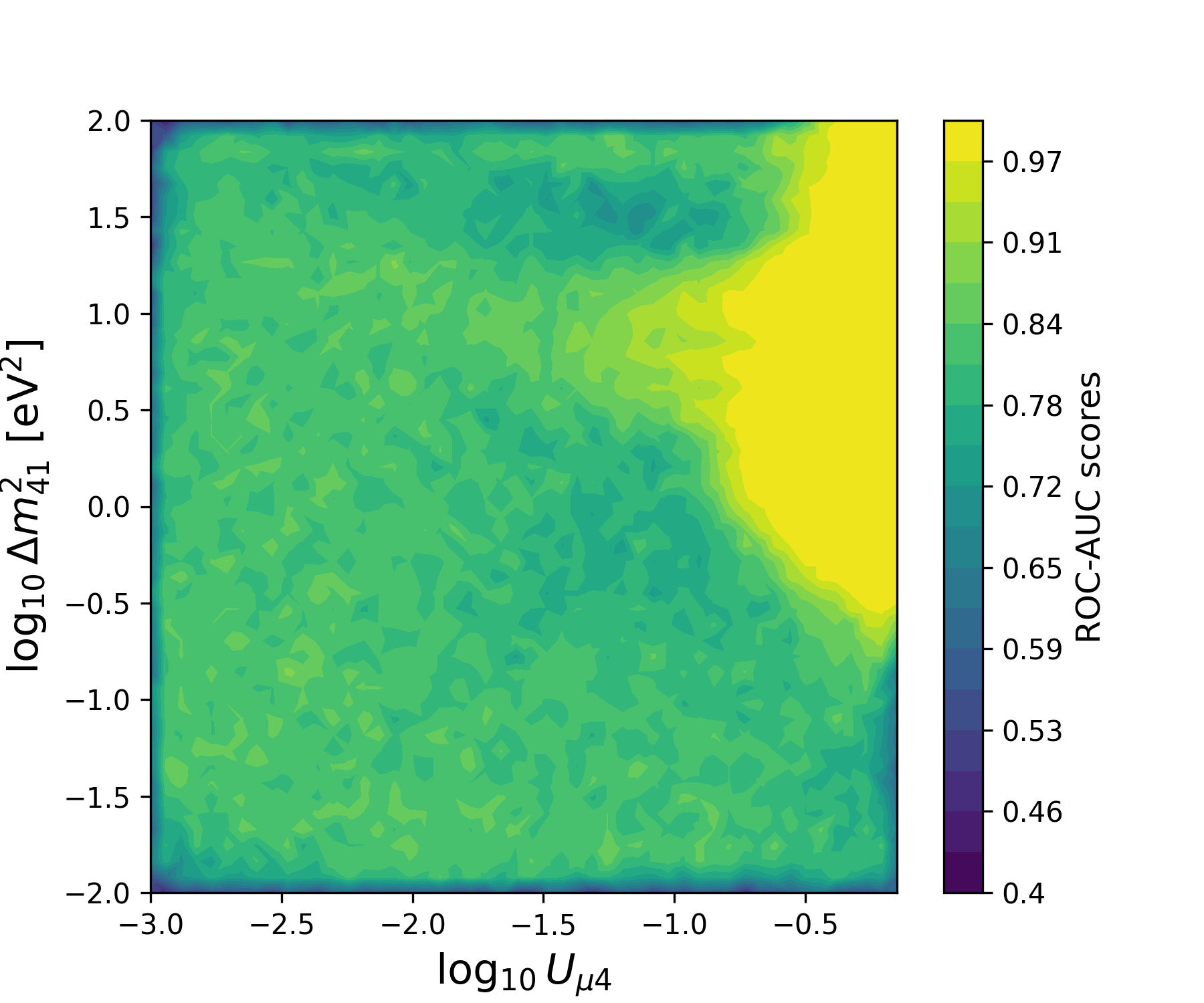}
        \label{fig:sbmbbar_roc-auc}
        \caption{SciBooNE-MiniBooNe $\bar{\nu}_\mu$}
    \end{subfigure}
    \begin{subfigure}[b]{0.45\textwidth}
        \includegraphics[width=\linewidth]{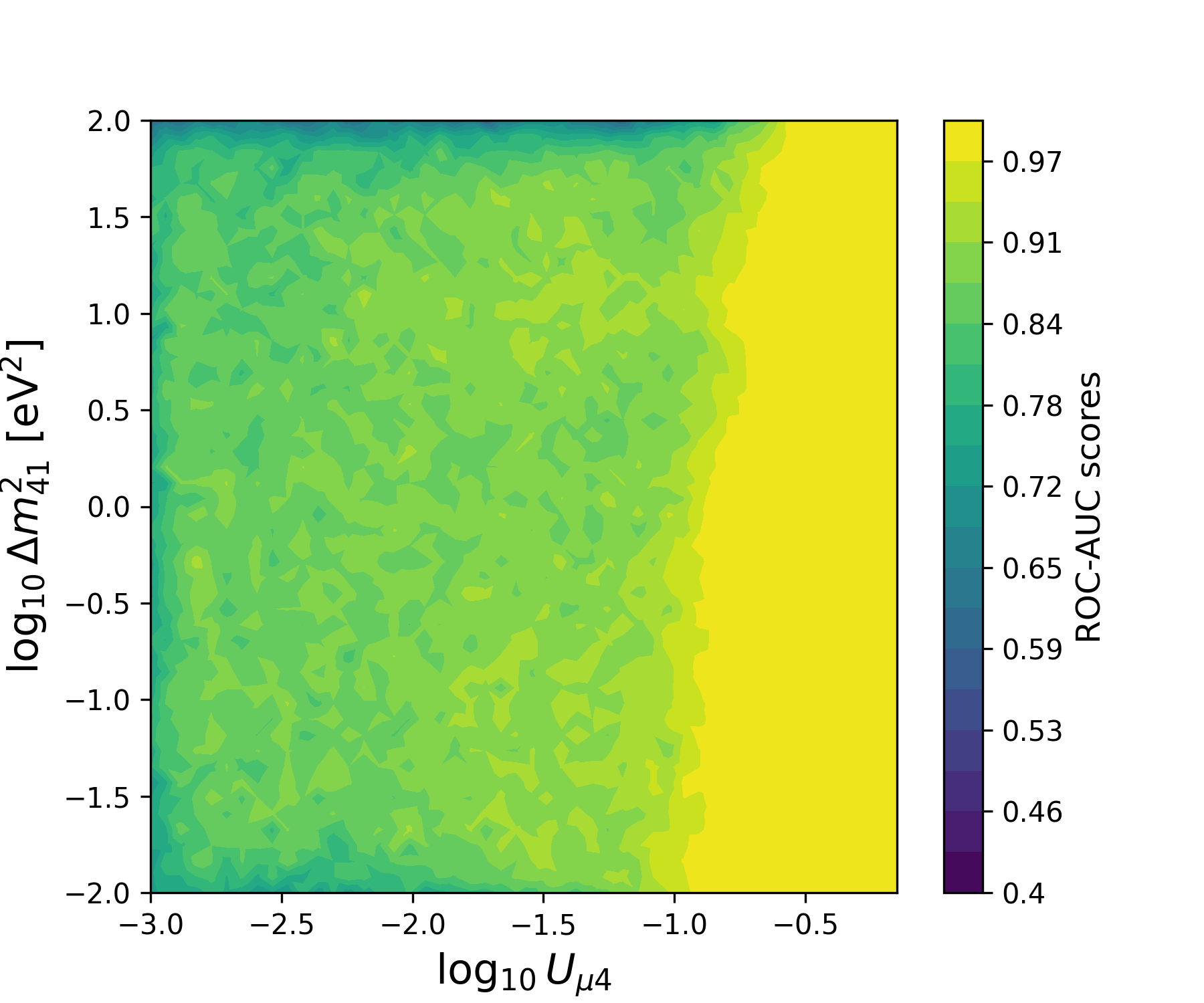}
        \label{fig:minos_roc-auc}
        \caption{MINOS}
    \end{subfigure}
    \begin{subfigure}[b]{0.45\textwidth}
        \includegraphics[width=\linewidth]{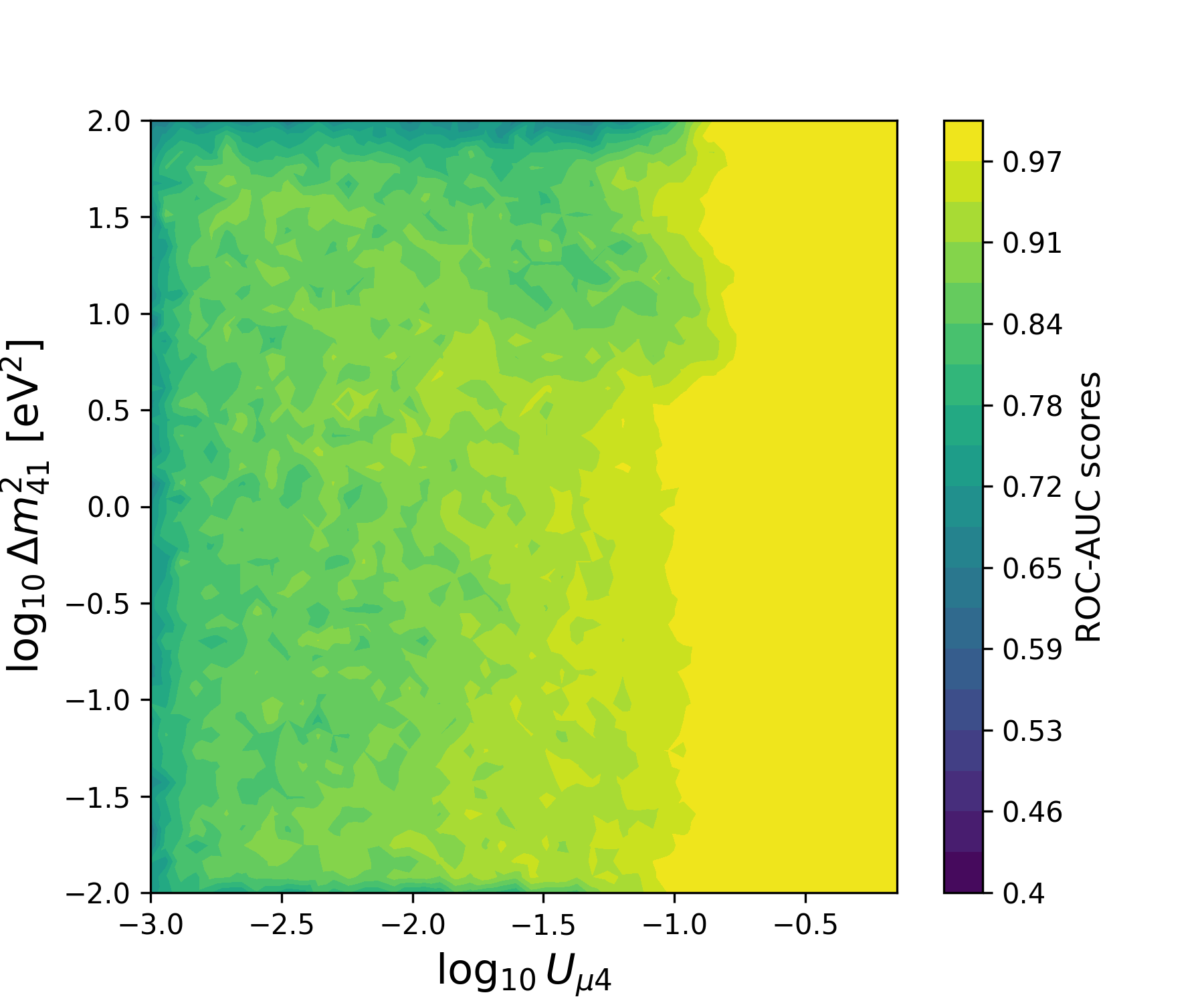}
        \label{fig:all_exp_roc-auc}
        \caption{All experiments}
    \end{subfigure}
    \caption{ROC-AUC score distributions for the DNRE trained on data.}
    \label{fig:roc-auc}
\end{figure}

\section{\label{sec:appB_coverages}Bayesian Coverages}
\begin{figure}[htbp]
    \centering
    \includegraphics[width=0.6\linewidth]{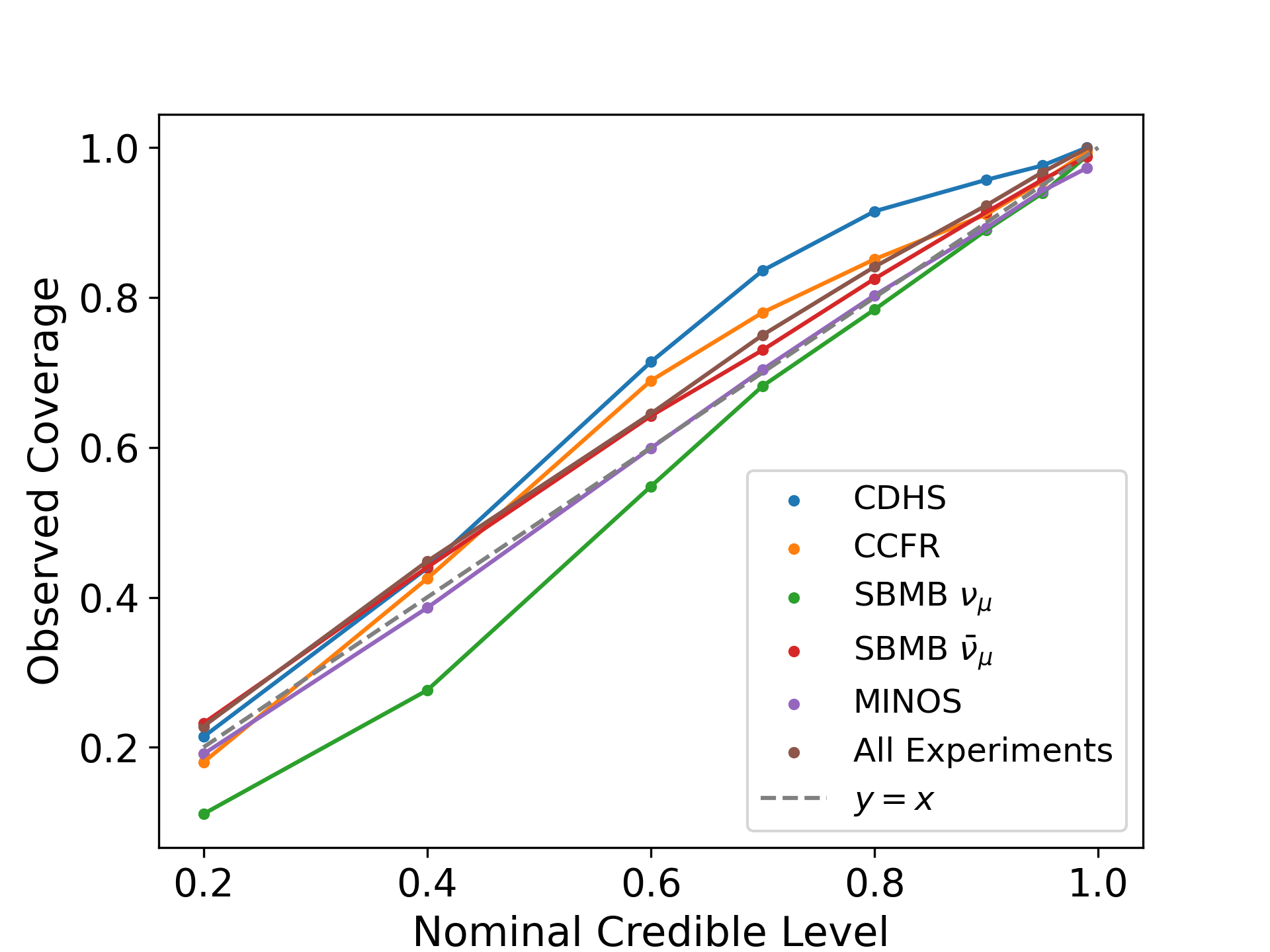}

    \caption{Observed vs Nominal Credible levels. $1000$ parameter points were randomly sampled to compute coverages. Optimal coverage falls along the gray $45 \degree$ line.}
    \label{fig:bayes_cov}
\end{figure}

In Figure \ref{fig:bayes_cov}, we display the Bayesian coverage of the estimated posterior distribution computed from Eq.~\ref{eq:posterior-estimate}. We observe  strong agreement between the nominal and observed credible levels.

\section{\label{appendixc:anscombe}Details of Direct Amortized Neural Likelihood Ratio Estimation on Anscombe's Quartet}

Here, we define priors on model parameters $m_i \sim \text{Unif} (-5, 5)$ and $b_i \sim \text{Unif} (-5,5)$, for $i \in 1, \dots 10000$. Given $m_i, b_i$, $\{x_j, y_j\}_{j=1}^{11} $ are gathered such that $x_j \sim \text{Unif}(0, 20)$, and $y_j | m_i, b_i \sim \mathcal{N} (m_i x_j + b_i, 1)$. The training strategy of Alg.~\ref{alg:dnre} was used for a neural network trained on BCE loss, having $3$ hidden layers with $64$ nodes each and ReLU activations. Batch size was held constant at $64$. Given that the Anscombe quartet shares a best-fit line at $y = 0.5x + 3$, we report the transformed test statistic $\mathbb{P} ((m, b) = (0.5, 3)) / \mathbb{P} ((m, b) = (0, 0))$ in Fig.~\ref{fig:sbi-anscombe}. Note that this test statistic is greatest for dataset I, where fluctuations are attributed to random (rather than systematic) effects. 

\section{\label{appendixc:icecube-sig}Relating Estimated Test Statistic to a Degree of Exclusion of IceCube}

In Sec.~\ref{subsec:icecube}, we report that our SBI+FC global fit analysis excludes the $3+1$ model for the IceCube best fit point $\theta^*_\text{IceCube}$ at $\sim 3\sigma$. To obtain this number, we use the trained neural likelihood ratio estimator to approximate a log-likelihood ratio $\hat{T} (\mathbf{x} | \theta^*_\text{IceCube}, \theta_0) = \log \mathbb{P} (\mathbf{x} | \theta^*_\text{IceCube}) - \log \mathbb{P} (\mathbf{x} | \theta_0) $, with $\theta_0$ parameters corresponding to no sterile mixing. We did not assume Wilks' theorem in obtaining the global exclusion reported in Fig.~\ref{fig:all_exp_real_data}, instead using the trials-based method outlined in Sec.~\ref{subsec:freq-cls}. In order to convert the estimated test statistic into a form with which we can draw comparison to IceCube's preference for a $3+1$ sterile neutrino, we temporarily assume Wilks' theorem to relate $\hat{T} (\mathbf{x} | \theta^*_\text{IceCube}, \theta_0)$ to a $\chi^2$ distribution with two degrees of freedom, from which we can assign a $p$-value and an associated degree of uncertainty.

%% file: main.bbl
\providecommand{\newblock}{}
\begin{thebibliography}{10}
\expandafter\ifx\csname url\endcsname\relax
  \def\url#1{{\tt #1}}\fi
\expandafter\ifx\csname urlprefix\endcsname\relax\def\urlprefix{URL }\fi
\providecommand{\eprint}[2][]{\url{#2}}

\bibitem{Wilks1938-yc}
Wilks S~S 1938 {\em Ann. Math. Stat.\/} {\bf 9} 60--62

\bibitem{Hardin:2022qdh}
Hardin J~M 2024 {\em Eur. J. Phys.\/} {\bf 45} 025806 (\textit{Preprint} \eprint{2211.06347})

\bibitem{Algeri2020-gv}
Algeri S, Aalbers J, Mor{\aa} K~D and Conrad J 2020 {\em Nat. Rev. Phys.\/} {\bf 2} 245--252

\bibitem{feldman-cousins}
Feldman G~J and Cousins R~D 1998 {\em Phys. Rev. D\/} {\bf 57}(7) 3873--3889 \urlprefix\url{https://link.aps.org/doi/10.1103/PhysRevD.57.3873}

\bibitem{icecube-sterile-prl}
Abbasi R, Ackermann M, Adams J, Agarwalla S~K, Aguilar J~A, Ahlers M, Alameddine J~M, Amin N~M, Andeen K, Arg\"uelles C, Ashida Y, Athanasiadou S, Ausborm L, Axani S~N, Bai X, Balagopal~V A, Baricevic M, Barwick S~W, Bash S, Basu V, Bay R, Beatty J~J, Becker~Tjus J, Beise J, Bellenghi C, Benning C, BenZvi S, Berley D, Bernardini E, Besson D~Z, Blaufuss E, Bloom L, Blot S, Bontempo F, Book~Motzkin J~Y, Boscolo~Meneguolo C, B\"oser S, Botner O, B\"ottcher J, Braun J, Brinson B, Brostean-Kaiser J, Brusa L, Burley R~T, Butterfield D, Campana M~A, Caracas I, Carloni K, Carpio J, Chattopadhyay S, Chau N, Chen Z, Chirkin D, Choi S, Clark B~A, Coleman A, Collin G~H, Connolly A, Conrad J~M, Coppin P, Corley R, Correa P, Cowen D~F, Dave P, De~Clercq C, DeLaunay J~J, Delgado D, Deng S, Desai A, Desiati P, de~Vries K~D, de~Wasseige G, Diaz A, D\'{\i}az-V\'elez J~C, Dierichs P, Dittmer M, Domi A, Draper L, Dujmovic H, Dutta K, DuVernois M~A, Ehrhardt T, Eidenschink L, Eimer A, Eller P, Ellinger E, El~Mentawi S, Els\"asser
  D, Engel R, Erpenbeck H, Evans J, Evenson P~A, Fan K~L, Fang K, Farrag K, Fazely A~R, Fedynitch A, Feigl N, Fiedlschuster S, Finley C, Fischer L, Fox D, Franckowiak A, Fukami S, F\"urst P, Gallagher J, Ganster E, Garcia A, Garcia M, Garg G, Genton E, Gerhardt L, Ghadimi A, Girard-Carillo C, Glaser C, Gl\"usenkamp T, Gonzalez J~G, Goswami S, Granados A, Grant D, Gray S~J, Gries O, Griffin S, Griswold S, Groth K~M, G\"unther C, Gutjahr P, Ha C, Haack C, Hallgren A, Halve L, Halzen F, Hamdaoui H, Ha~Minh M, Handt M, Hanson K, Hardin J, Harnisch A~A, Hatch P, Haungs A, H\"au\ss{}ler J, Helbing K, Hellrung J, Hermannsgabner J, Heuermann L, Heyer N, Hickford S, Hidvegi A, Hill C, Hill G~C, Hoffman K~D, Hori S, Hoshina K, Hostert M, Hou W, Huber T, Hultqvist K, H\"unnefeld M, Hussain R, Hymon K, Ishihara A, Iwakiri W, Jacquart M, Janik O, Jansson M, Japaridze G~S, Jeong M, Jin M, Jones B~J~P, Kamp N, Kang D, Kang W, Kang X, Kappes A, Kappesser D, Kardum L, Karg T, Karl M, Karle A, Katil A, Katz U, Kauer M, Kelley
  J~L, Khanal M, Khatee~Zathul A, Kheirandish A, Kiryluk J, Klein S~R, Kochocki A, Koirala R, Kolanoski H, Kontrimas T, K\"opke L, Kopper C, Koskinen D~J, Koundal P, Kovacevich M, Kowalski M, Kozynets T, Krishnamoorthi J, Kruiswijk K, Krupczak E, Kumar A, Kun E, Kurahashi N, Lad N, Lagunas~Gualda C, Lamoureux M, Larson M~J, Latseva S, Lauber F, Lazar J~P, Lee J~W, Leonard~DeHolton K, Leszczy\ifmmode~\acute{n}\else \'{n}\fi{}ska A, Liao J, Lincetto M, Liu Y~T, Liubarska M, Lohfink E, Love C, Lozano~Mariscal C~J, Lu L, Lucarelli F, Luszczak W, Lyu Y, Madsen J, Magnus E, Mahn K~B~M, Makino Y, Manao E, Mancina S, Marie~Sainte W, Mari\ifmmode~\mbox{\c{s}}\else \c{s}\fi{} I~C, Marka S, Marka Z, Marsee M, Martinez-Soler I, Maruyama R, Mayhew F, McNally F, Mead J~V, Meagher K, Mechbal S, Medina A, Meier M, Merckx Y, Merten L, Micallef J, Mitchell J, Montaruli T, Moore R~W, Morii Y, Morse R, Moulai M, Mukherjee T, Naab R, Nagai R, Nakos M, Naumann U, Necker J, Negi A, Neste L, Neumann M, Niederhausen H, Nisa M~U, Noda
  K, Noell A, Novikov A, Obertacke~Pollmann A, O'Dell V, Oeyen B, Olivas A, Orsoe R, Osborn J, O'Sullivan E, Pandya H, Park N, Parker G~K, Paudel E~N, Paul L, P\'erez de~los Heros C, Pernice T, Peterson J, Philippen S, Pizzuto A, Plum M, Pont\'en A, Popovych Y, Prado~Rodriguez M, Pries B, Procter-Murphy R, Przybylski G~T, Raab C, Rack-Helleis J, Ravn M, Rawlins K, Rechav Z, Rehman A, Reichherzer P, Resconi E, Reusch S, Rhode W, Riedel B, Rifaie A, Roberts E~J, Robertson S, Rodan S, Roellinghoff G, Rongen M, Rosted A, Rott C, Ruhe T, Ruohan L, Ryckbosch D, Safa I, Saffer J, Salazar-Gallegos D, Sampathkumar P, Sandrock A, Santander M, Sarkar S, Sarkar S, Savelberg J, Savina P, Schaile P, Schaufel M, Schieler H, Schindler S, Schl\"uter B, Schl\"uter F, Schmeisser N, Schmidt T, Schneider J, Schr\"oder F~G, Schumacher L, Sclafani S, Seckel D, Seikh M, Seo M, Seunarine S, Sevle~Myhr P, Shah R, Shefali S, Shimizu N, Silva M, Skrzypek B, Smithers B, Snihur R, Soedingrekso J, S\o{}gaard A, Soldin D, Soldin P, Sommani
  G, Spannfellner C, Spiczak G~M, Spiering C, Sponsler C, Stamatikos M, Stanev T, Stezelberger T, St\"urwald T, Stuttard T, Sullivan G~W, Taboada I, Ter-Antonyan S, Terliuk A, Thiesmeyer M, Thompson W~G, Thwaites J, Tilav S, Tollefson K, T\"onnis C, Toscano S, Tosi D, Trettin A, Turcotte R, Twagirayezu J~P, Unland~Elorrieta M~A, Upadhyay A~K, Upshaw K, Vaidyanathan A, Valtonen-Mattila N, Vandenbroucke J, van Eijndhoven N, Vannerom D, van Santen J, Vara J, Veitch-Michaelis J, Venugopal M, Vereecken M, Verpoest S, Veske D, Vijai A, Walck C, Wang A, Weaver C, Weigel P, Weindl A, Weldert J, Wen A~Y, Wendt C, Werthebach J, Weyrauch M, Whitehorn N, Wiebusch C~H, Williams D~R, Witthaus L, Wolf A, Wolf M, Wrede G, Xu X~W, Yanez J~P, Yildizci E, Yoshida S, Young R, Yu S, Yuan T, Zhang Z, Zhelnin P, Zilberman P and Zimmerman M (IceCube Collaboration) 2024 {\em Phys. Rev. Lett.\/} {\bf 133}(20) 201804 \urlprefix\url{https://link.aps.org/doi/10.1103/PhysRevLett.133.201804}

\bibitem{wherearewe}
Arguelles~Delgado C, Collin G, Conrad J and Shaevitz M 2020 {\em Physics Reports\/} {\bf 884} 1--59

\bibitem{DYDAK1984281}
Dydak F, Feldman G, Guyot C, Merlo J, Meyer H~J, Rothberg J, Steinberger J, Taureg H, {von Rüden} W, Wachsmuth H, Wahl H, Wotschack J, Blümer H, Buchholz P, Duda J, Eisele F, Kleinknecht K, Knobloch J, Pszola B, Renk B, Belusevic R, Falkenburg B, Flottmann T, {de Groot} J, Geweniger C, Keilwerth H, Tittel K, Debu P, Para A, Perez P, Peyaud B, Rander J, Schuller J, Turlay R, Abramowicz H and Królikowski J 1984 {\em Physics Letters B\/} {\bf 134} 281--286 ISSN 0370-2693 \urlprefix\url{https://www.sciencedirect.com/science/article/pii/0370269384906889}

\bibitem{Stockdale:1984ce}
Stockdale I~E {\em et~al.\/} 1984 {\em Conf. Proc. C\/} {\bf 841031} 258

\bibitem{minos-two-detector}
Adamson P, Anghel I, Aurisano A, Barr G, Bishai M, Blake A, Bock G, Bogert D, Cao S, Carroll T, Castromonte C, Chen R, Childress S, Coelho J, Corwin L, Cronin-Hennessy D, de~Jong J, De~Rijck S, Devan A, Devenish N, Diwan M, Escobar C, Evans J, Falk E, Feldman G, Flanagan W, Frohne M, Gabrielyan M, Gallagher H, Germani S, Gomes R, Goodman M, Gouffon P, Graf N, Gran R, Grzelak K, Habig A, Hahn S, Hartnell J, Hatcher R, Holin A, Huang J, Hylen J, Irwin G, Isvan Z, James C, Jensen D, Kafka T, Kasahara S, Koerner L, Koizumi G, Kordosky M, Kreymer A, Lang K, Ling J, Litchfield P, Lucas P, Mann W, Marshak M, Mayer N, McGivern C, Medeiros M, Mehdiyev R, Meier J, Messier M, Miller W, Mishra S, Moed~Sher S, Moore C, Mualem L, Musser J, Naples D, Nelson J, Newman H, Nichol R, Nowak J, O’Connor J, Orchanian M, Pahlka R, Paley J, Patterson R, Pawloski G, Perch A, Pfützner M, Phan D, Phan-Budd S, Plunkett R, Poonthottathil N, Qiu X, Radovic A, Rebel B, Rosenfeld C, Rubin H, Sail P, Sanchez M, Schneps J, Schreckenberger
  A, Schreiner P, Sharma R, Sousa A, Tagg N, Talaga R, Thomas J, Thomson M, Tian X, Timmons A, Todd J, Tognini S, Toner R, Torretta D, Tzanakos G, Urheim J, Vahle P, Viren B, Weber A, Webb R, White C, Whitehead L, Wojcicki S and Zwaska R 2019 {\em Physical Review Letters\/} {\bf 122} ISSN 1079-7114 \urlprefix\url{http://dx.doi.org/10.1103/PhysRevLett.122.091803}

\bibitem{PhysRevD.85.032007}
Mahn K~B~M, Nakajima Y, Aguilar-Arevalo A~A, Alcaraz-Aunion J~L, Anderson C~E, Bazarko A~O, Brice S~J, Brown B~C, Bugel L, Cao J, Catala-Perez J, Cheng G, Coney L, Conrad J~M, Cox D~C, Curioni A, Dharmapalan R, Djurcic Z, Dore U, Finley D~A, Fleming B~T, Ford R, Franke A~J, Garcia F~G, Garvey G~T, Giganti C, Gomez-Cadenas J~J, Grange J, Green C, Green J~A, Guzowski P, Hanson A, Hart T~L, Hawker E, Hayato Y, Hiraide K, Huelsnitz W, Imlay R, Johnson R~A, Jones B~J~P, Jover-Manas G, Karagiorgi G, Kasper P, Katori T, Kobayashi Y~K, Kobilarcik T, Kourbanis I, Koutsoliotas S, Kubo H, Kurimoto Y, Laird E~M, Linden S~K, Link J~M, Liu Y, Liu Y, Louis W~C, Loverre P~F, Ludovici L, Mariani C, Marsh W, Masuike S, Matsuoka K, Mauger C, McGary V~T, McGregor G, Metcalf W, Meyers P~D, Mills F, Mills G~B, Mitsuka G, Miyachi Y, Mizugashira S, Monroe J, Moore C~D, Mousseau J, Nakaya T, Napora R, Nelson R~H, Nienaber P, Nowak J~A, Orme D, Osmanov B, Otani M, Ouedraogo S, Patterson R~B, Pavlovic Z, Perevalov D, Polly C~C, Prebys
  E, Raaf J~L, Ray H, Roe B~P, Russell A~D, Sanchez F, Sandberg V, Schirato R, Schmitz D, Shaevitz M~H, Shibata T~A, Shoemaker F~C, Smith D, Soderberg M, Sorel M, Spentzouris P, Spitz J, Stancu I, Stefanski R~J, Sung M, Takei H, Tanaka H~A, Tanaka H~K, Tanaka M, Tayloe R, Taylor I~J, Tesarek R~J, Tzanov M, Uchida Y, Van~de Water R, Walding J~J, Wascko M~O, White D~H, White H~B, Wilking M~J, Yokoyama M, Yang H~J, Zeller G~P and Zimmerman E~D (MiniBooNE and SciBooNE Collaborations) 2012 {\em Phys. Rev. D\/} {\bf 85}(3) 032007 \urlprefix\url{https://link.aps.org/doi/10.1103/PhysRevD.85.032007}

\bibitem{PhysRevD.86.052009}
Cheng G, Huelsnitz W, Aguilar-Arevalo A~A, Alcaraz-Aunion J~L, Brice S~J, Brown B~C, Bugel L, Catala-Perez J, Church E~D, Conrad J~M, Dharmapalan R, Djurcic Z, Dore U, Finley D~A, Ford R, Franke A~J, Garcia F~G, Garvey G~T, Giganti C, Gomez-Cadenas J~J, Grange J, Guzowski P, Hanson A, Hayato Y, Hiraide K, Ignarra C, Imlay R, Johnson R~A, Jones B~J~P, Jover-Manas G, Karagiorgi G, Katori T, Kobayashi Y~K, Kobilarcik T, Kubo H, Kurimoto Y, Louis W~C, Loverre P~F, Ludovici L, Mahn K~B~M, Mariani C, Marsh W, Masuike S, Matsuoka K, McGary V~T, Metcalf W, Mills G~B, Mirabal J, Mitsuka G, Miyachi Y, Mizugashira S, Moore C~D, Mousseau J, Nakajima Y, Nakaya T, Napora R, Nienaber P, Orme D, Osmanov B, Otani M, Pavlovic Z, Perevalov D, Polly C~C, Ray H, Roe B~P, Russell A~D, Sanchez F, Shaevitz M~H, Shibata T~A, Sorel M, Spitz J, Stancu I, Stefanski R~J, Takei H, Tanaka H~K, Tanaka M, Tayloe R, Taylor I~J, Tesarek R~J, Uchida Y, Van~de Water R~G, Walding J~J, Wascko M~O, White D~H, White H~B, Wickremasinghe D~A, Yokoyama
  M, Zeller G~P and Zimmerman E~D (MiniBooNE and SciBooNE Collaborations) 2012 {\em Phys. Rev. D\/} {\bf 86}(5) 052009 \urlprefix\url{https://link.aps.org/doi/10.1103/PhysRevD.86.052009}

\bibitem{icecube-sterile-prd}
Abbasi R, Ackermann M, Adams J, Agarwalla S~K, Aguilar J~A, Ahlers M, Alameddine J~M, Amin N~M, Andeen K, Arg\"uelles C, Ashida Y, Athanasiadou S, Ausborm L, Axani S~N, Bai X, Balagopal A, Baricevic M, Barwick S~W, Bash S, Basu V, Bay R, Beatty J~J, Becker~Tjus J, Beise J, Bellenghi C, Benning C, BenZvi S, Berley D, Bernardini E, Besson D~Z, Blaufuss E, Bloom L, Blot S, Bontempo F, Book~Motzkin J~Y, Boscolo~Meneguolo C, B\"oser S, Botner O, B\"ottcher J, Braun J, Brinson B, Brostean-Kaiser J, Brusa L, Burley R~T, Butterfield D, Campana M~A, Caracas I, Carloni K, Carpio J, Chattopadhyay S, Chau N, Chen Z, Chirkin D, Choi S, Clark B~A, Coleman A, Collin G~H, Connolly A, Conrad J~M, Coppin P, Corley R, Correa P, Cowen D~F, Dave P, De~Clercq C, DeLaunay J~J, Delgado D, Deng S, Desai A, Desiati P, de~Vries K~D, de~Wasseige G, Diaz A, D\'{\i}az-V\'elez J~C, Dierichs P, Dittmer M, Domi A, Draper L, Dujmovic H, Dutta K, DuVernois M~A, Ehrhardt T, Eidenschink L, Eimer A, Eller P, Ellinger E, El~Mentawi S, Els\"asser
  D, Engel R, Erpenbeck H, Evans J, Evenson P~A, Fan K~L, Fang K, Farrag K, Fazely A~R, Fedynitch A, Feigl N, Fiedlschuster S, Finley C, Fischer L, Fox D, Franckowiak A, Fukami S, F\"urst P, Gallagher J, Ganster E, Garcia A, Garcia M, Garg G, Genton E, Gerhardt L, Ghadimi A, Girard-Carillo C, Glaser C, Gl\"usenkamp T, Gonzalez J~G, Goswami S, Granados A, Grant D, Gray S~J, Gries O, Griffin S, Griswold S, Groth K~M, G\"unther C, Gutjahr P, Ha C, Haack C, Hallgren A, Halve L, Halzen F, Hamdaoui H, Ha~Minh M, Handt M, Hanson K, Hardin J, Harnisch A~A, Hatch P, Haungs A, H\"au\ss{}ler J, Helbing K, Hellrung J, Hermannsgabner J, Heuermann L, Heyer N, Hickford S, Hidvegi A, Hill C, Hill G~C, Hoffman K~D, Hori S, Hoshina K, Hostert M, Hou W, Huber T, Hultqvist K, H\"unnefeld M, Hussain R, Hymon K, Ishihara A, Iwakiri W, Jacquart M, Janik O, Jansson M, Japaridze G~S, Jeong M, Jin M, Jones B~J~P, Kamp N, Kang D, Kang W, Kang X, Kappes A, Kappesser D, Kardum L, Karg T, Karl M, Karle A, Katil A, Katz U, Kauer M, Kelley
  J~L, Khanal M, Khatee~Zathul A, Kheirandish A, Kiryluk J, Klein S~R, Kochocki A, Koirala R, Kolanoski H, Kontrimas T, K\"opke L, Kopper C, Koskinen D~J, Koundal P, Kovacevich M, Kowalski M, Kozynets T, Krishnamoorthi J, Kruiswijk K, Krupczak E, Kumar A, Kun E, Kurahashi N, Lad N, Lagunas~Gualda C, Lamoureux M, Larson M~J, Latseva S, Lauber F, Lazar J~P, Lee J~W, Leonard~DeHolton K, Leszczy\ifmmode~\acute{n}\else \'{n}\fi{}ska A, Liao J, Lincetto M, Liu Y~T, Liubarska M, Lohfink E, Love C, Lozano~Mariscal C~J, Lu L, Lucarelli F, Luszczak W, Lyu Y, Madsen J, Magnus E, Mahn K~B~M, Makino Y, Manao E, Mancina S, Marie~Sainte W, Mari\ifmmode~\mbox{\c{s}}\else \c{s}\fi{} I~C, Marka S, Marka Z, Marsee M, Martinez-Soler I, Maruyama R, Mayhew F, McNally F, Mead J~V, Meagher K, Mechbal S, Medina A, Meier M, Merckx Y, Merten L, Micallef J, Mitchell J, Montaruli T, Moore R~W, Morii Y, Morse R, Moulai M, Mukherjee T, Naab R, Nagai R, Nakos M, Naumann U, Necker J, Negi A, Neste L, Neumann M, Niederhausen H, Nisa M~U, Noda
  K, Noell A, Novikov A, Obertacke~Pollmann A, O'Dell V, Oeyen B, Olivas A, Orsoe R, Osborn J, O'Sullivan E, Pandya H, Park N, Parker G~K, Paudel E~N, Paul L, P\'erez de~los Heros C, Pernice T, Peterson J, Philippen S, Pizzuto A, Plum M, Pont\'en A, Popovych Y, Prado~Rodriguez M, Pries B, Procter-Murphy R, Przybylski G~T, Raab C, Rack-Helleis J, Ravn M, Rawlins K, Rechav Z, Rehman A, Reichherzer P, Resconi E, Reusch S, Rhode W, Riedel B, Rifaie A, Roberts E~J, Robertson S, Rodan S, Roellinghoff G, Rongen M, Rosted A, Rott C, Ruhe T, Ruohan L, Ryckbosch D, Safa I, Saffer J, Salazar-Gallegos D, Sampathkumar P, Sandrock A, Santander M, Sarkar S, Sarkar S, Savelberg J, Savina P, Schaile P, Schaufel M, Schieler H, Schindler S, Schl\"uter B, Schl\"uter F, Schmeisser N, Schmidt T, Schneider J, Schr\"oder F~G, Schumacher L, Sclafani S, Seckel D, Seikh M, Seo M, Seunarine S, Sevle~Myhr P, Shah R, Shefali S, Shimizu N, Silva M, Skrzypek B, Smithers B, Snihur R, Soedingrekso J, S\o{}gaard A, Soldin D, Soldin P, Sommani
  G, Spannfellner C, Spiczak G~M, Spiering C, Sponsler C, Stamatikos M, Stanev T, Stezelberger T, St\"urwald T, Stuttard T, Sullivan G~W, Taboada I, Ter-Antonyan S, Terliuk A, Thiesmeyer M, Thompson W~G, Thwaites J, Tilav S, Tollefson K, T\"onnis C, Toscano S, Tosi D, Trettin A, Turcotte R, Twagirayezu J~P, Unland~Elorrieta M~A, Upadhyay A~K, Upshaw K, Vaidyanathan A, Valtonen-Mattila N, Vandenbroucke J, van Eijndhoven N, Vannerom D, van Santen J, Vara J, Veitch-Michaelis J, Venugopal M, Vereecken M, Verpoest S, Veske D, Vijai A, Walck C, Wang A, Weaver C, Weigel P, Weindl A, Weldert J, Wen A~Y, Wendt C, Werthebach J, Weyrauch M, Whitehorn N, Wiebusch C~H, Williams D~R, Witthaus L, Wolf A, Wolf M, Wrede G, Xu X~W, Yanez J~P, Yildizci E, Yoshida S, Young R, Yu S, Yuan T, Zhang Z, Zhelnin P, Zilberman P and Zimmerman M (IceCube Collaboration) 2024 {\em Phys. Rev. D\/} {\bf 110}(9) 092009 \urlprefix\url{https://link.aps.org/doi/10.1103/PhysRevD.110.092009}

\bibitem{doi:10.1073/pnas.1912789117}
Cranmer K, Brehmer J and Louppe G 2020 {\em Proceedings of the National Academy of Sciences\/} {\bf 117} 30055--30062 (\textit{Preprint} \eprint{https://www.pnas.org/doi/pdf/10.1073/pnas.1912789117}) \urlprefix\url{https://www.pnas.org/doi/abs/10.1073/pnas.1912789117}

\bibitem{Hardin2023}
Hardin J~M, Martinez-Soler I, Diaz A, Jin M, Kamp N~W, Arg{\"u}elles C~A, Conrad J~M and Shaevitz M~H 2023 {\em Journal of High Energy Physics\/} {\bf 2023} 58 ISSN 1029-8479 \urlprefix\url{https://doi.org/10.1007/JHEP09(2023)058}

\bibitem{10.1609/aaai.v38i18.30018}
Cobb A~D, Matejek B, Elenius D, Roy A and Jha S 2024 Direct amortized likelihood ratio estimation {\em Proceedings of the Thirty-Eighth AAAI Conference on Artificial Intelligence and Thirty-Sixth Conference on Innovative Applications of Artificial Intelligence and Fourteenth Symposium on Educational Advances in Artificial Intelligence\/} AAAI'24/IAAI'24/EAAI'24 (AAAI Press) ISBN 978-1-57735-887-9 \urlprefix\url{https://doi.org/10.1609/aaai.v38i18.30018}

\bibitem{4d625c73-e285-3ead-ae1f-415e8ad60311}
Cook S~R, Gelman A and Rubin D~B 2006 {\em Journal of Computational and Graphical Statistics\/} {\bf 15} 675--692 ISSN 10618600 \urlprefix\url{http://www.jstor.org/stable/27594203}

\bibitem{Talts:2018zdk}
Talts S, Betancourt M, Simpson D, Vehtari A and Gelman A 2018  (\textit{Preprint} \eprint{1804.06788})

\bibitem{toygithub}
\urlprefix\url{https://github.com/joshvillarreal/freq-sbi-toybox}

\bibitem{Neyman1933-uw}
Neyman J and Pearson E~S 1933 {\em Philos. Trans. R. Soc. Lond.\/} {\bf 231} 289--337

\bibitem{Ignarra:2014yqa}
Ignarra C~M 2014 {\em {Sterile Neutrino Searches in MiniBooNE and MicroBooNE}\/} Ph.D. thesis MIT, Cambridge, Dept. Phys.

\bibitem{GaryChengThesis}
Cheng G~C~L 2013 {\em {Precision Search for Muon Antineutrino Disappearance Oscillations Using a Dual Baseline Technique}\/} Ph.D. thesis Columbia U.

\bibitem{Anscombe1973-lj}
Anscombe F~J 1973 {\em Am. Stat.\/} {\bf 27} 17--21

\bibitem{Conrad:2016sve}
Conrad J~M and Shaevitz M~H 2018 {\em Adv. Ser. Direct. High Energy Phys.\/} {\bf 28} 391--442 (\textit{Preprint} \eprint{1609.07803})

\bibitem{Mikheyev:1985zog}
Mikheyev S~P and Smirnov A~Y 1985 {\em Sov. J. Nucl. Phys.\/} {\bf 42} 913--917

\bibitem{Mikheev:1986wj}
Mikheev S~P and Smirnov A~Y 1986 {\em Nuovo Cim. C\/} {\bf 9} 17--26

\bibitem{MSWPhysRevD.17.2369}
Wolfenstein L 1978 {\em Phys. Rev. D\/} {\bf 17}(9) 2369--2374 \urlprefix\url{https://link.aps.org/doi/10.1103/PhysRevD.17.2369}

\end{thebibliography}
